\RequirePackage[l2tabu,orthodox]{nag}
\documentclass
[letterpaper,12pt,]
{article}

\usepackage{etex}
\usepackage{verbatim}
\usepackage{xspace,enumerate}
\usepackage[dvipsnames]{xcolor}
\usepackage[T1]{fontenc}
\usepackage[full]{textcomp}
\usepackage[american]{babel}
\usepackage{mathtools}
\usepackage{amsthm}
\usepackage[
letterpaper,
top=1in,
bottom=1in,
left=1in,
right=1in]{geometry}
\usepackage{newpxtext} % T1, lining figures in math, osf in text
\usepackage{textcomp} % required for special glyphs
\usepackage[varg,bigdelims]{newpxmath}
\usepackage[scr=rsfso]{mathalfa}% \mathscr is fancier than \mathcal
\usepackage{bm} % load after all math to give access to bold math
\let\mathbb\varmathbb
\usepackage{microtype}
\usepackage[pagebackref,colorlinks=true,urlcolor=blue,linkcolor=blue,citecolor=OliveGreen]{hyperref}
\usepackage[capitalise,nameinlink]{cleveref}
\crefname{lemma}{Lemma}{Lemmas}
\crefname{fact}{Fact}{Facts}
\crefname{theorem}{Theorem}{Theorems}
\crefname{corollary}{Corollary}{Corollaries}
\crefname{claim}{Claim}{Claims}
\crefname{example}{Example}{Examples}
\crefname{algorithm}{Algorithm}{Algorithms}
\crefname{problem}{Problem}{Problems}
\crefname{definition}{Definition}{Definitions}
\crefname{exercise}{Exercise}{Exercises}
\usepackage{amsthm}

\newtheorem{theorem}{Theorem}[section]
\newtheorem*{theorem*}{Theorem}
\newtheorem{lemma}[theorem]{Lemma}
\newtheorem*{lemma*}{Lemma}
\newtheorem{fact}[theorem]{Fact}
\newtheorem*{fact*}{Fact}
\newtheorem{proposition}[theorem]{Proposition}
\newtheorem*{proposition*}{Proposition}
\newtheorem{corollary}[theorem]{Corollary}
\newtheorem*{corollary*}{Corollary}

\newtheorem*{hypothesis*}{Hypothesis}

\newtheorem*{conjecture*}{Conjecture}
\theoremstyle{definition}
\newtheorem{definition}[theorem]{Definition}
\newtheorem*{definition*}{Definition}

\newtheorem*{construction*}{Construction}

\newtheorem*{example*}{Example}

\newtheorem*{question*}{Question}
\newtheorem{algorithm}[theorem]{Algorithm}
\newtheorem*{algorithm*}{Algorithm}

\newtheorem*{assumption*}{Assumption}

\newtheorem*{problem*}{Problem}

\newtheorem*{openquestion*}{Open Question}

\newtheorem*{model*}{Model}
\theoremstyle{remark}

\newtheorem*{claim*}{Claim}
\newtheorem{remark}[theorem]{Remark}
\newtheorem*{remark*}{Remark}

\newtheorem*{observation*}{Observation}
\usepackage{paralist}
\let\originalleft\left
\let\originalright\right
\renewcommand{\left}{\mathopen{}\mathclose\bgroup\originalleft}
\renewcommand{\right}{\aftergroup\egroup\originalright}
\usepackage{turnstile}
\usepackage{mdframed}
\usepackage{tikz}
\usepackage{caption}
\DeclareCaptionType{Algorithm}
\usepackage{newfloat}
\usepackage{xparse}
\usepackage{amsthm} % for \@addpunct
\usetikzlibrary{positioning}
\makeatletter
\let\latexparagraph\paragraph
\RenewDocumentCommand{\paragraph}{som}{%
	\IfBooleanTF{#1}
	{\latexparagraph*{#3}}
	{\IfNoValueTF{#2}
		{\latexparagraph{\maybe@addperiod{#3}}}
		{\latexparagraph[#2]{\maybe@addperiod{#3}}}%
	}%
}
\newcommand{\maybe@addperiod}[1]{%
	#1\@addpunct{.}%
}
\makeatother

\newenvironment{algorithmbox}{\begin{mdframed}[nobreak=true]
		\begin{algorithm}}{\end{algorithm}\end{mdframed}}

\usepackage{boxedminipage}
\newcommand{\Paren}[1]{\left(#1\right)}

\newcommand{\brac}[1]{[#1]}
\newcommand{\Brac}[1]{\left[#1\right]}

\newcommand{\iverson}[1]{\llbracket #1 \rrbracket}

\newcommand{\Abs}[1]{\left\lvert#1\right\rvert}

\newcommand{\card}[1]{\lvert#1\rvert}
\newcommand{\Card}[1]{\left\lvert#1\right\rvert}

\newcommand{\set}[1]{\{#1\}}
\newcommand{\Set}[1]{\left\{#1\right\}}

\newcommand{\norm}[1]{\lVert#1\rVert}
\newcommand{\Norm}[1]{\left\lVert#1\right\rVert}

\newcommand{\Snorm}[1]{\Norm{#1}^2}

\newcommand{\normio}[1]{\norm{#1}_{\infty \rightarrow 1}}
\newcommand{\Normio}[1]{\Norm{#1}_{\infty \rightarrow 1}}

\newcommand{\iprod}[1]{\langle#1\rangle}

\newcommand{\Esymb}{\mathbb{E}}

\newcommand{\Vsymb}{\mathbb{V}}
\DeclareMathOperator*{\E}{\Esymb}
\DeclareMathOperator*{\Var}{\Vsymb}

\newcommand{\given}{\mathrel{}\middle\vert\mathrel{}}
\newcommand{\suchthat}{\;\middle\vert\;}
\newcommand{\tensor}{\otimes}
\newcommand{\sge}{\succeq}
\newcommand{\sg}{\succ}
\newcommand{\sle}{\preceq}
\newcommand{\tensorpower}[2]{#1^{\tensor #2}}
\newcommand{\from}{\colon}
\newcommand{\mper}{\,.}

\newcommand\bdot\bullet
\DeclareMathOperator{\Tr}{Tr}

\DeclareMathOperator{\poly}{poly}

\DeclareMathOperator{\polylog}{polylog}

\DeclareMathOperator{\sign}{sign}

\newcommand{\Erdos}{Erd\H{o}s\xspace}
\newcommand{\Renyi}{R\'enyi\xspace}

\newcommand{\N}{\mathbb N}
\newcommand{\R}{\mathbb R}

\newcommand{\cA}{\mathcal A}
\newcommand{\cB}{\mathcal B}
\newcommand{\cC}{\mathcal C}
\newcommand{\cD}{\mathcal D}
\newcommand{\cE}{\mathcal E}
\newcommand{\cF}{\mathcal F}
\newcommand{\cG}{\mathcal G}
\newcommand{\cH}{\mathcal H}
\newcommand{\cI}{\mathcal I}

\newcommand{\cP}{\mathcal P}
\newcommand{\cQ}{\mathcal Q}
\newcommand{\cR}{\mathcal R}
\newcommand{\cS}{\mathcal S}
\newcommand{\cT}{\mathcal T}

\newcommand{\cW}{\mathcal W}

\newcommand{\bbP}{\mathbb P}

\renewcommand{\leq}{\leqslant}

\renewcommand{\geq}{\geqslant}
\renewcommand{\ge}{\geqslant}
\let\epsilon=\varepsilon
\numberwithin{equation}{section}
\newcommand\MYcurrentlabel{xxx}
\newcommand{\MYstore}[2]{%
	\global\expandafter \def \csname MYMEMORY #1 \endcsname{#2}%
}
\newcommand{\MYload}[1]{%
	\csname MYMEMORY #1 \endcsname%
}
\newcommand{\MYnewlabel}[1]{%
	\renewcommand\MYcurrentlabel{#1}%
	\MYoldlabel{#1}%
}
\newcommand{\MYdummylabel}[1]{}
\newcommand{\torestate}[1]{%
	\let\MYoldlabel\label%
	\let\label\MYnewlabel%
	#1%
	\MYstore{\MYcurrentlabel}{#1}%
	\let\label\MYoldlabel%
}
\newcommand{\restatetheorem}[1]{%
	\let\MYoldlabel\label
	\let\label\MYdummylabel
	\begin{theorem*}[Restatement of \cref{#1}]
		\MYload{#1}
	\end{theorem*}
	\let\label\MYoldlabel
}
\newcommand{\restatelemma}[1]{%
	\let\MYoldlabel\label
	\let\label\MYdummylabel
	\begin{lemma*}[Restatement of \cref{#1}]
		\MYload{#1}
	\end{lemma*}
	\let\label\MYoldlabel
}
\newcommand{\restateprop}[1]{%
	\let\MYoldlabel\label
	\let\label\MYdummylabel
	\begin{proposition*}[Restatement of \cref{#1}]
		\MYload{#1}
	\end{proposition*}
	\let\label\MYoldlabel
}
\newcommand{\restatefact}[1]{%
	\let\MYoldlabel\label
	\let\label\MYdummylabel
	\begin{fact*}[Restatement of \cref{#1}]
		\MYload{#1}
	\end{fact*}
	\let\label\MYoldlabel
}
\newcommand{\restate}[1]{%
	\let\MYoldlabel\label
	\let\label\MYdummylabel
	\MYload{#1}
	\let\label\MYoldlabel
}
\newcommand{\eps}{\epsilon}

\allowdisplaybreaks
\newcommand*{\Id}{\mathrm{Id}}

\newcommand*{\Normf}[1]{\Norm{#1}_{\mathrm{F}}}

\newcommand*{\transpose}[1]{{#1}{}^{\mkern-1.5mu\mathsf{T}}}

\renewcommand{\ij}{{ij}}

\newcommand{\nbw}[2]{\text{NBW}^{#2}_{#1}}
\newcommand{\tgf}[2]{\text{TGF}^{#2}_{#1}}
\newcommand{\bnbw}[2]{\text{BNBW}^{#2}_{#1}}
\newcommand{\btgf}[2]{\text{BTGF}^{#2}_{#1}}
\newcommand{\ibtgf}[2]{\text{IBTGF}^{#2}_{#1}}
\newcommand{\hnbw}[2]{\text{HNBW}^{#2}_{#1}}
\newcommand{\hbnbw}[2]{\text{HBNBW}^{#2}_{#1}}
\newcommand{\htgf}[2]{\text{HTGF}^{#2}_{#1}}
\newcommand{\hbtgf}[2]{\text{HBTGF}^{#2}_{#1}}

\newcommand{\annoying}{\text{AN}}
\newcommand{\kxord}[1]{\cF_{{#1}\text{-XOR}(n,p)}}
\newcommand{\cspd}{\cF_{\textnormal{CSP(P)}}(n,p)}
\newcommand{\valI}[2]{\textnormal{Val}_{#1}\Paren{#2}}
\newcommand{\optI}[1]{\textnormal{Opt}_{#1}}

\newcommand{\Cov}[2]{\textnormal{Cov}_{#1}\Paren{#2}}
\newcommand{\LC}[2]{\textnormal{LC}_{#1}(#2)}
\newcommand{\GC}[1]{\textnormal{GC}(#1)}

\title{A Ihara-Bass Formula for Non-Boolean Matrices and  Strong Refutations of Random CSPs\thanks{This project has received funding from the European Research Council (ERC) under the European Union’s Horizon 2020 research and innovation programme (grant agreements No 815464 and 834861).}}

\author{ 
	Tommaso d'Orsi\thanks{ETH Z\"urich, tommaso.dorsi@inf.ethz.ch.}
	\and
	Luca Trevisan\thanks{Bocconi University, l.trevisan@unibocconi.it.}
}

\begin{document}

%%%% MAKE TITLE

\maketitle
\thispagestyle{empty} % seems to be required here to avoid page number on first page

% ABSTRACT

\begin{abstract}
	We define a novel notion of ``non-backtracking'' matrix associated to any symmetric matrix, and we prove a ``Ihara-Bass'' type formula for it. 

We use this theory to prove new results on polynomial-time strong refutations of random constraint satisfaction problems with $k$ variables per constraints (k-CSPs). For a random  k-CSP instance constructed out of a constraint that is satisfied by a $p$ fraction of assignments, if the instance contains $n$ variables and $n^{k/2} / \epsilon^2$ constraints, we can efficiently compute a certificate that the optimum satisfies at most a $p+O_k(\epsilon)$ fraction of constraints.

Previously, this was known for even $k$, but for odd $k$ one needed $n^{k/2} (\log n)^{O(1)} / \epsilon^2$ random constraints to achieve the same conclusion.

Although the improvement is only polylogarithmic, it overcomes a significant barrier to these types of results. Strong refutation results based on current approaches construct a certificate that a certain matrix associated to the k-CSP instance is quasirandom. Such certificate can come from a Feige-Ofek type argument, from an application of Grothendieck's inequality, or from a spectral bound obtained with a trace argument. The first two approaches require a union bound that cannot work when the number of constraints is $o(n^{\lceil k/2 \rceil})$ and the third one cannot work when the number of constraints is $o(n^{k/2} \sqrt{\log n})$. 

We further apply our techniques to obtain a new PTAS finding assignments for $k$-CSP instances with $n^{k/2} / \epsilon^2$ constraints in the semi-random settings where the constraints are random, but the sign patterns are adversarial.
\end{abstract}

\clearpage

%%%% TOC
% assumes microtype
\microtypesetup{protrusion=false}
\tableofcontents{}
\microtypesetup{protrusion=true}
\thispagestyle{empty}

\clearpage
\pagestyle{plain}
\setcounter{page}{1}

%%%% SECTION
%Include sections here
\section{Introduction}\label{sec:introduction}

If we take a random instance of 3SAT with $n$ variables and $m \geq cn$ clauses where $c$ is a sufficiently large constant, then almost surely the instance is not satisfiable. Indeed, an instance of random 3SAT with $n$ variables and $n/\epsilon^2$ clauses is almost surely such that at most a $7/8 + O(\epsilon)$ fraction of clauses can be simultanously satisfied by the best assignment. Finding a {\em certificate} that a specific random formula exhibits such behaviour is, however, believed to be quite hard.

In 2002, Feige \cite{Feige02} formulated the hypothesis that it is computationally intractable to find {\em strong refutations} of random 3-SAT formulas when the number of clauses is slightly superlinear in the number of variables. A {\em strong refutation} of a 3-SAT formula is a certificate, verifiable in polynomial time, that every assignment fails to satisfy a constant fraction of the clauses. Feige proved that his hypothesis has several consequences for the hardness of approximation of various problems.

Because of its centrality to the theories of proof complexity and of average-case complexity, and its connection to other questions in cryptography, computational complexity, and statistical physics, the complexity of strong refutations for random 3SAT and other random constraint satisfaction problems has been extensively studied since the 1980s. 

Among several important algorithmic milestones, we mention the idea of using spectral techniques to find refutations and strong refutations (introduced in \cite{GoerdtK01,FriedmanG01} and then refined in subsequent work) and a reduction from the problem of finding strong refutations for random 3SAT to the problem of finding strong refutations for random 3XOR (introduced in \cite{Feige02} and then refined in subsequent work). 

The state of the art concerning polynomial-time computable strong refutations of random constraint satisfaction problems is a 2015 paper by Allen, O'Donnell and Witmer \cite{AllenOW15}. We refer the reader to the introduction of \cite{AllenOW15} for an extended survey of algorithmic ideas and results related to refutations of random constraint satisfaction problems.  Allen, O'Donnell and Witmer \cite{AllenOW15} show how to obtain strong refutations for random $k$-XOR constraint satisfaction problems on $n$ variables and $n^{k/2}(\log n)^{O(1)}$ constraints. When $k$ is even, $O(n^{k/2})$ constraints suffice. Thanks to a reduction from arbitrary constraint satisfaction to $k$-XOR (of which we provide a self-contained simpler proof in \cref{section:csp-refutations}), similar bounds hold for any constraint satisfaction problem over $k$ variables.

To illustrate the difference between odd $k$ and even $k$, we briefly discuss how a strong refutation for random 4-XOR and random 3-XOR instances is constructed.

In general, if we have an instance of $k$-XOR with $m$ constraints and $n$ variables, a strong refutation is a certificate that

\[ \max_{x \in \{-1,1\}^n}
\sum_{i_1,\ldots,i_k} T_{i_1,\ldots,i_k} x_{i_1} \cdots x_{i_k} \leq \epsilon m \]
where $T$ is a symmetric tensor of order $k$ such that $T_{i_1,\ldots,i_k} = 0$ if there is no constraint on the $k$-tuple of variables $x_{i_1},\ldots,x_{i_k}$, and otherwise $T_{i_1,\ldots,i_k} = \pm 1$ depending on the right-hand-side of the constraint.

When $k=4$, we can flatten the tensor to an $n^2 \times n^2$ symmetric matrix $M$ (where $M_{(a,b), (c,d)} = T_{a,b,c,d}$) and we have 

\[  \max_{x \in \{-1,1\}^n}
\sum_{i_1,\ldots,i_4} T_{i_1,\ldots,i_4} x_{i_1} \cdots x_{i_4} = \max_{x \in \{-1,1\}^n}
\transpose{(x^{\otimes 2})} M x^{\otimes 2} \]

Now we can relax the right-hand side to a maximization over arbitrary $n^2$-dimensional Boolean vectors and further relax to the $\infty$-to-1 norm:

\[ \max_{x \in \{-1,1\}^n}
\transpose{(x^{\otimes 2})} M x^{\otimes 2}
\leq \max_{y \in \{-1,1\}^{n^2}}
\transpose{y} M y \leq \max_{y,z \in \{-1,1\}^{n^2}}
\transpose{y} M z = || M ||_{\infty \to 1} \]

Finally, the last expression above can be upper bounded by $\epsilon m$, by using Chernoff bounds and a union bound over all the $2^{2n^2}$ possible choices for $y$ and $z$, which is possible if $m$ is a sufficiently large constant times $n^2/\epsilon^2$. Finally, we can use Grothendieck's inequality to get us a certified upper bound of the $\infty \to 1$ norm in polynomial time up to a constant factor.

For 3-XOR, the idea is to apply a Cauchy-Schwarz step to reduce the problem of bounding a degree-4 problem, and then to flatten the resulting 4-tensor to an $n^2 \times n^2$ matrix $M$ such that

\[  \max_{x \in \{-1,1\}^n}
\sum_{i_1,i_2,i_4} T_{i_1,i_2,i_3} x_{i_1} x_{i_2} x_{i_3} \leq \sqrt n \cdot 
\sqrt{ \max_{x\in \{\pm 1\}^n} \ \ \transpose{(x^{\otimes 2})} M x^{\otimes 2} }
\leq \sqrt n \cdot 
\sqrt{ \max_{y,z\in \{\pm 1\}^{n^2}} \ \ \transpose{y} M z }
\]

Unfortunately, now it is not possible any more to bound the maximum on the above right-hand  via a union bound over $2^{2n^2}$ cases. Indeed, for this to be possible, we would need our distribution to have at least order of $n^2$ bits of entropy, and so we would need to have order of $n^2$ constraints. 

The alternative is to obtain a bound in terms of the spectral norm of $M$, using the fact that

\[ \max_{y,z\in \{\pm 1\}^{n^2}} \ \ \transpose{y} M z \leq n^2 \cdot || M || \ . \]

But for a sparse matrix to have a non-trivial bound on its spectral norm, we have to have at least ${\rm \poly}\log n$ non-zero entries per row on average\footnote{This is similar to the phenomenon that the quasirandomness of a $G_{n,p}$ random graphs can be certified in terms of the non-trivial eigenvalues of the adjacency matrix only if the average degree is at least logarithmic. We will return to the graph analogy shortly.}, and for this to happen the number of constraints has to be at least of the order of $n^{1.5} {\rm poly} \log n$.
In the regime of $n^{1.5} {\rm poly} \log n$ random 3-XOR constraints, a spectral norm bound on $M$ can be established via trace methods, and this is how the results of \cite{AllenOW15} are proved in the case of odd $k$.

\paragraph{Semi-random CSPs}
The complementary question to that of certifying strong refutations, concerns the design of algorithms that satisfy as-many-as-possible clauses in the given CSP instance. As for refutations, complexity theory paints a grim picture for (approximately) solving  worst case instances \cite{ Moshkovitz2008two, Chan2016approximation, Fotakis2015sub}. But, in the average case, polynomial time approximation schemes are known \cite{Barak2011rounding, Alev2019approximating} when the number of clauses is of the order $n^{k/2}(\log n)^{O(1)}\,.$ 

The algorithmic techniques behind these PTAS are closely related to those used for refutations and, in particular, again boils down to studying the spectrum of the flattened tensor representing the instance.

Remarkably, groundbreaking work \cite{Guruswami2022algorithms}, showed that a similar picture holds in the significantly more general settings of \textit{smoothed} CSPs: where both the literal negation patterns and clauses are chosen arbitrarily, but then signs are randomly flipped with a small, yet constant, probability.\footnote{Smoothed CSPs were first introduced in \cite{Feige2007refuting}}.

\subsection{Our Results}\label{sec:results}

\paragraph{Strong refutations} Our first result breaks the $n^{k/2} {\rm poly}\log n$ barrier for strong refutations of random $k$-XOR instances, with odd $k$.
%\Tnote{I changed the probability bounds.}

\begin{theorem}[Strong refutations of random $k$-XOR]\label{theorem:main-xor}
	There exists an efficient algorithm that, given an instance $\bm \cI$ of random $k$-XOR with $n^{k/2} / \epsilon^2$ constraints, with probability at least $0.99$, finds strong refutation of $\bm \cI$, that is, a certificate that  
	\begin{align*}
		 \optI{\bm \cI}\leq \frac{1}{2}+O(\epsilon)\,.
	\end{align*}
\end{theorem}

Using the known reduction of general $k$-CSP to $k$-XOR, of which we provide a simple self-contained proof, we have the following consequence.

\begin{theorem}[Strong refutations of random CSPs]\label{theorem:main-csp}
	Let $P:\{-1,+1\}^k\rightarrow\{0,1\}$ be a Boolean $k$-ary predicate, and call $\E P$ the probability that $P$ is satisfied by a random assignment. There exists a polynomial time algorithm that given a random instance $CSP(P)$ instances $\bm \cI$, over $n$ variables, with at least $n^{k/2}/\epsilon^2$ constraints, with probability at least $0.99$, finds a strong refutation of $\bm \cI$, that is, a certificate that
	%That is, it certifies 
\begin{align*}
	 \optI{\bm \cI}\leq \E P + O(\epsilon) \,.
	\end{align*}
\end{theorem}

\paragraph{Robust approximation algorithms against adversarial sign patterns} Our techniques can be further applied to design  efficient algorithms finding an assignment with value $\optI{}-O(\epsilon)$ beyond the $n^{k/2}\polylog n$ barrier.
Our sharp results not only works for random instances, but also in the semi-random settings where: \textit{first}, clauses are sampled randomly, and \textit{second}, given the instance, the sign pattern of \textit{each} clause is adversarially perturbed.
Such perturbations are not captured by the smooth models of \cite{Feige2007refuting, Guruswami2022algorithms} and hence require different algorithmic challenges. 
In the special case of even $k$, \cite{Kothari-personal} provided a PTAS whenever $p\geq n^{k/2}\polylog n\,.$

\begin{theorem}[Algorithm for k-XOR with adversarial patterns]\label{theorem:main-algorithm-xor}
Let $n\,,k$ be positive integers, $\epsilon > 0\,, n$ and $n^{-k/2}/\epsilon^2 < 1$.
Let $\cI$ be a $k$-XOR instance constructed through the following process:
\begin{itemize}
    \item Sample a random $k$-XOR instance $\bm \cI'$ with at least $n^{k/2}/\epsilon^2$ constraints.
    \item Given $\bm \cI'$, arbitrarily (possibly adversarially) replace the sign of each clause in $\bm \cI'\,.$
\end{itemize}
There exists a randomized algorithm, running in time $n^{O(k/\epsilon^2)}$, that returns an assignment $\hat{\mathbf{x}}$ with value
\begin{align*}
    \valI{\cI}{\hat{\mathbf{x}}}\geq \optI{\cI}-O(\epsilon)\,,
\end{align*}
with probability at least $0.99$.
\end{theorem}

As in the case of strong refutations, \cref{theorem:main-algorithm-xor} can be extended to $k$-CSPs. 

%\Tnote{Tentative formulation below, the proof part is only a sketch. Notice the running time is worse, this could be avoided.}
\begin{theorem}[Algorithm for semi-random k-CSPs]\label{theorem:main-algorithm-csp}
Let $n\,,k$ be positive integers, $\epsilon > 0\,, n$ and $n^{-k/2}/\epsilon^2 < 1$.
Let $P:\{-1,+1\}^k\rightarrow\{0,1\}$ be a Boolean $k$-ary predicate.
Let $\cI$ be a $CSP(P)$ instance constructed through the following process:
\begin{itemize}
   \item Sample a random $CSP(P)$ instance $\bm \cI'$ with at least $n^{k/2}/\epsilon^2$ constraints.
   \item  Given $\bm \cI'$, for each clause in $\bm \cI'$, replace the sign pattern with an arbitrary (possibly adversarial) sign pattern. 
\end{itemize}
There exists a randomized algorithm, running in time $n^{O(k/\epsilon^2)}$, that returns an assignment $\hat{\mathbf{x}}$ with value
\begin{align*}
    \valI{\cI}{\hat{\mathbf{x}}}\geq \optI{\cI}-O(\epsilon)\,,
\end{align*}
with probability at least $0.99$.
\end{theorem}
\subsection{Our Techniques}\label{sec:techniques}

We develop new techniques to bound\footnote{We use boldface to denote random variables.}

\begin{align}\label{eq:quadratic-form-hypercube}
    \max_{x \in \{ \pm 1 \}^N} \transpose{x} \mathbf{M} x
\end{align} 
when $\mathbf{M}$ is a random $N \times N$ matrix with only a constant expected number of non-zero entries per row and per column, and in which such entries are not independent.

\paragraph{A toy problem} Before we explain our ideas, consider the following question, which models some of the difficulties that we encounter: suppose that we are given a random graph on $N$ vertices, and such that every edge exists with probability $d/N$, where $d$ is a constant, but the edges are only known to be $\textit{poly} \log N$-wise independent, and not fully independent. Can we certify that the graph has interesting quasirandom properties, for example can we certify that the Max Cut optimum is at most a $1/2 + O(1/\sqrt d)$ fraction of edges?

One approach could be to bound $||\mathbf{A} - \E \mathbf{A} ||_{\infty \to 1}$ where $\mathbf{A}$ is the adjacency matrix of the graph. If the graph has mutually independent random edges, that is, if it is sampled from an Erd\H os-Reniy distribution $G_{N,\frac dN}$, then we can use a union bound over $2^{2N}$ cases to argue that with high probability

\[ ||\mathbf{A} - \E \mathbf{A} ||_{\infty \to 1} \leq O(\sqrt d N) \]
which is certifiable in polynomial time, up to a constant factor loss, using Grothendieck's inequality and which certifies that the Max Cut optimum is at most $1/2 + O(1/\sqrt d)$. Unfortunately, if the edges are only poly$\log N$-wise independent, then it is not possible to take such union bound.

Another option in the fully independent case is to use the results of Feige and Ofek \cite{FeigeO05}, which show that, after removing nodes of degree larger than, say, $10d$, the adjacency matrix of the residual graph has second eigenvalue at most $O(\sqrt d)$ with high probability. Unfortunately the proof of Feige and Ofek also relies on a union bound over $2^{O(N)}$ cases, and so it cannot work in the poly$\log N$-wise independent case.

A trace argument can be used to prove that, with high probability, we have

\[ || \mathbf{A}- \E  \mathbf{A} || \leq O(\sqrt{d \log N}) \]
which provides a polynomial time certificate that the Max Cut optimum is at most $1/2 + O(\sqrt{\log N} / \sqrt d)$, and the trace calculation only requires $O(\log N)$-wise independence. It does, however, introduce an extra logarithmic factor, which is unavoidable because the spectral norm of $||\mathbf{A} - \E \mathbf{A} ||$ is $\tilde\Omega (\sqrt{\log N})$ when $d$ is constant.

It is conceivable that one could prove the result of Feige and Ofek (that the adjacency matrix has second largest eigenvalue $O(\sqrt d)$ after the removal of high-degree vertices) through a trace bound on the adjacency matrix of the truncated graph, although it seems very difficult to deal with the conditional distribution of edges given that the edges survive the truncation.

\paragraph{A solution to the toy problem} Although all standard techniques fail, there is a way to combine certain recent results to solve our toy problem.
The starting point is the fact that, given an undirected graph $G=(V,E)$, 
we can define the ``non-backtracking'' $2|E| \times 2|E|$ matrix $B$ of $G$, and that this matrix satisfies the Ihara-Bass equation
\[ \det (\Id - x B) = (1-x^2)^{|E|-|V|} \cdot \det ( \Id - xA + x^2(D-\Id)) \]
where $A$ is the adjacency matrix of the graph, $D$ is the diagonal matrix of degrees, and the above equation holds as an identity of polynomials of degree $2|E|$ in $x$. See the survey of Horton \cite{horton2006zeta} for an exposition of these definitions and results.

Fan and Montanari \cite{FanM17} show that bounds on the spectral radius of $B$ imply useful PSD inequalities on $A$. In particular, if $\lambda_{\min}$ is the smallest real eigenvalue of $B$, then we have

\[ A \succeq - |\lambda_{\min}|\cdot  \Id - \frac1{|\lambda_{\min}|} \cdot (D-\Id) \]
In the context of their work on the Stochastic Block Model, Bordenave, Lalarge and Massouli\'e \cite{BordenaveLM15} use a trace argument to prove a result that implies   that $\lambda_{\min} \geq - (1+o(1)) \cdot \sqrt d$ in $G_{N,\frac dN}$ random graphs, and so all these results together imply that the Max Cut of  a
$G_{N,\frac dN}$ random graph is with high probability at most $1/2 + (1+o(1))/\sqrt d$, and that this upper bound is efficiently certifiable, for example by the dual of the Goemans-Williamson relaxation.

The key point is that there was never a union bound over $2^{O(N)}$ cases in the above argument and that, in fact, everything works assuming poly$\log N$-wise independence of the edges.\footnote{Incidentally, this combination of Fan-Montanari ideas and Bordenave-Lalarge-Massouli's bounds, also implies that if $A'$ is the adjacency matrix of a graph $G$ sampled from a distribution in which edges have probability $d/N$ and are poly$\log N$ wise independent, and then truncated by removing all vertices of degree more than, say, $10d$, then we have with high probability $A' \succeq - O(\sqrt d ) \cdot I$, proving a one-sided version of the result of Feige and Ofek.}

\paragraph{From unweighted graphs to general symmetric matrices}
Our goal is to develop an analog of this argument where we work with the $n^2 \times n^2$ matrix $M$ that comes up in the analysis of 3-XOR (or, in general, with the $n^{\lceil k/2 \rceil} \times n^{\lceil k/2 \rceil} $
matrix that comes up in the analysis of $k$-XOR when $k$ is odd) instead of the adjacency matrix $A$ of the pseudorandom graph analysed above.

The first challenge in carrying out this program is that the original notion of non-backtracking matrix is defined only with respect to 0/1 Boolean symmetric matrices, while we want to study matrices with positive and negative entries that can be arbitrary integers.

%\Tnote{Significant change here, we cannot really talk about "conceptual contributions" if a generalization of the Ihara-Bass formula was already known.}
A certain generalization of non-backtracking matrices was already introduced in \cite{Watanabe2009graph,FanM17}, however for technical reasons {\em we} were not able to use it to carry out our program.
We thus introduce a novel theory of ``non-backtracking'' matrices associated to any given symmetric matrix. In Section \ref{section:general-ihara-bass},
 given a symmetric $N\times N$ matrix $M$ with $Nz$ non-zero entries, we  define an $Nz\times Nz$ ``non-backtracking'' matrix $B_M$ associated to $M$, and we  prove (see Theorem \ref{theorem:general-ihara-bass}) an Ihara-Bass-type identity

\[ \det (\Id - x B_M + x (L_M-J_M)) =
(1-x^2)^{Nz/2 - N} \cdot \det(\Id - x M + x^2(D_M - \Id))
\]
where $D_M$, $L_M$ and $J_M$ are certain matrices that are associated to $M$. When $M$ is Boolean, $L_M = J_M$ and $D_M$ is the diagonal matrix such that $(D_M)_{i,i} = \sum_j M_{i,j}$, so our equation becomes the standard Ihara-Bass equation in the case of Boolean $M$. Conveniently, closed non-backtracking walks $W$ arising from the definition of $B_M$ take value in $\set{\pm \prod_{(i,j)\in W} M_{ij}}$, allowing one to easily mimic arguments used for standard non-backtracking matrices.

Now, given a bound on the spectral radius of $B_M-L_M+J_M$, it is possible, with an argument in the style of Fan and Montanari, to deduce a certifiable bound on the $\infty$-to-1 norm of $M$.

\paragraph{Bounding the spectral radius via weighted hyper-walks}
Studying  the spectral radius of $B_{\mathbf{M}}-L_{\mathbf{M}}+J_{\mathbf{M}}$ --matrices associated to the matrix $\mathbf{M}$ coming from random $k$-XOR instances-- is the main technical challenge of this work. 

Our bound (Theorem \ref{theorem:bound-preporcessed-A-via-Ihara-Bass}) relies on a trace argument of $B_{\mathbf{M}}$. However, compared to Bordenave, Lalarge and Massouli\'e \cite{BordenaveLM15} our setup presents a number of new technical challenges.

One challenge comes from the extra terms that we have in the non-Boolean case. In particular, our non-backtracking matrix $B_\mathbf{M}$ has entries that are the absolute values of certain entries of $\mathbf{M}$. To compute an expectation of the trace of the symmetrization of a power of $B_\mathbf{M}$, we replace absolute values with squares, and bound the error that we incur because of this. 

Perhaps the most important challenge comes from the fact that the trace bound ultimately boils down to a weighted count of certain closed ``hypergraph walks'' performed on the hypergraph corresponding to constraints of the $k$-XOR instance. 
These objects arise from our notion of non-backtracking walks on the symmetric matrix $\mathbf{M}$ obtained from the instance.
This count is performed by showing that such walks can be encoded with a small number of bits. It is enough to count walks in which every hyperedge is repeated at least twice, and the crux of the argument is that the second time we see a hyperedge we can encode that hyperedge in a compact way. A naive way of doing that would point back to the previous step in the walk in which that hyperedge appeared, and this costs $\log \ell$ bits where $\ell$ is the length of the walk. To obtain the right result, however, repeated hyperedges have to be represented with an amortized constant number of bits per occurrence. 
The argument of Bordenave, Lalarge and Massouli\'e \cite{BordenaveLM15} relies on the assumption, which is true with high probability, that the graph is ``tangle-free,'' meaning that small subgraphs have at most one cycle. We have to work with a looser notion of tangle-free hypergraph in order to prove that it holds with high probability, but we are still able to obtain the desired bound.

\paragraph{From spectral bounds to algorithms} It is clear that an algorithm certifying tight bounds on \cref{eq:quadratic-form-hypercube} for the matrix $M$ obtained from $k$-XOR instances can be used for strong refutations. Instead, to obtain \cref{theorem:main-algorithm-xor} additional ideas are needed.

Our starting point is the local-to-global rounding paradigm of \cite{Barak2011rounding}.
As it is often the case, the odd settings are significantly more challenging than the regimes with $k$ even. Hence consider first a $2$-XOR random instance $\bm \cI$. 
Up to the signs of the clauses, this may be represented as a graph $\mathbf{G}$ over $n$ vertices. Now, for a distribution $\nu$ over assignments, one may define the local and global correlations as
\begin{align*}
    \LC{\mathbf{G}}{\nu}&=\E_{(\mathbf{a}, \mathbf{b})\sim E(\mathbf{G})}\Abs{\Cov{\nu}{\mathbf{x^a, x^b}}}\\
    \GC{\nu} &= \E_{(\mathbf{a}, \mathbf{b})\sim [n]\times [n]}\Abs{\Cov{\nu}{\mathbf{x^a, x^b}}}\,.
\end{align*}
If the local correlation is bounded by $\epsilon$, it is possible to obtain an assignment with value $\optI{\bm I}- O(\epsilon)$ simply looking at the \textit{first moment} of $\nu$. 
Moreover, one can always transform $\nu$ into a distribution with small global correlation in polynomial time.

With these observations, the argument  of \cite{Barak2011rounding} comes down to: \textit{(i)}  bounding the difference between local and global correlation in terms of the spectral radius $\rho_{\mathbf{G}}$ of the centered adjacency matrix of the graph $\mathbf{G}$, \textit{(ii)} showing that one can always find, in time $n^{O(1/\eps^{2})}$, a degree $O(1)$ pseudo-distribution over the hypercube with global correlation at most $\epsilon$. As we only required low-degree moments to obtain the desired assignment, the argument goes through in this case as well.

To combine this approach with the bounds previously illustrated and extend the argument to random $k$-XOR instances with $m\geq \Omega(n^{k/2}/\epsilon^2)$ clauses, we need to introduce two novel ingredients.
\textit{First}, we need new notions of local and global correlations which difference can be bounded studying the matrix $\mathbf{M}$ arising from the instance.
\textit{Second}, we need to bound this difference not in term of the eigenvalues of $\mathbf{M}$ but rather in terms of \cref{eq:quadratic-form-hypercube}.

A careful Cauchy-Schwarz application allows us to formulate notions of local and global correlations in terms of $\mathbf{M}$. Its squaring step, further allows us to get rid of absolute values, thus providing an avenue to bound the difference between local and global correlation in terms of $\max_{x\in \set{\pm 1}^n} \transpose{x}\mathbf{M}x\,.$

Finally, since the adversarial perturbations in \cref{theorem:main-algorithm-xor} cannot alter the "hypergraph walks" required to prove our bound, we are able to generalize our result to these settings.

\subsection{Perspective}

Several results on the average-case complexity of Sum-of-Square relaxations rely on proving that sparse matrices with non-independent entries are ``quasirandom'' in an appropriate sense. We have introduced a new approach to prove results of this form, which applies to very sparse matrices that have only a constant expected number of non-zero entries per row and per column. We hope that our ideas will find further application, for example to the context of semi-random instances of constraint satisfaction problems \cite{GKM22} or of higher-degree Sum-of-Square relaxations of random constraint satisfaction problems \cite{RaghavendraRS17,WeinAM19}.

Our theory could also be useful to study problems on random weighted graphs.

Our certificates prove certain PSD inequalities, and can be seen as Semidefinite Duals of certain Sum-of-Squares relaxations, but the computation of the certificate only requires an eigenvalue computation of a certain matrix, and does not require the solution of an SDP. There might be other ways to apply our theory so that one uses SDP relaxations only in the analysis, but the algorithm itself is purely spectral.

\section{Preliminaries}\label{section:preliminaries}

We introduce some notation, useful facts and needed preliminary notions. We denote random variables in \textbf{bold}. We use lower case letters $a,b,c,d,\ldots$ to denote indices or scalars (the use willl be clear from context). We use the greek letters $\alpha, \beta,\eta$ to denote multi-indices. The cardinality of  a multi-index $\alpha$ is $\Card{\alpha}$. The $i$-th index in $\alpha$ is $\alpha(i)$. We may thus write a monomial (with coefficient $c$) in indeterminates $x_1,\ldots, x_n$ as $c\cdot x^{\alpha}$.
For two multi indices $\alpha, \beta \in [n]^k$  we denote by $(\alpha,\beta)$ the multi-index obtained concatenating $\alpha$ and $\beta$. Multi-indices $\alpha,\beta\in [n]^k$ satisfy $\alpha=\beta$ if at each position the corresponding indices are identical.
We use $S(\alpha)$ to denote the unordered multi-set of indices in $\alpha$. We use $n$ to denote our ambient dimension. For functions $f,g:\R\rightarrow\R$ we write $f=o(g)$ and $g=\omega(f)$ if $\lim_{n\rightarrow\infty}\frac{f(n)}{g(n)}=0$.

\paragraph{Matrices} For a matrix $M\in \R^{n\times n}$, we denote by $\lambda_1(M)\geq \ldots\geq \lambda_n(M)$ its eigenvalues. Then $\rho(M):=\max_i\Abs{\lambda_i(M)}$ is the spectral radius of $M$. W When the context is clear we simply write $\lambda_1,\ldots, \lambda_n$.
The spectral radius of a matrix satisfies the following inequality.
\begin{fact}[Gelfand's Formula]\label{fact:bound-spectral-radius}
	Let $M\in \R^{n\times n}$ and let $\Norm{\cdot}_*$ be a norm. Then for any positive integer $z$
	\begin{align*}
		\rho(M)\leq \Norm{M^z}_*^{1/z}.
	\end{align*}
\end{fact}
We write $\Norm{M}$ for the spectral norm of a matrix $M$, $\Normf{M}$ for its Frobenius norm %, 
%$\Normc{M}:=\max_{I, J \subseteq [n]}\Abs{\sum_{i\in I, j\in J} A_\ij}$ for its cut norm
$\Normio{M}:=\max_{x, y\in\Set{\pm 1}^n}\iprod{M, x \transpose{y}}$ and $\Norm{M}_{\textnormal{max}}:=\max_{ij}\Abs{M_\ij}\,.$  
Furthermore, we let $\Norm{M}_{\textnormal{Gr}}=\max\Set{\iprod{M, X}\given X\sge 0, X_{ii}\leq 1\,, \forall i \in [n]}\,.$
We denote by $\Abs{M}$ the matrix with entries $\Paren{\Abs{M}}_\ij:=\Abs{M_\ij}$. We write $\Id_t$ for the $t$-by-$t$ identity matrix, $\mathbf{0}$ for the zero matrix and $J$ for the all-ones matrix.
%The following relation is well-known.

%\begin{fact}\label{fact:relation-cut-norm-infty}
	%Let $M\in \R^{n\times n}$. Then
	%\begin{align*}
		%\Normc{M}\leq \Normio{M}\leq 4 \Normc{M}\,.
	%\end{align*}
%\end{fact}

\paragraph{Graphs}
For a graph $G$, $V(G)$ and $E(G)$ denotes respectively its set of vertices and edges.  $\vec{E}(G):=\Set{(u,v)\,:\, u\neq v \in V(G)\,, uv\in E(G)}$ is the set of all its ordered pairs such that $\Set{u,v}\in E(G)$.
For $e\in \vec{E}(G)$, $s(e)$ and $t(e)$ are respectively the source and target of the oriented edge. We write $e^{-1}$ for its inverse.
We also write $K_n$ for the complete graph over $n$ vertices.  
For a graph $G$ with $n$ vertices, we write $A(G)\in \R^{n\times n}$ for its adjacency matrix.
 For a vertex $v\in V(G)$, we denote by $\deg_G(v)$ its degree. We denote by $N_{G,t}(v)$ the set of vertices in $G$ at distance $t$ from $v$. We  and drop the subscript $G$ when the context is clear.
%\Tnote{We could think about the complex case too}  
If the graph $G$ is weighted with weights given by $w:V(G)\times V(G)\rightarrow \R$, then $A_{uv}= w(\Set{uv})$. If $e\neq E(G)$, then we assume $w(e)=0$.

\paragraph{Walks}
A walk $W$ in a graph $G$ is a sequence of vertices $(v_1,\ldots, v_{z+1})$. We refer to the directed edges of a walk $W$ as $e_1(W),\ldots, e_z(W)$. When the context is clear we simply write $e_1,\ldots, e_z$. We use $G(W)$ to denote the subgraph of $G$ traversed by the walk $W$. The set of vertices and distinct edges in $G(W)$ are denoted by $V(W)$ and $E(W)$. The multi-set of edges in $W$ is denoted by $M(W)$. For $e \in E(W)$, $m_W(e)$ is the multiplicity of $e$ in $M(W)$. We write $H(V(W), M(W))$ for the multigraph generated by $W$. 
A walk $v_1,\ldots, v_{z+1}$ is said to be non-backtracking if for any $i\leq z-1$, $v_i\neq v_{i+2}$.
For $e,f \in \vec{E}(G)$, $\nbw{ef}{z}(G)$ denotes the set of non-backtracking walks in $G$ starting with $e$ and ending with $f$ of length $k$. For simplicity we let $\nbw{ef}{z} := \nbw{ef}{z}(K_n)$.

\paragraph{Hypergraphs} We use the notation $H(V, E)$ for an hyper-graphs over $V(H)$ with hyper-edge set $E(H)$. We only consider hypergraphs in which edges have the same arity. The arity of the edges will be clear from context. A multi-hyper-graph is a hyper-graph where edges may have multiplicity more than $1$, we denote its multi-set of hyper-edges by $M(H)$ and its set of distinct hyper-edges by $E(H)$. When clear from context, we will  refer to multi-hyper-graphs simply as hyper-graphs.
Given hyper-graphs $H, H'$ we denote by $H^*=H\oplus H'$ the multi-hyper-graph obtained by taking $V(H^*)$ as  the union of the vertex sets and $M(H^*)$ as the multi-set of elements either in $M(H)$ or $M(H')$.

%We do introduce specific notation to describe the arity of edges as this will be clear from the context at hand.

\subsection{CSPs, k-XOR and strong refutations}\label{section:preliminaries-csps}

%Throughout the paper we assume $k$ to be an \textit{odd} integer as for the even case sharp refutation algorithms are already known (e.g see \cite{AllenOW15}).

\paragraph{k-XOR}
A random $k$-XOR instance $\bm{\cI}$ with $n$ variables and $p\binom{n}{k}(1\pm o(1))$ clauses can be generated by picking a random symmetric tensor $\mathbf{T}$, with independent entries, such that $\mathbf{T}_{\alpha}=0$ if the indices in the multi-index $\alpha\in [n]^{k}$ are not distinct and otherwise:%\Tnote{add notation of multi-index to preliminaries}
\begin{align*}
	\mathbf{T}_{\alpha} = \begin{cases}
		0& \text{with probability }1-p\,,\\
		+1&\text{with probability }p/2\,,\\
		-1&\text{with probability }p/2\,.
	\end{cases}
\end{align*}
We denote by $m$ the exact number of clauses in the instance.
Then $\bm \cI$ consists of the $m$ $k$-XOR predicates $k\text{-XOR}(\alpha)=\frac{1-x^{\alpha}(-\mathbf{T})_\alpha}{2}$ where $\mathbf{T}_{\alpha}$ is non-zero.
We  use $\kxord{k}$ to denote such distribution and $\bm \cI\sim \kxord{k}$ to denote a random instance. 
We let $\valI{\bm \cI}{x}$ be the fraction of constrained satisfied by the assignment $x\in \Set{\pm 1}^n$ and  $\optI{\bm \cI}:=\max_{x\in \Set{\pm 1}^n}\valI{\bm I}{x}$.
For any assignment $x\in \Set{\pm 1}^n$ we have
\begin{align*}
	\valI{\bm \cI}{x}=\frac{1}{2}+ \frac{1}{m(\bm \cI)}\sum_{\alpha\in [n]^k}\frac{x^{\alpha}\mathbf{T}_\alpha}{2}\,.
\end{align*}
Notice that since $m$ will be $(1\pm o(1))p\binom{n}{k}$ with overwhelming probability, we blur the distinction between these parameters.
Then the  max $k$-XOR problem  is that of finding an assignment with value
\begin{align}\label{eq:preliminaries-max-value}
	\max_{x\in \Set{\pm 1}^n}\underset{\alpha\in [n]^k}{\sum} \mathbf{T}_{\alpha}x^{\alpha}\,.
\end{align}
This is captured by the following proposition.
\begin{proposition}\label{proposition:max-value-xor}
	Let $\bm \cI \sim \kxord{k}$ and let $\mathbf{T}$ be the associated $k$-th order tensor. Then with overwhelming probability
	\begin{align*}
		\optI{\bm \cI} \leq \frac{1}{2}+ \Paren{1+o(1)} \Paren{\binom{n}{k}\cdot p}^{-1}\cdot \underset{\alpha\in [n]^k}{\sum} \mathbf{T}_{\alpha}x^{\alpha}\,.
	\end{align*}
\end{proposition}
Throughout the paper we assume $k$ to be an \textit{odd} integer as for the even case sharp refutation algorithms are already known (e.g see \cite{AllenOW15}).

A random $k$-XOR instance $\bm{\cI}$ with  $n$ variables and exactly $m$ clauses can be generated by picking $m$ times a clause at random out of the $\binom{n}{k}$ possible $k$-XOR-clause. It is possible to show that a refutation algorithm for $\bm \cI\sim \kxord{k}$ can also be used for refutation of $k$-XOR instances sampled through this second process. For this reason, we blur the distinction between these two processes. We direct the  reader interested in a formal reduction to \cite{AllenOW15} (Appendix D).

\paragraph{CSPs} 
 Given a predicate $P:\{-1, 1\}^k \rightarrow \{0, 1\}$, an instance $\cI$ of the CSP(P) problem over variables $x_1 , \ldots, x_n$ is a multi-set of pairs $(c,\alpha)$ representing constraints of the form $P(c\circ x^\alpha):=P(c_{1} x^{\alpha(1)},\ldots, c_k x^{\alpha(k)})=1$  where $\alpha\in  [n]^k$ is the scope and $c\in \{\pm 1 \}^k$ is the negation pattern.
We can represent the predicate $P$ as a multi-linear polynomial of degree $k$ in indeterminates $c_1x^{\alpha(1)},\ldots, c_kx^{\alpha(k)}$,
 \begin{align*}
 	P(c\circ x^{ \alpha}) =\sum_{d\leq k} P_d(c\circ x^{ \alpha})\,,
 \end{align*}
 where $P_d$ denotes the degree $d$ part of the predicate. In particular $P_0:=P_0(c\circ x^{ \alpha})$ denotes the constant part of the polynomial, which does not depend on the assignment.
 
The fraction of all possible assignments that satisfy $P$ is given by  $\E_{\mathbf{z}\overset{u.a.r}{\sim}\Set{\pm 1}^k}\Brac{P(\mathbf{z})}$.
 For any assignment $x\in\Set{\pm 1}^n$ and an instance $\cI$ over $m$ constraints we  have
 \begin{align*}
 	\valI{\cI}{x} &=  \frac{1}{m}\sum_{(c,\alpha)\in \cI} P(c\circ x^{\alpha})\\
 	\text{and }\quad\optI{\cI} &= \max_{x\in \Set{\pm 1}^n}\valI{\cI}{x}\,.
 \end{align*}
 A \textit{random} CSP(P) instance $\bm \cI$ with %\footnote{Here $n^{\underline{k}}$ is the fallen factorial.} 
 $(1+o(1))m=p\cdot 2^k\cdot n^k$  constraints can be generated as follows: 
 \begin{enumerate}[(i)]
 	\item Pick independently with probability $p$ each pair $(\mathbf{c}, \bm \alpha)$ where $\mathbf{c}$ is a random negation pattern  from $\Set{-1,+1}^k$ and  $\mathbf{\alpha}$  is a multi-index from $[n]^k$,
 	\item For each such pair $(\mathbf{c}, \bm \alpha)$ add the constraint $P(\mathbf{c}\circ x^{\bm\alpha})=1$ to $\bm \cI$.
 \end{enumerate} 
Notice that we do not rule out predicates with same multi-index but different negation pattern as multi-indices in which an index appears multple time. We also do not assume $P$ to be symmetric.
We denote such distribution by $\cspd$.

As in the case of $k$-XOR a random CSP(P) instance $\bm{\cI}$ with  $n$ variables and exactly $m$ clauses can be generated by picking $m$ times a clause and a negation pattern at random. Again it is possible to show that a refutation algorithm for $\bm \cI\sim \cspd$ can also be used for refutation of instances sampled through this second process  (see Appendix D in  \cite{AllenOW15}).

\paragraph{Refutation and certification}
We say that $\cA$ is a $\delta$-\textit{refutation algorithm} for random CSP(P)  if 
$\cA$ has the following properties:
\begin{enumerate}[(i)]
	\item on all instances $\cI$ the output of $\cA$ si either $\optI{\cI}\leq 1-\delta$ or "fail",
	\item if $\optI{\cI}>1-\delta$ then $\cA$ \textit{never} outputs $\optI{\cI}\leq 1-\delta$.
\end{enumerate}
More generally, for an set of possible inputs $\cS$ and a property $p$ over instances in $\cS$, we say that an algorithm $\cA$ \textit{certifies} $p$  if:
\begin{enumerate}[(i)]
	\item on all inputs $\cI\in \cS$ the output of $\cA$ is either "$\cI$ satisfies $p$" or "fail",
	\item if $\cI\in \cS$ does not satisfy $p$ then $\cA$ \textit{never} outputs "$\cI$ satisfies $p$".
\end{enumerate}

In the context of random CSP(P) (and hence $k$-XOR), a \textbf{\textit{strong refutation}} is a $\delta$-refutation for $1-\delta\leq \E_{\mathbf{x}\overset{u.a.r}{\sim}\Set{\pm 1}^k}\Brac{P(\mathbf{x})}+o(1)$.

\subsection{Sum-of-squares and pseudo-distributions}\label{section:preliminaries-sos-pseudodistributions}
We briefly introduce the sum-of-squares proof system. We direct the interested reader to \cite{BoazNotes}.

Let $x = (x_1, x_2, \ldots, x_n)$ be a tuple of $n$ indeterminates and let $\R[x]$ be the set of polynomials with real coefficients and indeterminates $x_1,\ldots,x_n$.
We say that a polynomial $p\in \R[x]$ is a \emph{sum-of-squares (sos)} if there are polynomials $q_1,\ldots,q_r$ such that $p=q_1^2 + \cdots + q_r^2$.

\subsubsection{Pseudo-distributions}\label{section:preliminaries-pseudodistributions}
Pseudo-distributions are generalizations of probability distributions.
We can represent a discrete (i.e., finitely supported) probability distribution over $\R^n$ by its probability mass function $\mu\from \R^n \to \R$ such that $\mu \geq 0$ and $\sum_{x \in \mathrm{supp}(\mu)} \mu(x) = 1$.
Similarly, we can describe a pseudo-distribution by its mass function.
Here, we relax the constraint $\mu\ge 0$ and only require that $\mu$ passes certain low-degree non-negativity tests.

Concretely, a \emph{level-$\ell$ pseudo-distribution} is a finitely-supported function $\mu:\R^n \rightarrow \R$ such that $\sum_{x} \mu(x) = 1$ and $\sum_{x} \mu(x) f(x)^2 \geq 0$ for every polynomial $f$ of degree at most $\ell/2$.
(Here, the summations are over the support of $\mu$.)
A straightforward polynomial-interpolation argument shows that every level-$\infty$-pseudo distribution satisfies $\mu\ge 0$ and is thus an actual probability distribution.
We define the \emph{pseudo-expectation} of a function $f$ on $\R^d$ with respect to a pseudo-distribution $\mu$, denoted $\tilde{\E}_{\mu(x)} f(x)$, as
\begin{equation}
	\tilde{\E}_{\mu(x)} f(x) = \sum_{x} \mu(x) f(x) \,\mper
\end{equation}
The degree-$\ell$ moment tensor of a pseudo-distribution $\mu$ is the tensor $\E_{\mu(x)} (1,x_1, x_2,\ldots, x_n)^{\otimes \ell}$.
In particular, the moment tensor has an entry corresponding to the pseudo-expectation of all monomials of degree at most $\ell$ in $x$.
The set of all degree-$\ell$ moment tensors of probability distribution is a convex set.
Similarly, the set of all degree-$\ell$ moment tensors of degree $d$ pseudo-distributions is also convex.
Key to the algorithmic utility of pseudo-distributions is the fact that while there can be no efficient separation oracle for the convex set of all degree-$\ell$ moment tensors of an actual probability distribution, there's a separation oracle running in time $n^{O(\ell)}$ for the convex set of the degree-$\ell$ moment tensors of all level-$\ell$ pseudodistributions.

\begin{fact}[\cite{MR939596-Shor87,parrilo2000structured,MR1748764-Nesterov00,MR1846160-Lasserre01}]
	\label[fact]{fact:sos-separation-efficient}
	For any $n,\ell \in \N$, the following set has a $n^{O(\ell)}$-time weak separation oracle (in the sense of \cite{MR625550-Grotschel81}\footnote{Note that in general there may be bit complexity issues for running sum-of-squares algorithms, see \cite{o2017sos}}):
	\begin{equation}
		\Set{ \tilde{\E}_{\mu(x)} (1,x_1, x_2, \ldots, x_n)^{\otimes d} \mid \text{ degree-d pseudo-distribution $\mu$ over $\R^n$}}\,\mper
	\end{equation}
\end{fact}
This fact, together with the equivalence of weak separation and optimization \cite{MR625550-Grotschel81} allows us to efficiently optimize over pseudo-distributions (approximately)---this algorithm is referred to as the sum-of-squares algorithm.

The \emph{level-$\ell$ sum-of-squares algorithm} optimizes over the space of all level-$\ell$ pseudo-distributions that satisfy a given set of polynomial constraints---we formally define this next.

\begin{definition}[Constrained pseudo-distributions]
	Let $\mu$ be a level-$\ell$ pseudo-distribution over $\R^n$.
	Let $\cA = \{f_1\ge 0, f_2\ge 0, \ldots, f_m\ge 0\}$ be a system of $m$ polynomial inequality constraints.
	We say that \emph{$\mu$ satisfies the system of constraints $\cA$ at degree $r$}, denoted $\mu \sdtstile{r}{} \cA$, if for every $S\subseteq[m]$ and every sum-of-squares polynomial $h$ with $\deg h + \sum_{i\in S} \max\set{\deg f_i,r}\leq \ell$,
	\begin{displaymath}
		\tilde{\E}_{\mu} h \cdot \prod _{i\in S}f_i  \ge 0\,.
	\end{displaymath}
	We write $\mu \sdtstile{}{} \cA$ (without specifying the degree) if $\mu \sdtstile{0}{} \cA$ holds.
	Furthermore, we say that $\mu\sdtstile{r}{}\cA$ holds \emph{approximately} if the above inequalities are satisfied up to an error of $2^{-n^\ell}\cdot \norm{h}\cdot\prod_{i\in S}\norm{f_i}$, where $\norm{\cdot}$ denotes the Euclidean norm\footnote{The choice of norm is not important here because the factor $2^{-n^\ell}$ swamps the effects of choosing another norm.} of the coefficients of a polynomial in the monomial basis.
\end{definition}

We remark that if $\mu$ is an actual (discrete) probability distribution, then we have  $\mu\sdtstile{}{}\cA$ if and only if $\mu$ is supported on solutions to the constraints $\cA$.

We say that a system $\cA$ of polynomial constraints is \emph{explicitly bounded} if it contains a constraint of the form $\{ \|x\|^2 \leq M\}$.
The following fact is a consequence of \cref{fact:sos-separation-efficient} and \cite{MR625550-Grotschel81},

\begin{fact}[Efficient Optimization over Pseudo-distributions]\label{fact:running-time-sos}
	There exists an $(n+ m)^{O(\ell)} $-time algorithm that, given any explicitly bounded and satisfiable system\footnote{Here, we assume that the bit complexity of the constraints in $\cA$ is $(n+m)^{O(1)}$.} $\cA$ of $m$ polynomial constraints in $n$ variables, outputs a level-$\ell$ pseudo-distribution that satisfies $\cA$ approximately. 
\end{fact}

\subsubsection{Sum-of-squares proof}\label{section:preliminaries-sos}
Let $f_1, f_2, \ldots, f_r$ and $g$ be multivariate polynomials in $x$.
A \emph{sum-of-squares proof} that the constraints $\{f_1 \geq 0, \ldots, f_m \geq 0\}$ imply the constraint $\{g \geq 0\}$ consists of  sum-of-squares polynomials $(p_S)_{S \subseteq [m]}$ such that
\begin{equation}
	g = \sum_{S \subseteq [m]} p_S \cdot \Pi_{i \in S} f_i
	\mper
\end{equation}
We say that this proof has \emph{degree $\ell$} if for every set $S \subseteq [m]$, the polynomial $p_S \Pi_{i \in S} f_i$ has degree at most $\ell$.
If there is a degree $\ell$ SoS proof that $\{f_i \geq 0 \mid i \leq r\}$ implies $\{g \geq 0\}$, we write:
\begin{equation}
	\{f_i \geq 0 \mid i \leq r\} \sststile{\ell}{}\{g \geq 0\}
	\mper
\end{equation}

Sum-of-squares proofs satisfy the following inference rules.
For all polynomials $f,g\colon\R^n \to \R$ and for all functions $F\colon \R^n \to \R^m$, $G\colon \R^n \to \R^k$, $H\colon \R^{p} \to \R^n$ such that each of the coordinates of the outputs are polynomials of the inputs, we have:

\begin{align}
	&\frac{\cA \sststile{\ell}{} \{f \geq 0, g \geq 0 \} } {\cA \sststile{\ell}{} \{f + g \geq 0\}}, \frac{\cA \sststile{\ell}{} \{f \geq 0\}, \cA \sststile{\ell'}{} \{g \geq 0\}} {\cA \sststile{\ell+\ell'}{} \{f \cdot g \geq 0\}} \tag{addition and multiplication}\\
	&\frac{\cA \sststile{\ell}{} \cB, \cB \sststile{\ell'}{} C}{\cA \sststile{\ell \cdot \ell'}{} C}  \tag{transitivity}\\
	&\frac{\{F \geq 0\} \sststile{\ell}{} \{G \geq 0\}}{\{F(H) \geq 0\} \sststile{\ell \cdot \deg(H)} {} \{G(H) \geq 0\}} \tag{substitution}\mper
\end{align}

Low-degree sum-of-squares proofs are sound and complete if we take low-level pseudo-distributions as models.

Concretely, sum-of-squares proofs allow us to deduce properties of pseudo-distributions that satisfy some constraints.

\begin{fact}[Soundness]
	\label{fact:sos-soundness}
	If $\mu \sdtstile{r}{} \cA$ for a level-$\ell$ pseudo-distribution $\mu$ and there exists a sum-of-squares proof $\cA \sststile{r'}{} \cB$, then $\mu \sdtstile{r\cdot r'+r'}{} \cB$.
\end{fact}

If the pseudo-distribution $\mu$ satisfies $\cA$ only approximately, soundness continues to hold if we require an upper bound on the bit-complexity of the sum-of-squares $\cA \sststile{r'}{} B$  (number of bits required to write down the proof).

In our applications, the bit complexity of all sum of squares proofs will be $n^{O(\ell)}$ (assuming that all numbers in the input have bit complexity $n^{O(1)}$).
This bound suffices in order to argue about pseudo-distributions that satisfy polynomial constraints approximately.

The following fact shows that every property of low-level pseudo-distributions can be derived by low-degree sum-of-squares proofs.

\begin{fact}[Completeness]
	\label{fact:sos-completeness}
	Suppose $d \geq r' \geq r$ and $\cA$ is a collection of polynomial constraints with degree at most $r$, and $\cA \vdash \{ \sum_{i = 1}^n x_i^2 \leq B\}$ for some finite $B$.
	
	Let $\{g \geq 0 \}$ be a polynomial constraint.
	If every degree-$d$ pseudo-distribution that satisfies $\mu \sdtstile{r}{} \cA$ also satisfies $\mu \sdtstile{r'}{} \{g \geq 0 \}$, then for every $\epsilon > 0$, there is a sum-of-squares proof $\cA \sststile{d}{} \{g \geq - \epsilon \}$.
\end{fact}

\subsubsection{Sum of Squares toolkit}

We introduce here several statements involving sum of squares that will be used throughout the paper.
We start with a certificate on the infinity-to-one  norm of a matrix.

\begin{fact}[\cite{AlonN04}]\label{fact:sos-grothendieck-inequality}
	There exists an absolute constant $K_G$ (Grothendieck's constant) and a polynomial time algorithm (based on sum-of-squares)  that, for every $A\in \R^{n\times m}$, certifies an upper bound to $\Normio{A}$ tight up to a factor $K_G$. 
    More specifically, for any degree-$2$ pseudo-distribution $\mu:\Set{\pm 1}^n\times \Set{\pm 1}^m\rightarrow \R$ 
	\begin{align*}
		\tilde{\E}_{\mu(x,y)} \iprod{A, x \transpose{y}} 
		\leq K_G \cdot \Normio{A}\,.
	\end{align*}
\end{fact}

The next fact establishes a Cauchy-Schwarz inequality for pseudo-distributions.

\begin{fact}[Cauchy-Schwarz for pseudo-distributions \cite{Barak2012hypercontractivity}]\label{fact:cauchy-schwarz}
Let $f,g$ be vector polynomials of degree at most $d$ in indeterminate $x\in \R^n$. Then, for any degree $d^2$ pseudo-distribution $\mu$,
\begin{align*}
    \tilde{\E}_\mu\Brac{\iprod{f,g}}\leq \sqrt{\tilde{\E}_\mu\brac{\norm{f}^2} }\cdot \sqrt{\tilde{\E}_\mu\brac{\norm{g}^2} }\,.
\end{align*}
\end{fact}

We will use the notions of pseudo-covariance and conditional pseudo-distributions.

\begin{definition}[Pseudo-covariance]
Let $\alpha, \beta$ be multi-indices over $[n]$. Let $d/2\geq \card{\alpha} + \card{\beta}$. Let $\mu$ be a degree-$d$ pseudo-distribution in indeterminates  $x_1,\ldots, x_n\,.$ .
Then we write
\begin{align*}
    \Cov{\mu}{x^\alpha, x^\beta} = \tilde{\E}_\mu \brac{x^\alpha x^\beta}-\tilde{\E}_\mu \brac{x^\alpha}\tilde{\E}_\mu \brac{x^\beta}\,.
\end{align*}
Similarly, we define $\Var_\mu(x^{\alpha}) = \Cov{\mu}{x^{\alpha}, x^{\alpha}}\,.$
\end{definition}

\begin{definition}[Conditional pseudo-distribution]\label{definition:conditional-pseudo-distribution}
Let $\mu$ be a degree-$d$ pseudo-distribution in indeterminates $x_1,\ldots, x_n\,.$ Let $t\geq 0$. Suppose $\mu$ satisfies $\Set{x_i^2=1\,, \forall i \in [n]}\,.$ Then for any $\alpha\in [n]^t$ such that $\tilde{\E} \brac{\frac{1+x^{\alpha}}{2}}> 0$ we may define the conditional pseudo-distribution of degree $d-t$ as:
\begin{align*}
    \tilde{\E}_\mu \brac{p(x) \,|\, x^\alpha=1} = \frac{\tilde{\E}_\mu \brac{p(x) \frac{1+x^\alpha}{2}}}{\tilde{\E}_\mu \brac{\frac{1+x^\alpha}{2}}}\,.
\end{align*}
Similarly, if $\tilde{\E} \brac{\frac{1+x^\alpha}{2}} < 1$, we may define the conditional pseudo-distribution of degree $d-t$ as: 
\begin{align*}
    \tilde{\E}_\mu \brac{p(x) \,|\, x^\alpha=-1} = \frac{\tilde{\E}_\mu \brac{p(x) \frac{1-x^\alpha}{2}}}{\tilde{\E}_\mu \brac{\frac{1-x^\alpha}{2}}}\,.
\end{align*}
\end{definition}

It is straightforward to see that, after conditioning, the result is a valid pseudo-distribution of degree $d-t$.
Notice also that, when $\mu$ is an actual distribution, then we simply recover the corresponding conditional distribution.

Last, we introduce the following crucial observation about pseudo-distributions over the hypercube.

\begin{lemma}[E.g. see \cite{tselil-notes}]\label{lemma:moment-matching-distribution}
    Let $\mu$ be a degree $d$ pseudo-distribution over indeterminates $x_1,\ldots, x_n$ satisfying $\Set{x^2=1\,, \forall i \in [n]}$. 
    Then, for any $S\subseteq [n]$ with $\card{S}\leq d$, there exists a distribution $\nu$ over $\Set{\pm 1}^n$ such that, for all multi-indices $\alpha$ over $S$, 
    \begin{align*}
        \tilde{E}_{\mu}\brac{x^{\alpha}} = \tilde{E}_{\nu}\brac{x^{\alpha}}
    \end{align*}
\end{lemma}

In other words, \cref{lemma:moment-matching-distribution} states that, for any degree-$d$ pseudo-distribution $\mu$ over the hypercube and any subset $S$ of $d$ indeterminates, there exists an actual distribution $\nu$ over the hypercube matching its first $d$ moments on $S$.
Notably, combining this results with \cref{definition:conditional-pseudo-distribution}, one gets that these low-degree moments of $\mu$ and $\nu$ match even after conditioning.

\section{A generalized Ihara-Bass formula}\label{section:general-ihara-bass}

In this section we present an extension of the Ihara-Bass theorem (see \cite{horton2006zeta} and references therein) to arbitrary real symmetric matrices. We remark that our extension differs from the one in \cite{FanM17}.

Throughout the section we assume to be given a \textit{symmetric} matrix $A\in \R^{n\times n}$ with $2m$ non-zero entries and zeroed diagonal. We use the following notation.
We will use letters $u,v$ to denote indices in $[n]$ and $e,f$ for indices in $[2m]$.
We conveniently think of $A$ as the adjacency matrix of a weighted undirected graph $G$ with $n$ vertices and $2m$ oriented edges. Then $uv \in E(G)$ if $A_{uv}\neq 0$, moreover then the inverse edge $vu$ is also in $E(G)$ since $A_{uv}=A_{vu}$ by definition. Recall for an edge $e\in E(G)$ we write $e^{-1}$ for its inverse and for a vertex $v\in V(G)$ we write $N^+(v)$ (respectively $N^-(v)$) for its set of outgoing (resp. incoming) oriented edges in $G$.
We write $\sigma_{uv}=\sign(A_{uv})$.
To reason about the spectrum of $A$, we introduce several matrices: the diagonal matrices
\begin{align*}
	&D(A)\in \R^{n\times n}\,, \quad \text{with } D_{uv}(A) = \begin{cases}
		\sum_w \Abs{A_{uw}} & u=v\\
	    0&\text{otherwise.}
	\end{cases}\\
	&Q(A)\in \R^{m\times m}\,, \quad \text{with } Q_{ef}(A)= \begin{cases}
		\Abs{A_e}& e=f\\
		0&\text{otherwise.}
	\end{cases}
\end{align*}
the block matrices
\begin{align*}
	& J(A) = 
	\begin{pmatrix}
		0 & \Id_m\\
		\Id_m & 0
	\end{pmatrix}
	\in \R^{2m\times 2m}\\
	&L(A) = 
	\begin{pmatrix}
		0 & Q(A)\\
		Q(A) & 0
		\end{pmatrix}
	\in \R^{2m\times 2m}
\end{align*}
and the source, target and non-backtracking matrices
\begin{align*}
	& S(A)\in \R^{n\times 2m}\,, \quad \text{with } S_{ue}(A) =
	\begin{cases}
		\sigma_{uv}\sqrt{\Abs{A_{uv}}}& \text{if $u$ is the source of } e=uv \text{ and } u< v\\
		\sqrt{\Abs{A_{uv}}}& \text{if $u$ is the source of } e=uv \text{ and } u> v\\
		0& \text{ otherwise.}
	\end{cases}\\
	& T(A)\in \R^{n\times 2m}\,, \quad \text{with } T_{ue}(A) = 
	\begin{cases}
		\sigma_{uv}\sqrt{\Abs{A_{uv}}}& \text{if $u$ is the target of } e=vu \text{ and } u< v\\
		\sqrt{\Abs{A_{uv}}}& \text{if $u$ is the target of } e=vu \text{ and } u> v\\
		0& \text{ otherwise.}
	\end{cases}\\
	&B(A)\in\R^{2m\times 2m}\,, \quad \text{with }  B_{ef}(A) = 
	\begin{cases}
		\sigma_e\sigma_f\sqrt{\Abs{A_e A_f}}& \text{if $ef$ is a non-backtracking walk}\\
		&\text{ $e=uv$, f=$vw$ and } v< u,w \\
		\sigma_e\sqrt{\Abs{A_e A_f}}& \text{if $ef$ is a non-backtracking walk}\\
		&\text{ $e=uv$, f=$vw$ and } w < v < u \\
		\sigma_f\sqrt{\Abs{A_e A_f}} & \text{if $ef$ is a non-backtracking walk}\\
		&\text{ $e=uv$, f=$vw$ and } u< v<w \\
		\sqrt{\Abs{A_e A_f}}\\& \text{if $ef$ is a non-backtracking walk}\\
		&\text{ $e=uv$, f=$vw$ and } v> u,w \\
		0& \text{ otherwise.}
	\end{cases}
\end{align*}
When the context is clear we simply write $B$ for $B(A)$ (analogously for the other matrices). To gain intuition on these linear maps, it is instructive to consider the case when $A$ is the adjacency matrix of an unweighted graph $G$. Then $D$ is the degree diagonal matrix with $D_{uu} = deg_G(u)$, $L=J$ and $B$ corresponds to the non-backtracking matrix of $G$.

Throughout the other sections of the paper, for a given non-backtracking matrix $B\in \R^{2m\times 2m}$, we will consider the related extension matrix $B^*\in \R^{2n^2\times 2n^2}$ with entries
\begin{align*}
	B^*_{ef} = 
	\begin{cases}
		B_{ef}&\text{ if }e,f\in E(G)\\
		0&\text{ otherwise.}
	\end{cases}
\end{align*}
For simplicity of the notation, we will often denote  $B^*$ simply by $B$. The context will always be clarified by the ambient dimension. 
We can now state the main result of the section.

\begin{theorem}[Generalized Ihara-Bass Theorem]\label{theorem:general-ihara-bass}
	Let $n,m$ be integers and let $A\in \R^{n\times n}$ be  a symmetric matrix with $m$ non-zero entries, all off-diagonal. Let $B,L,J,D$ defined as above.
	Then, for any $u\in \R$,
	\begin{align*}
		\det\Paren{\Id_{2m}-u(B+ L-J)} = (1-u^2)^{m-n}\Paren{\Id_n-uA+u^2D-u^2\Id_n}\,.
	\end{align*}
\end{theorem}

Our proof of \cref{theorem:general-ihara-bass} closely resembles the proof of Bass  \cite{bass1992ihara}. We first observe that the matrices above satisfy several useful identities, than tackle the theorem.
\begin{lemma}\label{lemma:identities-graph-matrices}
	Using the definitions above:
	\begin{enumerate}[i)]
		\item $SJ = T$ and $TJ = S\,,$
		\item $A= S\transpose{T}\,,$
		\item $D = S\transpose{S} = T \transpose{T}\,,$
		\item $B+ L = \transpose{T}S\,.$
	\end{enumerate}
	\begin{proof}
		For i), notice that $SJ\in \R^{n\times 2m}$ and $SJ_{ue} = \iprod{S_{u,-}, J_{-,e}}= S_{u e^{-1}} = T_{u e}$, where in the third step we used symmetry of $A$. A similar argument can be made to show $TJ=S$.
		For ii) observe that
		\begin{align*}
			A_{uv} = \iprod{S_{u,-}, T_{v,-}} = \sum_e S_{ue} T_{ve}
		\end{align*}
		which is nonzero only when $e=uv$. In that case, by definition $A_{uv} = \sigma_{uv}\Abs{A_{uv}} = S_{ue} T_{ve}$ since either $u< v$ or $u> v$.
		Consider now $S\transpose{S}$, the matrix is diagonal since each edge has at most one source vertex, then
		\begin{align*}
			(S\transpose{S})_{uu} = \sum_e S_{ue}^2 = \sum_{v \in N^+(u)} \Abs{A_{uv}}= D_{uu}\,.
		\end{align*}
		A symmetric derivation  shows $D_{uu}= (T\transpose{T})_{uu}$.
		It remains to prove iv).
		It is trivial to check that
		\begin{align*}
			(\transpose{T}S)_{ee} = \iprod{T_{-, e}, S_{-,e}} = \sum_{u} T_{ue} S_{ue}=0\,,
		\end{align*}
		since there are no self-loops in the graph.
		For distinct $e, f\in [2m]$
		\begin{align*}
			(\transpose{T}S)_{ef} = \sum_{u} T_{ue} S_{uf}\,.
		\end{align*}
		There is at most one non-zero element in the sum, corresponding to the case when $u$ is the target vertex of $e$ and the source of $f$, which means $ef$ is a walk of length $2$ in $G$.
		If $ef$ is a non-backtracking walk (that is, $e\neq f^{-1}$) then $B_{ef} = 	(\transpose{T}S)_{ef}$ and $L_{ef}=0$.
		Conversely, if $e=f^{-1}$ then $B_{ef} =0$ and $L_{ef} = 	(\transpose{T}S)_{ef}$. Finally, signs can be checked case by case.
	\end{proof}
\end{lemma}

We are now ready to prove \cref{theorem:general-ihara-bass}.

\begin{proof}[Proof of \cref{theorem:general-ihara-bass}]
	In the following identities all matrices are $(n+2m)\times (n+2m)$ block matrices where the first block has size $n\times n$. Let $u\in \R$,
	\begin{align}
		\label{eq:ihara-bass-proof-m1}
		&\begin{pmatrix}
			\Id_n & 0\\
			\transpose{T} & \Id_{2m}
		\end{pmatrix}
		\begin{pmatrix}
			\Id_n(1-u^2) & Su\\
			0 & \Id_{2m} - (B+ L-J)u
		\end{pmatrix}\\
		&=\begin{pmatrix}
			\Id(1-u^2) & Su\\
			\transpose{T}(1-u^2) & \transpose{T}Su + \Id_{2m} -(B+L-J)u
		\end{pmatrix}\nonumber\\
		&=\begin{pmatrix}
			\Id(1-u^2) & Su\\
			\transpose{T}(1-u^2) &\Id_{2m} +Ju
		\end{pmatrix}\,.\nonumber
	\end{align}
	On the other hand
	\begin{align}
		\label{eq:ihara-bass-proof-m2}
		&\begin{pmatrix}
			\Id_n(1-u^2) - Au + D u^2 & Su\\0 & \Id_{2m} + Ju
		\end{pmatrix}
		\begin{pmatrix}
			\Id_n & 0\\
			\transpose{T}-\transpose{S}u& \Id_{2m}
		\end{pmatrix}\\
		&= \begin{pmatrix}
			\Id_n(1-u^2)  -Au+Du^2 +S\transpose{T}u-S\transpose{S}u^2& Su\\
			\transpose{T}-\transpose{S}u + J\transpose{T}u-J\transpose{S}u^2& \Id_{2m}+ Ju
		\end{pmatrix}\nonumber\\
		&= \begin{pmatrix}
			\Id_n(1-u^2) & Su\\
			\transpose{T}(1-u^2)& \Id_{2m} +Ju
		\end{pmatrix}\,.\nonumber
	\end{align}
	Putting \cref{eq:ihara-bass-proof-m1} and \cref{eq:ihara-bass-proof-m2} together and taking determinants we get
	\begin{align*}
		(1-u^2)^n\det\Paren{\Id_{2m}-(B+L-J)u}=\det\Paren{\Id_n(1-u^2)-Au+Du^2}\det\Paren{\Id_{2m}+Ju}\,.
	\end{align*}
	Now notice that 
	\begin{align*}
		\Id_{2m}+Ju = 
		\begin{pmatrix}
			\Id_m & \Id_mu\\
			\Id_mu & \Id_m\\
		\end{pmatrix}
	\end{align*}
	and thus $\det\Paren{\Id_{2m}+Ju} = (1-u^2)^{m}$. Rearranging, the result follows. 
\end{proof}

\subsection{Norm bounds via the Ihara-Bass formula}\label{section:norm-bounds-via-ihara-bass}

In this section we show how \cref{theorem:general-ihara-bass} can be used to study the spectrum of a real symmetric matrix $A$ via the spectrum of related matrices. The central tool is the theorem below.

\begin{theorem}\label{theorem:lowner-order-A}
	Let $A\in \R^{n\times n}$ a  symmetric matrix with zero diagonal. Let $B\,, L\,, J\,, D$ be as defined in \cref{section:general-ihara-bass}. Let $\lambda_{\min}$ be the smallest eigenvalue of the matrix $B+L-J\in \R^{2m\times 2m}$. Then for any $\lambda \leq \lambda_{\min}$
	\begin{align*}
		A \sge - \Abs{\lambda}\Id_n  - \Abs{\lambda}^{-1}(D-\Id_n)\,.
	\end{align*}
	\begin{proof}
		Let $\lambda_{\min}$ be the smallest real eigenvalue of $B+L-J$.	By \cref{theorem:general-ihara-bass} we know $-1$ is a real eigenvalue of  $B+L-J$ and thus $\lambda_{\min}\leq -1$. Moreover, for every $\lambda< \lambda_{\min}$ we have $\det\Paren{\Id_{2m}-\lambda^{-1}B+ \lambda^{-1}L-\lambda^{-1}J}\neq 0$ otherwise $\lambda$ would be an eigenvalue smaller than $\lambda_{\min}$.
		Define the matrix
		\begin{align*}
			M_{\lambda}:= \Id_n - \lambda^{-1}A+\lambda^{-2}(D-\Id_n)\,.
		\end{align*}
		By the same reasoning as in \cref{theorem:general-ihara-bass}, $\det(M_\lambda)\neq 0$ as long as $\lambda < \lambda_{\min}$. We make the stronger claim
		\begin{align*}
			\forall \lambda,\,\lambda_{\min}\,:\, M_\lambda \sg \mathbf{0}\,.
		\end{align*}
		To prove the above claim, suppose toward a contradiction that $\lambda' < \lambda_{\min}$  is such that $M_{\lambda'}$ has a negative eigenvalue. Since $M_\lambda$ tends to $\Id_n$ when $\lambda \rightarrow -\infty$, there is a value $\lambda_{PD} < \lambda'$  such that $M_{\lambda_{PD}}$ is strictly positive definite.
		Consider now the smallest eigenvalue of $M_\lambda$ for values of $\lambda$ in the range $(\lambda_{PD}, \lambda')$. The smallest eigenvalue of $M_\lambda$ varies continuously with $\lambda$, it is positive for $\lambda = \lambda_{PD}$ and it is negative for $\lambda = \lambda'$, so it must be equal to zero for some $\lambda^*\leq \lambda' < \lambda_{\min}$. But this means that $\det(M_{\lambda^*} ) = 0$ and so $\lambda^*$ is an eigenvalue of $B+L-J$, which contradicts the definition of $\lambda_{\min}$. We have thus established our claim. Rearranging the result follows.
	\end{proof}
\end{theorem}

A crucial consequence of \cref{theorem:lowner-order-A} is that, exploiting the diagonal structure of the matrices $D\,,\Id_n$ one can bound the norm $\Normio{A}$ as a function of the smallest eigenvalue of the associated non-backtracking matrix.

\begin{corollary}\label{corollary:bound-norm-A}
		Let $A\in \R^{n\times n}$ a symmetric matrix with zero diagonal. 
		Let $\lambda_{\min}$ and $\lambda_{\min}'$  be respectively the smallest eigenvalue of the matrix $B(A)+L(A)-J(A)\in \R^{2m\times 2m}$ and  $B(-A)+L(A)-J(A)\in \R^{2m\times 2m}$, for $B\,, L\,, J\,, D$ as defined in \cref{section:general-ihara-bass}. Then, for any $\lambda\geq \max\Set{\Abs{\lambda_{\min}}\,, \Abs{\lambda_{\min}'}}$,
		\begin{align*}
			\Normio{A}\leq 2\Tr\Abs{\Paren{\lambda\Id_n+\lambda^{-1}(D(A)-\Id_n)}}
			%\Normio{A}\leq 3\max&\left\{\Tr\Abs{\Paren{\Abs{\lambda_{\min}}\Id_n+\Abs{\lambda_{\min}}^{-1}(D(A)-\Id_n)}}\right.\,,\\
			%&\,\,\left. \Tr\Abs{\Paren{\Abs{\lambda_{\min}'}\Id_n+\Abs{\lambda_{\min}'}^{-1}(D(A)-\Id_n)}}\right\}\,.
		\end{align*}
		\begin{proof}
			Define
			\begin{align*}
				R&:=\Abs{\lambda\Id_n+\lambda^{-1}(D(A)-\Id_n)}\,.%\\
				%R^*& :=\Abs{\Abs{\lambda_{\min}'}\Id_n+\Abs{\lambda_{\min}^*}^{-1}(D(A)-\Id_n)}\,.
			\end{align*}
			By  \cref{theorem:lowner-order-A} for any $x\in \Set{\pm 1}^n$ we have $\Abs{\transpose{x}Ax}\leq \Abs{\transpose{x}Rx}$.
			For any $y\in\Set{\pm 1}^n$ we can write
			\begin{align*}
				2\Abs{\transpose{x}Ay}&\leq \Abs{\transpose{(x+y)}A(x+y)-\transpose{x}Ax-\transpose{y}Ay}\\
				&\leq  				\Abs{\transpose{(x+y)}A(x+y)}+\Abs{\transpose{x}Ax}+\Abs{\transpose{y}Ay}\,.
			\end{align*}
		Now $x+y\in\Set{-2,0,+2}^n$ and thus
		\begin{align*}
			\Abs{\transpose{(x+y)}A(x+y)}\leq 4\max_{z\in \Set{\pm 1}^n}\transpose{z}Rz\,,%2\max\Set{\max_{z\in \Set{\pm 1}^n}\transpose{z}Rz\,, \max_{z\in \Set{\pm 1}^n}\transpose{z}R^*z}\,,
		\end{align*}
		the result follows by definition of $R$.
		\end{proof}
\end{corollary}

\section{Warm-up: spectrum of  binary matrices with dependencies}\label{section:warm-up}

In preparation to a proof of \cref{theorem:main-xor}, we show here how to use the ideas of \cref{section:norm-bounds-via-ihara-bass} to study the spectrum of random  symmetric matrices with entries in $\Set{-1\,,0\,, 1}$, even when dependencies between the entries appear.
As we may view any symmetric matrix as the adjacency matrix of a weighted graph (up to the diagonal entries) we partially shift to the language of graphs. In particular, we study graphs sampled from the following family of distributions.

\begin{definition}[$\gamma$-wise independent random binary-weighted graphs]\label{definition:almost-independent-graph-distribution}
	$\cD_{d,\gamma}$ is the family of distributions $P_{d,\gamma}$ over graphs with vertex set $[n]$ and adjacency matrix satisfying:
	\begin{enumerate}
		\item for all $uv \in E(K_n)\,,$ $\bbP_{\mathbf{G}\sim P_{d,\gamma}} \Brac{A_{uv}(\mathbf{G})\neq 0 }=\bbP_{\mathbf{G}\sim P_{d,\gamma}} \Brac{uv\in E(\mathbf{G}) }=d/n\,.$
		\item for all $uv \in E(K_n)\,,$ the distribution of $A_{uv}(\mathbf{G})$ is symmetric with support $\{-1\,,0\,,+1\}\,.$
		\item edges in $\mathbf{G}$ are $\gamma$-wise independent. %the entries of $A(\mathbf{G})$ are $\gamma$-wise independent.%random variables $A_e(\mathbf{G}),\ldots w_{\mathbf{G}}(e_{\binom{n}{2}})$ 
	\end{enumerate}
\end{definition}
Notice that the adjacency matrix of $\mathbf{G}$ is a random binary symmetric matrix with $\gamma$-wise independent entries (up to symmetries).
With a slight abuse of notation, we  write  $\mathbf{G}\sim P_{d,\gamma}$ and $A(\mathbf{G})\sim P_{d,\gamma}$, respectively for a graph sampled from $ P_{d,\gamma}$ in  $\cD_{d,\gamma}$ and for the adjacency matrix of a graph sampled from this distribution. 
We consider the associated matrices $D(\mathbf{A})\,, B(\mathbf{A})$ as defined in \cref{section:general-ihara-bass}.
We prove the following theorem.

\begin{theorem}\label{theorem:main-xor-binary}
	Let $n$ be an integer, $d>0$ and $\gamma \geq C\log_d^2 n$ for a large enough universal constant $C>0$. Consider a distribution $P_{d,\gamma}\in \cD_{d,\gamma}$. Then for $\mathbf{A}\sim P_{d,\gamma}$ 
	\begin{align*}
		\Normio{\mathbf{A}}\leq O\Paren{n\sqrt{d}}\,, %\mathbf{A}\sge \Omega\Paren{-\sqrt{d}\cdot \Id_n + \frac{1}{\sqrt{d}}(D(\mathbf{A})-\Id_n)}
	\end{align*}
	with probability $0.99$.%\Tnote{To improve the probability, improve the concentration of the average degree result}
\end{theorem}
Observe that \cref{theorem:main-xor-binary} shows how the adjacency matrix of a graph sampled from some distribution in $\cD_{d,\gamma}$  behaves --up to a universal constant-- as the adjacency matrix of an \Erdos-\Renyi graph with expected degree $d$. As we are oblivious to the specific correlations in the graph, the result only relies on local independence of the edges.
The central tool behind \cref{theorem:main-xor-binary} is the following lemma.

\begin{lemma}\label{lemma:spectrum-power-binary-non-backtracking}
	Consider the settings of \cref{theorem:main-xor-binary}.  Let $ z\leq \frac{\log_d n}{10}$. For $\mathbf{A}\sim P_{d,\gamma}$, the associated non-backtracking matrix satisfies
	\begin{align*}
		\Norm{B^{z-1}(\mathbf{A})}\leq O(d^{z/2})\cdot (\log n)^{O(\log\log n)}\,.\\
	\end{align*}
	with probability $1-o(1)$.
\end{lemma}
The choice of the non-backtracking matrix $B^{z-1}(\mathbf{A})$ stems from the observation that is contains more structural information about the graph, compared to $B(\mathbf{A})$. Indeed, the singular values of $B(\mathbf{A})$ contain only information on the degree sequence of the underlying graph. Another perspective on this can be obtained recalling that for any natural norm $\Norm{\cdot}_*$, by Gelfand's formula \cref{fact:bound-spectral-radius}, $\rho(B(\mathbf{A}))=\lim_{z\rightarrow \infty}\Norm{B(\mathbf{A})^z}^{1/z}_*$. Hence one can expect that a careful analysis on non-backtracking matrices of sufficiently large length, would yield a tighter bound on the spectral radius (see \cite{BordenaveLM15} for a more in-depth discussion).
Indeed, by \cref{fact:bound-spectral-radius}, a consequence of \cref{lemma:spectrum-power-binary-non-backtracking} is the following result regarding the spectral radius of $B(\mathbf{A})$.

\begin{corollary}\label{corollary:radius-binary-non-backtracking}
	Consider the settings of \cref{theorem:main-xor-binary}. For $\mathbf{A}\sim P_{d,\gamma}$, the associated non-backtracking matrix satisfies
	\begin{align*}
		\rho\Paren{B(\mathbf{A})}\leq O(\sqrt{d})\,.
	\end{align*}
	with probability $1-o(1)$.
\end{corollary}

We can combine \cref{corollary:radius-binary-non-backtracking} with \cref{corollary:bound-norm-A} to obtain the theorem.

\begin{proof}[Proof of \cref{theorem:main-xor-binary}]
	By \cref{corollary:radius-binary-non-backtracking} for both $\mathbf{A}$ and $-\mathbf{A}$ we get $\rho\Paren{B(\mathbf{A})}\leq O(\sqrt{d})$ and $\rho\Paren{B(-\mathbf{A})}\leq O(\sqrt{d})$.  By \cref{lemma:avg-degree-sparse-dependent-graph} we have $\Tr D(\mathbf{A})\leq O(nd)$ with probability at least $0.999$, where $D(\mathbf{A})$ is the associated degree matrix as defined in \cref{section:general-ihara-bass}. Finally applying \cref{corollary:bound-norm-A} the result follows.
\end{proof}

The rest of the section is devoted to the proof of \cref{lemma:spectrum-power-binary-non-backtracking}. 
We will prove  the result via the trace method.

\begin{proposition}\label{proposition:trace-method}
	Let $\mathbf{M}$ be an $n$-by-$n$ random matrix and $c>0$. Then
	\begin{align*}
		\E_{\mathbf{M}}\Brac{\Tr \Paren{\mathbf{M}\transpose{\mathbf{M}}}^q}\leq \gamma \Longrightarrow \bbP \Paren{\Norm{\mathbf{M}}\geq c\cdot \gamma^{1/2q}}\leq c^{-2q}\,.
	\end{align*}
\end{proposition}

In \cref{section:block-non-backtracking-matrices} we illustrate how the trace computation can be reduced to a path counting problem. In \cref{section:expectation-block-non-backtracking} we compute the expectation of such paths. Finally we put things together  in \cref{section:proof-binary-theorem}.

\subsection{Powers of non-backtracking matrices}\label{section:block-non-backtracking-matrices}

%Our object of interest is thus the matrix $B(\mathbf{A})$, which we simply denote by $\mathbf{B}$.
For simplicity, we will  write $B$ in place of $B(A)$ for an adjacency matrix $A$.
Moreover we will consider the $2n^2$-by-$2n^2$ extension of $B(A)$ as described in \cref{section:general-ihara-bass}. Indeed this extension has the same eigenvalues and singular values (up to some additional zero-value ones). We will overload our notation and simply denote it by $B$.
%For simplicity, we will simply write $B$ in place of $B(A)$ for an adjacency matrix $A$.
%For a given non-backtracking matrix $B\in \R^{2m\times 2m}$, we consider the related matrix $B^*\in \R^{2n^2\times 2n^2}$ with entries
%\begin{align*}
%	B^*_{ef} = 
%	\begin{cases}
%		B_{ef}&\text{ if }e,f\in E(G)\\
%		0&\text{ otherwise.}
%	\end{cases}
%\end{align*}
%Recall here $G$ is the graph associated with $A$.
%As eigenvalues and singular values of $B^*, B$ coincide (up to some additional zero-value ones). It suffices to study the spectrum of $B^*$. For this reason, we will henceforth denote  $B^*$ simply by $B$. The context will be clarified by the ambient dimension.
We now explain how to reduce the computation of the trace of powers of non-backtracking walks to a graph counting argument.
We start by introducing additional notions.\bigskip

For a graph $G$ and $e,f \in \vec{E}(G)$, we write $\nbw{ef}{z}$ to denote the set of length $z$ non-backtracking walks in $G$ starting with $e$ and ending with $f$. 
Then $\nbw{}{z}(G) = \underset{e,f \in \vec{E}(G)}{\bigcup} \nbw{ef}{z}(G)$.
For simplicity we let $\nbw{ef}{z} := \nbw{ef}{z}(K_n)$.
A walk $W\in \nbw{}{2q,z}$ over $e$ distinct edges is said $t$-\textit{tangle}-free if the number of vertices in the walk is at least $e-t+1$. That is, any minimum spanning tree of the subgraph traversed by the walk has $e-t$ edges. For $t=0$ the definition implies the walk is a path. We remark that this definition differs from the one in \cite{BordenaveLM15} (which corresponds to $t< 2$). We adopt this different definition in preparation of our arguments in \cref{section:random-3xor}.
A graph $G$ is said $t$-\textit{tangle}-free if every walk in  $ \nbw{}{2q,z}(G)$ is $t$-tangle-free.
We use $\tgf{ef}{z, t}(G)\subseteq \nbw{ef}{z}(G)$ to denote the subset of $t$-tangle-free walks.
When the value of $t$ is clear from context we will simply say $W$ is tangle-free and write $\tgf{ef}{z}(G)\subseteq \nbw{ef}{z}(G)$ for the corresponding set. %$\btgf{e}{q,k}$.

We can extend the notion of closed non-backtracking matrices by introducing closed block non-backtracking matrices. $\bnbw{e}{q,z}$ is the set of walks $W$ of length $q\cdot z$ with starting edge $e\in \vec{E}(G)$ and ending edge $e^{-1}\in \vec{E}(G)$ satisfying:
\begin{enumerate}[(i)]
	\item $W$ can be partitioned into $q$ non-backtracking walks $W_1,\ldots, W_q$ of length $z$.
	\item For each pair of walks $W_{i}\,, W_{i+1}$ (with the convention $W_{q+1}=W_1$) the starting edge of $W_{i+1}$ is the inverse of the ending edge of $W_i$. I.e. we have $e_z(W_i) = e_1^{-1}(W_{i+1})$.
\end{enumerate}
Similarly, $\btgf{e}{q,z, t}\subseteq \bnbw{e}{q,z}$ is the set of closed block non-backtracking walks such that each block is $t$-tangle-free.
Given $W \in \bnbw{e}{q,z}$, we use $W_i\in \nbw{}{z}$ to denote the $i$-th block of $W$ (where the first block maybe chosen arbitrarily among  the ones starting at $e$). 
For $uv \in \vec{E}(G)$, recall we write $\sigma_{uv}=\sign(A_{uv})$, we further let 
\begin{align*}
    \tilde{\sigma}_{uv} = \begin{cases}
    \sigma_{uv}&\text{ if }u<v\,,\\
    1&\text{ otherwise.}
    \end{cases}
\end{align*}

This notation will be used to describe the $k$-th power of the non-backtracking matrix.

\begin{fact}\label{fact:expansion-power-non-backtracking-matrix}
	Let $G$ be a weighted graph with $m$ edges. Let $A\in \R^{n\times n}$ be its adjacency matrix and $B\in \R^{2n^2\times n^2}$ the associated non-backtracking matrix.
	Let $z$ be a positive integer and let $e,f \in \vec{E}(G)$. Then for $e_1=e\,, e_z=f$
	\begin{align*}
		(B^{z-1})_{ef} = \sum_{W\in \nbw{ef}{z}} \tilde{\sigma}_{e_1^{-1}}\cdot\tilde{\sigma}_{e_z} \cdot \sqrt{\Abs{A_{e_1}\cdot A_{e_{z}}}}\cdot \prod_{s=2}^{z-1}A_{e_{s}}\,.
	\end{align*}
\end{fact}
\cref{fact:expansion-power-non-backtracking-matrix} can be checked simply by expanding $B^{z-1}$.
We can now define the tangle-free $2n^2$-by-$2n^2$ non-backtracking matrix $B^{(z-1)_t}$ associated to $B^{z-1}$ (we will drop the subscript $t$ when the context is clear). For $e,f\in [2n^2]$,
\begin{align*}
	(B^{(z-1)_t})_{ef} := \sum_{W\in \tgf{ef}{z,t}}
	\tilde{\sigma}_{e_1^{-1}}\cdot\tilde{\sigma}_{e_z}\cdot\sqrt{\Abs{A_{e_1}\cdot A_{e_{z}}}}\cdot \prod_{s=2}^{z-1}A_{e_{s}}\,.
\end{align*}
 It will be convenient to decompose the terms $(B^{z-1})_{ef}$ as follows:
\begin{align*}
	(B^{z-1})_{ef} =& \sum_{W\in \nbw{ef}{z}} \tilde{\sigma}_{e_1^{-1}}\cdot\tilde{\sigma}_{e_z}\sqrt{\Abs{A_{e_1}\cdot A_{e_{z}}}}\cdot \prod_{s=1}^{z-1}A_{e_{s}}\\
	=& \sum_{W\in \tgf{ef}{z,t}}\tilde{\sigma}_{e_1^{-1}}\cdot\tilde{\sigma}_{e_z} \sqrt{\Abs{A_{e_1}\cdot A_{e_{z}}}}\cdot \prod_{s=2}^{z-1}A_{e_{s}} \\
	&+ \sum_{W\in \nbw{ef}{z}\setminus\tgf{ef}{z,t}}\tilde{\sigma}_{e_1^{-1}}\cdot\tilde{\sigma}_{e_z}\sqrt{\Abs{A_{e_1}\cdot A_{e_{z}}}}\cdot \prod_{s=2}^{z-1}A_{e_{s}}\\
	&= (B^{(z-1)_t})_{ef}\\
	&+ \sum_{W\in \nbw{ef}{z}\setminus\tgf{ef}{z,t}}\tilde{\sigma}_{e_1^{-1}}\cdot\tilde{\sigma}_{e_z}\sqrt{\Abs{A_{e_1}\cdot A_{e_{z}}}}\cdot \prod_{s=2}^{z-1}A_{e_{s}}\,.
\end{align*}
%We write 
%\begin{align*}
	%(T^k)_{ef}:= \sum_{W\in \tgf{ef}{k+1}}\alpha_{e,s}\cdot \alpha_{e,t} \cdot \sqrt{\Abs{A_{e_1}\cdot A_{e_{k+1}}}}\cdot \prod_{s=1}^{k-1}A_{e_{s+1}}\,.
%\end{align*}
Notice that for a $t$-tangle-free graph, $B^{z-1}=B^{(z-1)_t}\,.$ The trace of powers of $B^{z-1}$ can also be written as a sum over block non-backtracking walks. In particular, we can obtain the fact below observing that even though $B$ is not a normal matrix, it has some symmetry as $$\Brac{\transpose{\Paren{B^{(z-1}}}}_{ef} = \Paren{B^z}_{fe} = \Paren{B^z}_{e^{-1}f^{-1}}\,.$$ Conveniently, the factors $\tilde{\sigma}_{e_1^{-1}}\cdot\tilde{\sigma}_{e_z}$ disappear. 
 
\begin{fact}\label{fact:trace-expansion-non-backtracking-power}
	Let $G$ be a weighted graph with $m$ edges. Let $A\in \R^{n\times n}$ be its adjacency matrix and $B\in \R^{2n^2\times 2n^2}$ the associated non-backtracking matrix. Let $q, z\geq 2$ and $t$ be integers. Then 
	\begin{align*}
		\Tr \Brac{B^{z-1}\transpose{\Paren{B^{z-1}}}}^q = \sum_{W\in \bnbw{}{2q,z}} \prod_{i=1}^{2q}\Brac{\sqrt{\Abs{A_{e_1(W_i)}\cdot A_{e_z(W_i)}}}\cdot\Paren{\prod_{s=2}^{z-1}A_{e_{s}(W_i)}}}
	\end{align*}
	and
	\begin{align*}
		\Tr \Brac{B^{(z-1)_t}\transpose{\Paren{B^{(z-1)_t}}}}^q = \sum_{W\in \btgf{}{2q,z,t}} \prod_{i=1}^{2q}\Brac{ \sqrt{\Abs{A_{e_1(W_i)}\cdot A_{e_z(W_i)}}}\cdot\Paren{\prod_{s=2}^{z-1}A_{e_{s}(W_i)}}}\,.
	\end{align*}
\end{fact}

\subsection{Expectation of block non-backtracking walks}\label{section:expectation-block-non-backtracking}

By \cref{proposition:trace-method} and \cref{fact:trace-expansion-non-backtracking-power} we need to  study the expectation of closed block non-backtracking walks under distributions in $\cD_{d, \gamma}$. First we observe that, if an edge in a walk in $\bnbw{}{2q,z}$ has multiplicity one, then by symmetry the expectation of the whole walk is zero.

\begin{fact}
	Let $n$ be an integer, $d>0\,,$,$0<\gamma\leq n$. Consider a distribution $P_{d, \gamma}\in \cD_{d,\gamma}$.
	Let $W\in \bnbw{}{2q,z}$ with $z>1$ and $2q\cdot z\leq \gamma$. If there exists $e\in E(W)$ with multiplicity $m_W(e)=1$, then
	\begin{align}
		\E_{\mathbf{A}\sim P_{d, \gamma}} \prod_{i=1}^{2q}\Brac{ \sqrt{\Abs{\mathbf{A}_{e_1(W_i)}\cdot \mathbf{A}_{e_z(W_i)}}}\cdot\Paren{\prod_{s=2}^{z-1}\mathbf{A}_{e_{s}(W_i)}}}=0\,.
	\end{align}
	\begin{proof}
		By assumption there must be $e\in E(W)$ with $m_W(e)=1$ and by construction this cannot be any of the starting edges $e_{t\cdot z+1}$, with $0\leq t\leq 2q-1$ in $W$. 
		Let's denote this edge by $f\in \E(W)$.
		Then we may write
		\begin{align*}
		\E_{\mathbf{A}\sim P_{d, \gamma}}& \prod_{i=1}^{2q}\Brac{ \sqrt{\Abs{\mathbf{A}_{e_1(W_i)}\cdot \mathbf{A}_{e_z(W_i)}}}\cdot\Paren{\prod_{s=2}^{z-1}\mathbf{A}_{e_{s}(W_i)}}} \\
		&= \E_{\mathbf{A}\sim P_{d, \gamma}} \Brac{\prod_{i=1}^{2q}\Brac{ \sqrt{\Abs{\mathbf{A}_{e_1(W_i)}\cdot \mathbf{A}_{e_z(W_i)}}}\cdot\Paren{\prod_{s=2}^{z-1}\mathbf{A}_{e_{s}(W_i)}}}\cdot \frac{1}{\mathbf{A}_{f}}}\cdot \E_{\mathbf{A}\sim P_{d, \gamma}}  \mathbf{A}_{f}\,,
		\end{align*}
		where we used $\gamma$-wise independence of the edges.
		By symmetry of the distribution $\E  \mathbf{A}_{f}=0$. The result follows.
	\end{proof}
\end{fact}

For any $t\geq 0$, we say that a tangle-free walk $W\in \btgf{}{2q,z, t}$ is \textit{interesting} if there is no edge in $W$ with multiplicity one. We denote the set of interesting walks in $\btgf{}{2q,z, t}$ by $\ibtgf{}{2q,z, t}$. The next fact bounds the expectation of interesting closed block non-backtracking walks.

\begin{fact}\label{fact:expectation-binary-block-nbw}
	Let $n$ be an integer, $d>0\,,$ and $0<\gamma\leq n$. Consider a distribution $P_{d, \gamma}\in \cD_{d,\gamma}$.
	Let $W\in \ibtgf{}{2q,z, t}$ with $z>1, t\geq 0$ and $2q\cdot z\leq \gamma$. Then
	\begin{align*}
		\E_{\mathbf{A}\sim P_{d, \gamma}} \prod_{i=1}^{2q}\Brac{ \sqrt{\Abs{\mathbf{A}_{e_1(W_i)}\cdot \mathbf{A}_{e_z(W_i)}}}\cdot\Paren{\prod_{s=2}^{z-1}\mathbf{A}_{e_{s}(W_i)}}} \leq \Paren{\frac{d}{n}}^{\Card{E(W)}}\,.
	\end{align*}
	\begin{proof}
		By $\gamma$-wise independence,
		\begin{align*}
			\E_{\mathbf{A}\sim P_{d, \gamma}} \prod_{i=1}^{2q}\Brac{ \sqrt{\Abs{\mathbf{A}_{e_1(W_i)}\cdot \mathbf{A}_{e_z(W_i)}}}\cdot\Paren{\prod_{s=2}^{z-1}\mathbf{A}_{e_{s}(W_i)}}} &\leq \prod_{e\in E(W)} \E_{\mathbf{A}\sim P_{d, \gamma}} \Abs{\mathbf{A}_{e}}^{m_W(e)}\\
			&=\prod_{e\in E(W)}  \E_{\mathbf{A}\sim P_{d, \gamma}}  \Abs{\mathbf{A}_{e}}\\
			&=\prod_{e\in E(W)} \Paren{\frac{d}{n}}\,.
		\end{align*}
	\end{proof}
\end{fact}

\subsection{Bound on the spectrum of non-backtracking matrices}\label{section:proof-binary-theorem}

We are ready to tackle \cref{lemma:spectrum-power-binary-non-backtracking}. Our proof consists of two main ingredients.
First we shows that, for a wide range of parameters, graphs sampled from distributions in $\cD_{d, \gamma}$ are tangle-free. Then we prove that the number of closed, tangle-free, block non-backtracking walks is not large. Combining these results with
\cref{fact:expectation-binary-block-nbw} will conclude the proof.

\begin{lemma}\label{lemma:tangle-free-sparse-graph}
	Let $n$ be a large enough integer. Let  $d>0$, $z \leq \log_d n\leq  \gamma\leq n$. Consider a distribution $P_{d, \gamma}\in \cD_{d,\gamma}$.
	Then for $t\geq 100\log\log_d n$, $\mathbf{G}\sim  P_{d,\gamma}$ is $t$-tangle-free with probability at least $1-o(1)$.
	\begin{proof}
		For $t'\geq t$, the number of non-backtracking walks over $v$ vertices of length $z$ with  $t'$-tangles is upper bounded by $n^v v^z$.
		Each such walk $W$ appears in the graph with probability $\Paren{\frac{d}{n}}^{e}\leq \Paren{\frac{d}{n}}^{t'+v-1}$. For $t'\geq t$, by union bound the probability that such a walk appears in the graph is at most $\Paren{vd}^{\log_d n}n^{-t+1}=o(1)$.
	\end{proof}
\end{lemma}

The fact that our graphs of interest are with high probability $O(\log\log_d n)$-tangle-free means we need only to focus on this small subset of closed block non-backtracking walks. For the rest of the section we will fix
\begin{align*}
	t=100\log\log_d n
\end{align*}  and drop the superscript $t$. We will refer to $O(\log\log n)$-tangle-free walks simply by tangle-free walks and write $\mathbf{B}^{(z-1)}$ in place of $\mathbf{B}^{(z-1)_t}$.
We say that a walk $W\in \ibtgf{}{2q,z}$ over $v$ vertices is \textit{canonical} if its set of vertices is $[v]\subseteq[n]$ and the vertices are first visited in order.
The use of canonical paths is convenient as, by construction, no two canonical paths are isomorphic (so we are in fact choosing an arbitrary element for each equivalence class). We denote the set of canonical paths by $\cW^{2q,z}\subseteq\ibtgf{}{2q,z}$. Notice that for $P \in \cW^{2q,z}$, there are $\binom{n}{v}(v)!$ isomorphic walks in $\ibtgf{}{2q,z}$.

\begin{lemma}[Enumeration of canonical paths]\label{lemma:enumeration-binary-canonical-paths}
	Let $\cW^{2q,z}(v,e)$ be the set of canonical paths with $v$ vertices and $e$ distinct edges. We have
	\begin{align*}
		\Card{\cW^{2q,z}(v,e)}\leq z^{4tq}\cdot (2zq)^{6tq\cdot (e-v+1)}\,.
	\end{align*}
\end{lemma}

We defer the proof of \cref{lemma:enumeration-binary-canonical-paths} to \cref{section:deferred-proofs} and use it here to prove the main lemma of the section.

\begin{proof}[Proof of \cref{lemma:spectrum-power-binary-non-backtracking}]
	First notice that by \cref{lemma:tangle-free-sparse-graph}, with high probability $\mathbf{B}^{z-1}=\mathbf{B}^{(z-1)}$, thus it suffices to bound the spectral norm of $\mathbf{B}^{(z-1)}$.
	Then
	\begin{align*}
		\E&\Brac{ \Tr \Brac{\mathbf{B}^{(z-1)}\transpose{\Paren{\mathbf{B}^{(z-1)}}}}^q } \\
		&= \sum_{W\in \ibtgf{}{2q,z}} \E \Brac{\prod_{i=1}^{2q}\Brac{ \sqrt{\Abs{\mathbf{A}_{e_1(W_i)}\cdot \mathbf{A}_{e_z(W_i)}}}\cdot\Paren{\prod_{s=2}^{z-1}\mathbf{A}_{e_{s}(W_i)}}}}\\
		&\leq \sum_{W\in \ibtgf{}{2q,z}} \E \prod_{i=1}^{2q}\Brac{ \sqrt{\Abs{\mathbf{A}_{e_1(W_i)}\cdot \mathbf{A}_{e_z(W_i)}}}\cdot\Paren{\prod_{s=2}^{z-1}\mathbf{A}_{e_{s}(W_i)}}}\,.
	\end{align*}
	Now by \cref{fact:expectation-binary-block-nbw} and \cref{lemma:enumeration-binary-canonical-paths}
	\begin{align*}
		\sum_{W\in \ibtgf{}{2q,z}} \E &\prod_{i=1}^{2q}\Brac{ \sqrt{\Abs{\mathbf{A}_{e_1(W_i)}\cdot \mathbf{A}_{e_z(W_i)}}}\cdot\Paren{\prod_{s=2}^{z-1}\mathbf{A}_{e_{s}(W_i)}}}\\
		&\leq 	\sum_{W\in \ibtgf{}{2q,z}} \Paren{\frac{d}{n}}^{E(W)}\\
		&\leq \sum_{v\geq 3}^{z\cdot q+1}\sum_{e=v-1}^{z\cdot q}\Card{\cW^{2q,z}(v,e)}\cdot \binom{n}{v}(v)!\cdot \Paren{\frac{d}{n}}^e\\
		&\leq  \sum_{v\geq 3}^{z\cdot q+1}\sum_{e=v-1}^{z\cdot q} (3\cdot n)^{v} \cdot \Paren{2^{2t}z}^{2qt}\cdot (2zq)^{6tq\cdot (e-v+1)}\cdot \Paren{\frac{d}{n}}^e\\
		&\leq \sum_{v\geq 3}^{z\cdot q+1}\sum_{e=v-1}^{z\cdot q}\Paren{2^{2t}z}^{2qt}\cdot d^{e}\cdot n\cdot  \Paren{\frac{(6zq)^{6tq}}{n}}^{e-v+1}\\
		&\leq \Paren{2^{2t}z}^{2qt}\cdot d^{zq}\cdot n\cdot\sum_{v\geq 3}^{z\cdot q+1}\sum_{e=v-1}^{z\cdot q}  \Paren{\frac{(6zq)^{6tq}}{n}}^{e-v+1}\\
		&\leq z^{O(qt)}\cdot d^{zq}\cdot n\cdot\sum_{v\geq 3}^{z\cdot q+1}\sum_{e=v-1}^{z\cdot q}  \Paren{\frac{(6zq)^{6tq}}{n}}^{e-v+1}\,.
	\end{align*}
	For $t=100\log\log_d n$,  $q=\frac{\log n}{10^3\log^2\log_d n}$ and $z\leq \frac{\log_d n}{6}$ the series converges.
	Thus
	\begin{align*}
		\E\Brac{ \Tr \Brac{\mathbf{B}^{(z-1)}\transpose{\Paren{\mathbf{B}^{(z-1)}}}}^q }  &\leq O\Paren{z^{O(qt)}\cdot d^{zq}\cdot n}
	\end{align*}
	Finally by \cref{proposition:trace-method} with probability at least $1-o(1)$
	\begin{align*}
		\Norm{\mathbf{B}^{z-1}}\leq O(d^{z/2})\cdot \Paren{z^{2t}\cdot n^{1/q}}\leq  O(d^{z/2})\cdot\Paren{\log n}^{\poly(\log\log n)}\,,
	\end{align*}
	concluding the proof.
\end{proof}

%\begin{lemma}
%	Let $G$ be $O(\polylog(n))$-tangle-free with probability $1-p$ \Tnote{and log-n wise independence, define $\ibtgf{arg1}{arg2}$}. Then
%	\begin{align*}
%		\E \Tr \Brac{B^{k-1}\transpose{\Paren{B^{k-1}}}}^q \leq \frac{1}{1-p} \sum_{W\in \ibtgf{}{2q,k}} \prod_{i=1}^{2q}\Brac{\E\sqrt{\Abs{A_{e_1(W_i)}}}\cdot\E \sqrt{\Abs{ A_{e_k(W_i)}}}\cdot\Paren{\prod_{s=2}^{k-1}\E A_{e_{s+1}(W_i)}}}\,.
%	\end{align*}
%\end{lemma}

\section{Strong refutations for random k-XOR}\label{section:random-3xor}
%With a slight abuse of notation we also write $\mathbf{T}\sim \kxord{k}$ to denote the associated random tensor.

We prove here a result on strong refutation of random $k$-XOR instances.

\begin{theorem}\label{theorem:refutation-k-xor}
	Consider a random $k$-XOR instance $\bm \cI\sim  \kxord{k}$ for $k\geq 3$. 
	% such that $k\log k \leq o(\log n)$. 
	For $n$ large enough,  there exists a universal constant $C>0$ and a polynomial time algorithm that, if
	\begin{align*}
		p\geq \frac{C\cdot n^{-k/2}}{\eps^2}
		%m\geq \frac{C\cdot n^{k/2}}{\gamma^2}
	\end{align*}
	certifies with probability at least $0.99$
	\begin{align*}
		\optI{\bm \cI} \leq \frac{1}{2} +O(\eps)\,.
	\end{align*}
	%\Tnote{Check if there is a$2^{O(k)}$ factor.}
\end{theorem}
An efficient algorithm for $k$-XOR refutation was already known to hold for $p\geq \Omega\Paren{n^{-k/2}\log^{3/2}n}$ or when $k$ is even \cite{AllenOW15}. Thus we only need to design an algorithm for the settings $p\leq O\Paren{n^{-k/2}\polylog(n)}$ and $k$ odd. We restrict our analysis to those. 
Following our discussion in \cref{section:preliminaries-csps} and using \cref{proposition:max-value-xor}, the theorem can be directly obtained as a corollary to the result below.

\begin{theorem}[Sharp bounds for random polynomials]\label{theorem:main-xor-technical}
	Let $\mathbf{T}$ be a random %symmetric\Tnote{Remove symmetry, for this I need to update indices throughout the proof :( } 
	tensor, with independent entries, such that $\mathbf{T}_{\alpha}=0$ if the indices in the multi-index $\alpha\in [n]^{k}$ are not distinct and otherwise:
	\begin{align*}
		\E \Brac{\mathbf{T}_{\alpha}} &=0\\
		\bbP \Brac{\mathbf{T}_{\alpha}\neq 0}&\leq p\\
		%\text{if }\mathbf{T}_{\alpha}\neq 0\text{ then }\Omega(1)\leq \Abs{\mathbf{T}_\alpha}&\leq 1
		\bbP \Brac{\Omega(1)\leq \Abs{\mathbf{T}_{\alpha}}\leq 1\given \mathbf{T}_{\alpha}\neq 0}&=1
		%\bbP \Brac{\Abs{\mathbf{T}_\alpha}>1\given \mathbf{T}_\alpha\neq 0}&=0\\
		%\bbP \Brac{\Abs{\mathbf{T}_\alpha}\leq \Omega(1)\given \mathbf{T}_\alpha\neq 0}&=0
	\end{align*}
	%\begin{align*}
	%\mathbf{T}_{\alpha} = \begin{cases}
		%0& \text{with probability }1-p\,,\\
		%+1&\text{with probability }p/2\,,\\
		%-1&\text{with probability }p/2\,.
	%\end{cases}
	%\end{align*}
	For $k\geq 3$ and $n$ large enough,  %such that $k\log k \leq o(\log n)$,
	there exists a universal constant $C>0$ and a polynomial time algorithm that, if
	\begin{align*}
		C\cdot n^{-k/2}\leq p\leq n^{-k/2}\cdot O\Paren{\log^{10}n}\,,
	\end{align*}
	certifies with probability larger than $0.99$
	\begin{align*}
		\max_{x\in \Set{\pm 1}^n}\underset{\alpha\in [n]^{k}}{\sum} \mathbf{T}_{\alpha}\cdot x^{\alpha} 
		\leq O\Paren{\sqrt{p}\cdot n^{3k/4}}\,.%\frac{p\cdot n^k}{2}\,.
	\end{align*}
	%\Tnote{At the end of the proof a term of the form $k!$ or $2^k$ could appear}
\end{theorem}

\begin{remark}
	\cref{theorem:main-xor-technical} is not strictly about strong refutations. In fact, it states that the value of a random polynomial evaluated over the hypercube is concentrated around its expectation.
	%\Tnote{If we want to generalize to $\iprod{T, x_1\otimes\cdots\otimes x_k}$ we need to drop the symmetry assumption.}
\end{remark}

%\Tnote{To change after we get rid of symmetry}
For the reminder of the section, we  assume without loss of generality that $\mathbf{T}$ is symmetric. Indeed we may have this assumption without loss of generality by \cref{fact:symmetric-tensor-equivalent}.
%Our analysis will work both in the asymmetric and in the symmetric case, so we do not need to distinguish as the bounds are the same (up to constant factors).
For any assignment of $x\in \Set{\pm 1}^n$, by Cauchy-Schwarz,
%\begin{align*}
%    \iprod{\mathbf{T}, x_1\otimes\cdots\otimes x_k} &\leq \Paren{\sum_{i\in [n]}(x_k)_i^2}^{1/2}\Paren{\sum_{\ell\in [n]}\Paren{\sum_{\alpha\in [n]^{k-1}}\mathbf{T}_{(\alpha, \ell)}\cdot (x_1)_{\alpha(1)}\cdots (x_{k-1})_{\alpha(k-1)}}^2}^{1/2}\\
%	&\leq \sqrt{n}\cdot\Paren{\sum_{\ell\in [n]}\Paren{\sum_{\alpha\in [n]^{k-1}}\mathbf{T}_{(\alpha, \ell)}\cdot (x_1)_{\alpha(1)}\cdots (x_{k-1})_{\alpha(k-1)}}^2}^{1/2}\,.
%\end{align*}
\begin{align*}
	\underset{\alpha\in [n]^k}{\sum} \mathbf{T}_{\alpha}\cdot x^{\alpha} &\leq \Paren{\sum_{i\in [n]}x_i^2}^{1/2}\Paren{\sum_{\ell\in [n]}\Paren{\sum_{\alpha'\in [n]^{k-1}}\mathbf{T}_{(\alpha', \ell)}\cdot x^{\alpha'}}^2}^{1/2}\\
	&\leq \sqrt{n}\cdot\Paren{\sum_{\ell\in [n]}\Paren{\sum_{\alpha' \in [n]^{k-1}}\mathbf{T}_{(\alpha',\ell)}\cdot x^{\alpha'}}^2}^{1/2}\,.
\end{align*}
Consider the $n^{k-1}$-by-$n^{k-1}$ matrix with entries, for $\alpha_1, \alpha_2, \beta_1, \beta_2\in [n]^{(k-1)/2}$,
\begin{align}\label{eq:matrix-A}
	\mathbf{A}_{(\alpha, \beta),(\alpha',\beta')}:= \sum_{\ell\in [n]}\mathbf{T}_{(\alpha \alpha'\ell)}\cdot \mathbf{T}_{(\beta\beta'\ell)}
\end{align}
we may use the rewriting
\begin{align*}
    %\sum_{\ell\in [n]}\Paren{\sum_{\alpha\in [n]^{k-1}}\mathbf{T}_{(\alpha, \ell)}\cdot (x_1)_{\alpha(1)}\cdots (x_{k-1})_{\alpha(k-1)}}^2 &= \iprod{\mathbf{A}, x_1\otimes \cdots\otimes x_{k-1}}\\
	\sum_{\ell\in [n]}\Paren{\sum_{\alpha, \beta \in [n]^{k-1}}\mathbf{T}_{(\alpha, \beta, \ell)}x^{(\alpha,\beta)}}^2&=\sum_{\ell\in [n]}\sum_{\alpha, \alpha', \beta, \beta'\in [n]^{k-1}} \mathbf{T}_{\alpha,\beta,\ell}\cdot \mathbf{T}_{\alpha',\beta',\ell} \cdot x^{(\alpha, \beta, \alpha', \beta')}\\
	&= \iprod{\tensorpower{x}{k-1}, \mathbf{A}\tensorpower{x}{k-1}}\\
	&\leq \max_{z, y\in\Set{\pm 1}^{n^{k-1}}}\iprod{z, \mathbf{A}y}= \Normio{\mathbf{A}}\,.
\end{align*}
Thus we obtain
\begin{align*}
	\max_{x\in \Set{\pm 1}^n}\underset{\alpha\in [n]^{k}}{\sum} \mathbf{T}_{\alpha}x^{\alpha} 
	%\max_{x_1,\ldots, x_k\in \Set{\pm 1}^n}\iprod{\mathbf{T}, x_1\otimes\cdots\otimes x_k}
	\leq \sqrt{n\cdot \normio{\mathbf{A}}}\,,
\end{align*}
which means that any algorithm computing an upper bound to $ \normio{\mathbf{A}}$ \textit{immediately yields} a refutation algorithm for  $k$-XOR. In particular, by \cref{fact:sos-grothendieck-inequality} we get that there is an  algorithm (based on sum-of-squares)  that certifies
\begin{align}
	%\max_{x\in \Set{\pm 1}^n}\underset{\alpha\in [n]^{k}}{\sum} \mathbf{T}_{\alpha}x^{\alpha} 
	\max_{x_1,\ldots, x_k\in \Set{\pm 1}^n}\iprod{\mathbf{T}, x_1\otimes\cdots\otimes x_k}\leq \sqrt{K_G\cdot n\cdot \normio{\mathbf{A}}}\,,
\end{align}
 in time $O(n^{O(1)})$, where $K_G$ is Grothendieck's constant.
Now, \cref{theorem:main-xor-technical} follows combining the above reasoning with the result below.

\begin{lemma}\label{lemma:cut-norm-A}
	Consider the settings of \cref{theorem:main-xor-technical}. 
	Consider the $n^{k-1}$-by-$n^{k-1}$ matrix defined in \cref{eq:matrix-A}.%, for $\cI_1, \cI_2, \cJ_1, \cJ_2\in [n]^{(k-1)/2}$,
	%\begin{align}
	%	\mathbf{A}_{(\cI, \cI'),(\cJ,\cJ')}:= \sum_{\ell\in [n]}\mathbf{T}_{(\cI,\cJ,\ell)}\cdot \mathbf{T}_{(\cI',\cJ',\ell)}\,.
	%\end{align}
	Then with probability at least $0.998$
	\begin{align*}
		\Normio{\mathbf{A}}\leq n^{k-1}\cdot O\Paren{p \cdot n^{k/2}}\,.
	\end{align*}
\end{lemma}
%\subsection{Norm bounds for the constraints matrix}
We may rewrite $\mathbf{A}=\mathbf{A}'+\mathbf{A}''$ where  for any $\alpha_1, \alpha_2, \beta_1, \beta_2\in [n]^{(k-1)/2}$
\begin{align}\label{eq:preprocessed-A}
		\mathbf{A}'_{(\alpha_1, \beta_1),(\alpha_2,\beta_2)}:= 
		\begin{cases}
			\mathbf{A}_{(\alpha_1,\beta_1),(\alpha_2,\beta_2)}& \text{if } %\Card{S(\alpha_1,\alpha_2)\cap S(\beta_1,\beta_2)}<= \frac{k-3}{2}\\
            \Card{S(\alpha_1,\alpha_2)\cap S(\beta_1,\beta_2)}=0\\
			0&\text{ otherwise}
		\end{cases}
\end{align}
and $\mathbf{A}''=\mathbf{A}-\mathbf{A}'$. This decomposition is convenient as, for example, now $\E 	\mathbf{A}'_{(\alpha_1, \beta_1),(\alpha_2,\beta_2)}=0$ for all multi-indices $\alpha_1,\alpha_2,\beta_1,\beta_2$. By triangle inequality, to prove \cref{lemma:cut-norm-A} it suffices to bound the norm $\Normio{\cdot}$ of both $\mathbf{A}'$ and $\mathbf{A}''$.

\begin{lemma}\label{lemma:cut-norm-residual-A}
	Consider the settings of \cref{lemma:cut-norm-A}. 
	Let $\mathbf{A}'$ as defined in \cref{eq:preprocessed-A} and let $\mathbf{A}''= \mathbf{A}-\mathbf{A}'$.
	Then with probability $1-o(1)$ 
	\begin{align*}
		\Normio{\mathbf{A}''}\leq n^{k-1-\Omega(1)}\cdot (k-1)!\cdot O(p\cdot n^{k/2})\,.
	\end{align*}
\end{lemma}

The proof of \cref{lemma:cut-norm-residual-A} is straightforward as the required bound is very loose. We defer it to \cref{section:deferred-proofs}. The next result bounds $\Normio{\mathbf{A}'}$, this is the main technical challenge of this work and the subsequent sections are dedicated to its proof.

\begin{lemma}\label{lemma:cut-norm-preprocessed-A}
	Consider the settings of \cref{lemma:cut-norm-A}.
	Let $\mathbf{A}'$ as defined in \cref{eq:preprocessed-A} 
	Then with probability at least $0.999$
	\begin{align*}
		\Normio{\mathbf{A}'}\leq n^{k-1}\cdot O(p\cdot n^{k/2})\,.
	\end{align*}
	%\Tnote{Check if a term $k!$ is needed}
\end{lemma}

\subsection{Bounding the norm of A'}\label{section:bounding-norm-preprocessed-A}

Using a reasoning similar in spirit to that shown in \cref{section:warm-up},  we prove \cref{lemma:cut-norm-preprocessed-A} studying the associated non-backtracking matrix.

\begin{theorem}\label{theorem:bound-preporcessed-A-via-Ihara-Bass}
	Consider the settings of \cref{lemma:cut-norm-preprocessed-A}. Let $\mathbf{B}, \mathbf{L}, \mathbf{J}$ be the matrices associated to $\mathbf{A}'$ as defined in \cref{section:general-ihara-bass}.
	Then with probability $1-o(1)$
	\begin{align*}
		\rho(\mathbf{B}+\mathbf{L}-\mathbf{J})\leq O\Paren{p\cdot n^{k/2}}\,.
	\end{align*}
	%\Tnote{Check if a term $k!$ is needed}
\end{theorem}

We can use \cref{theorem:bound-preporcessed-A-via-Ihara-Bass} to prove  \cref{lemma:cut-norm-preprocessed-A}. The argument closely resembles that used for \cref{theorem:main-xor-binary}

\begin{proof}[Proof of  \cref{lemma:cut-norm-preprocessed-A}]
	By \cref{theorem:bound-preporcessed-A-via-Ihara-Bass} for both $\mathbf{A}'$ and $-\mathbf{A}'$ we get $\rho\Paren{B(\mathbf{A}')}\leq O\Paren{p\cdot n^{k/2}}$ and $\rho\Paren{B(-\mathbf{A}')}\leq O\Paren{p\cdot n^{k/2}}$.  By \cref{lemma:avg-degree-sparse-hyper-graph} we have $\Tr D(\mathbf{A}')\leq O\Paren{p^2\cdot n^{2k-1}}$ with probability at least $1-10^4$, where $D(\mathbf{A}')$ is the associated degree matrix as defined in \cref{section:general-ihara-bass}. Finally applying \cref{corollary:bound-norm-A} the result follows.
\end{proof}

By definition of spectral radius, there exists a unit vector $v\in \R^m$ such that
\begin{align*}
	\rho(\mathbf{B}+\mathbf{L}-\mathbf{J})\leq \transpose{v}(\mathbf{B}+\mathbf{L}-\mathbf{J})v\leq \rho(\mathbf{B})+ \Norm{\mathbf{L}}+ \Norm{\mathbf{J}}.
\end{align*}
Thus, we can use next two results to obtain \cref{theorem:bound-preporcessed-A-via-Ihara-Bass}.

\begin{lemma}
	Consider the settings of \cref{theorem:bound-preporcessed-A-via-Ihara-Bass} and suppose $p\leq n^{-1}$. Then with probability $1-o(1)$
	\begin{align*}
		\Norm{\mathbf{L}}+\Norm{\mathbf{J}}\leq O(1)\,.
	\end{align*}
	\begin{proof}
		By construction $\Norm{\mathbf{J}}\leq 1$ and  $\Norm{\mathbf{L}}\leq \max_{\alpha_1,\beta_1,\alpha_2,\beta_2\in [n]^{k-1}}\Abs{\mathbf{A}'_{(\alpha_1,\beta_1)(\alpha_2,\beta_2)}}$.
		Now, for fixed $\alpha_1,\beta_1,\alpha_2,\beta_2\in [n]^{k-1}$ and $t\geq 1$
		\begin{align*}
			\bbP \Paren{\Abs{\mathbf{A}'_{(\alpha_1,\beta_1)(\alpha_2,\beta_2)}}\geq t} %&= \bbP \Paren{\Abs{\mathbf{A}'_{(\alpha_1,\alpha_2)(\beta_1,\beta_2)}} \geq t}\\
			%&=\sum_{q\geq t}\bbP \Paren{\Paren{\mathbf{A}'_{(a,b)(c,d)}}^2 = q^2}\\
			%&=\sum_{q\geq t}\bbP \Paren{\Paren{\sum_{\ell\in[n]}T_{a,c,\ell}T_{b,d,\ell}}^2 = q^2}\\
			&= \bbP \Paren{\Abs{\sum_{\ell\in[n]}\mathbf{T}_{(\alpha_1\alpha_2\ell)}\mathbf{T}_{(\beta_1\beta_2\ell)}}\geq t}\\
			&\leq \sum_{q\geq t}\bbP \Paren{\sum_{\ell\in[n]}\Abs{\mathbf{T}_{(\alpha_1\alpha_2\ell)}\mathbf{T}_{(\beta_1\beta_2\ell)}}= q}\\
			&\leq \sum_{q\geq t} \binom{n}{q}\cdot p^{2q}\cdot (1-p)^{n-2q}\\
			&\leq  \sum_{q\geq t} \binom{n}{q}\cdot p^{2q}\\
			&\leq O(1)\cdot\Paren{\frac{e\cdot n\cdot p^2}{t}}^{t}	\,.
		\end{align*}
		Thus, as $p\leq n^{-k/2}\poly\log(n)$ and $\mathbf{A}'$ has $n^{2k-2}$ entries, by union bound $\max_{\alpha_1,\alpha_2,\beta_1,\beta_2\in [n]^{k-1}}\Abs{\mathbf{A}'_{(\alpha_1, \beta_1)(\alpha_2,\beta_2)}}\leq 1000$, with probability  $1-n^{-\Omega(1)}$.
	\end{proof}
\end{lemma}

\begin{lemma}\label{lemma:spectral-radius-non-backtracking}
	Consider the settings of \cref{theorem:bound-preporcessed-A-via-Ihara-Bass}. Then with probability $1-o(1)$
	\begin{align*}
		\rho(\mathbf{B})\leq O\Paren{p\cdot n^{k/2}}\,.
	\end{align*}
	%\Tnote{Check if additional terms of the form $(k-1)!$ or $2^k$ appear.}
\end{lemma}

The proof of  \cref{lemma:spectral-radius-non-backtracking} follows a recipe similar to the one used in \cref{section:warm-up}. We consider the extended non-backtracking $(2n^{k-1})$-by-$(2n^{k-1})$ matrix $\mathbf{B}$. We use the trace method \cref{proposition:trace-method} to bound the spectral norm of the closely related non-backtracking matrix $\mathbf{B}^{z-1}$ for large enough $z$, and then use Gelfand's formula \cref{fact:bound-spectral-radius} to relate the result to the spectral radius of $\mathbf{B}$.
In particular, \cref{lemma:spectral-radius-non-backtracking} is an immediate consequence of the following result.

\begin{lemma}\label{lemma:bound-trace-xor}
	Consider the settings of \cref{theorem:bound-preporcessed-A-via-Ihara-Bass}. Let $q\leq  \frac{\log n}{\Paren{10^3\log\log n}^2}$ and $z\leq\frac{\log n}{50}$. Then with probability $1-o(1)$
	\begin{align*}
		\Tr \Brac{\Paren{\mathbf{B}^{z-1}}\transpose{\Paren{\mathbf{B}^{z-1}}}}^q \leq \Paren{p\cdot n^{k/2}}^{2qz}\cdot O\Paren{z^{\frac{\log n}{10\log\log n}+2}}\,.
	\end{align*}
	%\Tnote{Could be Explicit bound on $k$}
\end{lemma}

The rest of the section is dedicated to proving \cref{lemma:bound-trace-xor}.

\subsubsection{From block non-backtracking walks to block hyper non-backtracking walks}\label{section:hyper-nbw}
We start our analysis by opening up the terms in $\Brac{\Paren{\mathbf{B}^{z-1}}\transpose{\Paren{\mathbf{B}^{z-1}}}}^q$.
By \cref{fact:expansion-power-non-backtracking-matrix} and \cref{fact:trace-expansion-non-backtracking-power} we know how to represent these terms using the entries of $\mathbf{A}'$. As these terms contain absolute values of sums, it will be more convenient to work with simpler products in the entries of $\mathbf{T}$. 
In order to achieve this we need to manipulate our walks further.
For each walk in $W\in \bnbw{}{2q,z}$  with $W=(\alpha_1^1, \beta_1^1),(\alpha_2^1,\beta_2^1),\ldots, (\alpha_{z+1}^1,\beta_{z+1}^1),\ldots, (\alpha^{2q}_{z+1},\beta^{2q}_{z+1})$  (using multi-indices in $\alpha^i_j\,,\beta^i_j\in [n]^{(k-1)/2}$), %and assuming, only for simplicity of the notation, that $z$ is odd), 
we use the rewriting  
\begin{align}
	&\prod_{i=1}^{2q} \Brac{\sqrt{\Abs{\mathbf{A}_{(\alpha_1^i, \beta_1^i),(\alpha_2^i, \beta_2^i)}'\mathbf{A}_{(\alpha_z^i, \beta_z^i),(\alpha_{z+1}^i, \beta_{z+1}^i)}'}}\cdot \Paren{\prod_{\substack{s=2}}^{z-1}\mathbf{A}'_{(\alpha_s^i, \beta_{s}^i),(\alpha_{s+1}^i, \beta_{s+1}^i)}}}\nonumber\\
	&=\prod_{i=1}^{2q} \Brac{\Abs{\mathbf{A}_{(\alpha_1^i, \alpha_2^i),(\beta_1^i, \beta_2^i)}'}\cdot \Paren{\prod_{\substack{s=2}}^{z-1}\mathbf{A}'_{(\alpha_s^i, \beta_{s}^i),(\alpha_{s+1}^i, \beta_{s+1}^i)}}}\nonumber\\
	&=\sum_{\ell_2^1,\ldots,\ell_z^1,\ldots,\ell_z^{2q}\in [n]}\prod_{i=1}^{2q} \Brac{\Abs{\sum_{\ell_1^i\in [n]}\mathbf{T}_{(\alpha_1^i\alpha_2^i\ell_1^i)}\mathbf{T}_{(\beta_1^i\beta^i_2\ell^i_1)}}\Paren{\prod_{\substack{s=2}}^{z-1}\mathbf{T}_{(\alpha^i_s \alpha_{s+1}^i\ell_{s}^i)} \mathbf{T}_{( \beta_{s}^i \beta_{s+1}^i\ell_{s}^i)}}}\,,\label{eq:walk-expansion}
\end{align}
where in the first step we used the fact that the last edge of each $W_i$ in $W$ is the first edge of $W_{i+1}$.
%We use $T(W)$ to denote the terms in \cref{eq:walk-expansion}.
%For a given $W\in \bnbw{}{2q,k}$, t
Elements in the sum of the form
\begin{align*}
	\prod_{i=1}^{2q} \Brac{\Abs{\sum_{\ell_1^i\in [n]}\mathbf{T}_{(\alpha_1^i\alpha_2^i\ell_1^i)}\mathbf{T}_{(\beta_1^i\beta^i_2\ell^i_1)}}\Paren{\prod_{\substack{s=2}}^{z-1}\mathbf{T}_{(\alpha^i_s \alpha_{s+1}^i\ell_{s}^i)} \mathbf{T}_{( \beta_{s}^i \beta_{s+1}^i\ell_{s}^i)}}}
\end{align*}
such that every term with odd degree in 
\begin{align*}
\prod_{i=1}^{2q} \Brac{\Paren{\prod_{\substack{s=2}}^{z-1}\mathbf{T}_{(\alpha^i_s \alpha_{s+1}^i\ell_{s}^i)} \mathbf{T}_{( \beta_{s}^i \beta_{s+1}^i\ell_{s}^i)}}}
\end{align*}
also appears in 
\begin{align*}
	\prod_{i=1}^{2q}\Abs{\sum_{\ell_1^i\in [n]}\mathbf{T}_{(\alpha_1^i\alpha_2^i\ell_1^i)}\mathbf{T}_{(\beta_1^i\beta^i_2\ell^i_1)}}
\end{align*}
are called \textit{annoying}. We denote the set of annoying terms by $\annoying(W)$. If a term is not annoying it is said to be \textit{nice}.
$\overline{\annoying}(W)$ is the set of nice terms in \cref{eq:walk-expansion}. We can upper bound the expectation of nice terms with the expectation of a related polynomial.
The next result formalizes this idea, we defer its proof to \cref{section:deferred-proofs}.
%The next result provides convenient upper bounds to both nice and annoying terms in $T(W)$, for any $W\in \bnbw{}{2q,k}$. We deferred its proof to \cref{section:deferred-proofs}.

\begin{lemma}\label{lemma:rewriting-backtracking-xor}
	Consider the settings of \cref{theorem:bound-preporcessed-A-via-Ihara-Bass}. Let $W\in \bnbw{}{2q,z}$. Then for any term in $\overline{\annoying}(W)$
	\begin{align}
		\E &\prod_{i=1}^{2q} \Brac{\Abs{\sum_{\ell_1^i}\mathbf{T}_{(\alpha_1^i\alpha_2^i\ell_1^i)}\mathbf{T}_{(\beta_1^i\beta^i_2\ell^i_1)}}\Paren{\prod_{\substack{s=2}}^{z-1}\mathbf{T}_{(\alpha^i_s \alpha_{s+1}^i\ell_{s}^i)} \mathbf{T}_{(\beta_{s}^i \beta_{s+1}^i\ell_{s}^i)}}}\nonumber\\
		&\leq O(1)^{2q} \cdot \E \prod_{i=1}^{2q} \Brac{\Paren{\sum_{\ell_1^i}\mathbf{T}_{(\alpha_1^i\alpha_2^i\ell_1^i)}\mathbf{T}_{(\beta_1^i\beta^i_2\ell^i_1)}}^2\Paren{\prod_{\substack{s=2}}^{z-1}\mathbf{T}_{(\alpha^i_s\alpha_{s+1}^i\ell_{s}^i)} \mathbf{T}_{(\beta_{s}^i \beta_{s+1}^i\ell_{s}^i)}}} \label{eq:expansion-nice}\,.
	\end{align}
	%for some constant $C>0$.
	%\Tnote{Already incorporating difference for CSPs}
\end{lemma}
Instead, for each annoying term in $\annoying(W)$ we bound the expectation as 
\begin{align}
	&\E \prod_{i=1}^{2q} \Brac{\Abs{\sum_{\ell_1^i}\mathbf{T}_{(\alpha_1^i\alpha_2^i\ell_1^i)}\mathbf{T}_{(\beta_1^i\beta^i_2\ell^i_1)}}\Paren{\prod_{\substack{s=2}}^{z-1}\mathbf{T}_{(\alpha^i_s a_{s+1}^i\ell_{s}^i)} \mathbf{T}_{\beta_{s}^i b_{s+1}^i\ell_{s}^i}}}\nonumber\\
	&\leq  \sum_{\ell^i_1,\ldots, \ell^{2q}_1} \E\Abs{\prod_{i=1}^{2q} \Brac{\mathbf{T}_{(\alpha_1^i\alpha_2^i\ell_1^i)}\mathbf{T}_{(\beta_1^i\beta^i_2\ell^i_1)}\Paren{\prod_{\substack{s=2}}^{z-1}\mathbf{T}_{(\alpha^i_s\alpha_{s+1}^i \ell_{s}^i)} \mathbf{T}_{(\beta_{s}^i \beta_{s+1}^i\ell_{s}^i)}}}}\nonumber\\
	&\leq  \sum_{\ell^i_1,\ldots, \ell^{2q}_1} \E\prod_{i=1}^{2q} \Brac{\Abs{\mathbf{T}_{(\alpha_1^i\alpha_2^i\ell_1^i)}}\Abs{\mathbf{T}_{(\beta_1^i\beta^i_2\ell^i_1)}}\Paren{\prod_{\substack{s=2}}^{z-1}\Abs{\mathbf{T}_{(\alpha^i_s \alpha_{s+1}^i\ell_{s}^i)}} \Abs{\mathbf{T}_{(\beta_{s}^i \beta_{s+1}^i\ell_{s}^i)}}}}\nonumber
	\\
	&\leq O(1)^{2q} \sum_{\ell^i_1,\ldots, \ell^{2q}_1} \E\prod_{i=1}^{2q} \Brac{\mathbf{T}_{(\alpha_1^i\alpha_2^i\ell_1^i)}^2\mathbf{T}_{(\beta_1^i\beta^i_2\ell^i_1)}^2\Paren{\prod_{\substack{s=2}}^{z-1}\Abs{\mathbf{T}_{(\alpha^i_s\alpha_{s+1}^i \ell_{s}^i)}} \Abs{\mathbf{T}_{(\beta_{s}^i \beta_{s+1}^i\ell_{s}^i)}}}}\label{eq:expansion-annoying}
\end{align}
We will encode every term in \cref{eq:expansion-nice} and \cref{eq:expansion-annoying} as a sequence $Z$ of multi-indices  $(\alpha_1^1,\beta^1_1,\ell_1^1, \alpha_1^2,\beta^1_2,\ell^1_2,\ldots, \ell^{2q}_z, \alpha^{2q}_{z+1}, \beta^{2q}_{z+1})$, where the $\ell_s^i$'s are indices and $\alpha^i_s, \beta^i_s$'s are multi-indices of cardinality $(z-1)/2$. For a subsequence $Z$, we will use $S(Z)$ to denote the set of its indices (without repetitions).
\cref{lemma:rewriting-backtracking-xor} together with these observations allows us to upper bound the contribution of each walk $W\in \bnbw{}{2q,z}$ to the expectation of the trace of $\Brac{\Paren{\mathbf{B}^{z-1}}\transpose{\Paren{\mathbf{B}^{z-1}}}}^q$ as
\begin{align}
	\E\prod_{i=1}^{2q}& \Brac{\Abs{\mathbf{A}_{(\alpha_1^i, \beta_2^i),(\alpha_2^i, \beta_2^i))}'}\cdot \Paren{\prod_{\substack{s=2}}^{z-1}\mathbf{A}'_{(\alpha_s^i\beta_s^i ),(\alpha_{s+1}^i \beta_{s+1}^i)}}}\nonumber\\
	\leq& O(1)^{2q}\cdot \E \sum_{\substack{\alpha^1_1,\ldots, \alpha^1_{z+1},\ldots \alpha^{2q}_{z+1},\\ \beta^1_1,\ldots, \beta^1_{z+1},\ldots \beta^{2q}_{z+1},\\\ell^1_1,\ldots,\ell^{2q}_z\\\text{in }\cC}}\mathbf{T}_{(\alpha^1_1\alpha^1_2\ell^1_1)}\cdots \mathbf{T}_{(\beta^{2q}_{z}\beta^{2q}_{z+1} \ell^{2q}_{z})}\,,\label{eq:nice-hyper-bnbw}\\
	&+O(1)^{2q}\cdot\E \sum_{\substack{\alpha^1_1,\ldots, \alpha^1_{z+1},\ldots \alpha^{2q}_{z+1},\\ \beta^1_1,\ldots, \beta^1_{z+1},\ldots \beta^{2q}_{z+1},\\\ell^1_1,\ldots,\ell^{2q}_z\\\text{in }\cC^*}}\Abs{\mathbf{T}_{(\alpha^1_1\alpha^1_2\ell^1_1)}}\cdots\Abs{\mathbf{T}_{(\beta^{2q}_{z}\beta^{2q}_{z+1} \ell^{2q}_{z})}}\,.\label{eq:annoying-hyper-bnbw}
\end{align}
Here $\cC$ is the set of conditions (using the convention $i+1=1$ for $i=2q$)%\Tnote{Check preprocessing, seems unnecessary strong}
\begin{align}\label{eq:conditions-hw}\tag{\(\cC\)}
	\left \{
	\begin{aligned}
		&\text{pre-processing:} %&\Card{S(\alpha^i_j\alpha^i_{j+1})\cap S(\beta^i_{j}\beta^i_{j+1})}<=\frac{k-3}{2}& \\
  &\Card{S(\alpha^i_j\alpha^i_{j+1})\cap S(\beta^i_{j}\beta^i_{j+1})}=0& \\
		%&&\Card{S(\alpha^i_{j+2}\beta_j^i)\cap S(\alpha^i_{j+3}\beta^i_{j+1})}<\frac{k-3}{2}&\\
		&&\text{all distinct in } (\alpha^i_j, \alpha^i_{j+1}, \ell^i_j )&\\
		&&\text{all distinct in } (\beta^i_{j}, \beta^i_{j+1}, \ell^i_j)&\\
		%&&\text{all distinct in } (\alpha^i_{j}, \alpha^i_{j+1}, \ell^i_{j}) &\\
		%&&\text{all distinct in } (\beta^i_{j}, \beta^i_{j+1}, \ell^i_{j+1})&\\
		&\text{block:}& \alpha^i_{z}=\alpha^{i+1}_{2}&\\
		&& \alpha^i_{z+1}=\alpha^{i+1}_{1}&\\
		&& \beta^i_{z+1}=\beta^{i+1}_{1}&\\
		&& \beta^i_{z}=\beta^{i+1}_{2}&\\
		%&\text{no self-loop:}&\Set{a^i_j, b^i_j}\neq \Set{a^i_{j+1}, b^i_{j+1}}&\\
		%&&\Set{a^i_{j+2}b^i_j}\neq \Set{a^i_{j+3}b^i_{j+1}}&\\
		&\text{non-backtracking:} &\alpha^i_j \neq \alpha^i_{j+2} \text{ or }\beta^i_j \neq \beta^i_{j+2}%\alpha^i_{j+1} \neq \alpha^i_{j+3}&\\
		%&\Set{a^i_j, a^i_{j+1}}\neq \Set{a^i_{j+2}, a^i_{j+3}}&\\
		%&&\beta^i_j \neq \beta^i_{j+2} \text{ or }\beta^i_{j+1} \neq \beta^i_{j+3}&\\
		%\Set{b^i_j, b^i_{j+1}}\neq \Set{b^i_{j+2}, b^i_{j+3}}&\\
	\end{aligned}
	\right \}
\end{align}
and $\cC^*$ is the set of conditions in $\cC$ with the addition of $\Set{\forall i\in [2q]\,, \ell_z^i = \ell_1^{i+1}}$ and
\begin{itemize}
	\item If a tuple $Z'$ in $\Set{(\alpha_j^i, \alpha^i_{j+1},\ell_j^i)\,, (\beta_j^i,\beta^i_{j+1}, \ell_j^i)} $ appears an odd number of times in the whole sequence  then there exists $i'$ such that $\Card{S(Z)\cap S(\alpha_1^{i'}, \alpha_1^{i'+1}, \ell_1^{i'})}\geq k-1$ or $\Card{S(Z)\cap S(\beta_1^{i'}, \beta_2^{i'}, \ell_1^{i'})}\geq k-1$.
	%\Card{S\cap\Set{a_2^i, b_2^i, \ell_2^i}}\geq 2$.
\end{itemize}
Let's try to unravel the meaning of $\cC$ and $\cC^*$.
The set of conditions \ref{eq:conditions-hw} are meant to captures products arising from terms in $\overline{\annoying}(W)$ for $W\in \bnbw{}{2q,z}$. Notice the \textit{block} condition correspond to the observation that in each $W\in \bnbw{}{2q,z}$ partitioned in non-backtracking walks $W_1,\ldots, W_{2q}$, for any consecutive  $W_i, W_{i+1}$ the first edge traversed in $W_{i+1}$ is the last edge traversed in $W_i$. The pre-processing condition allows to exclude corner cases\footnote{The experienced reader may recognize the \textit{pre-processing} condition in \ref{eq:conditions-hw} implies  the  \textit{no self-loop} condition from \cite{AllenOW15}. } as these where handled with $\mathbf{A}''$. A consequence of this condition is that the underlying hyper-graph cannot have self-loops.
Finally, the crucial non-backtracking conditions will help us upper bound the number of terms in the sum.

The terms appearing in $\cC^*$ correspond to the ones generated by $\annoying(W)$ for $W\in \bnbw{}{2q, z}$. 
As such the first additional condition captures the fact that we don't take the square of the first and last edges in each non-backtracking walk $W_i$ in $W$. The second condition captures the fact that these terms are annoying, and thus only few specific elements in the product may have odd degree. While our bound for annoying terms is tough, the number of such terms will be small. We remark that every sequence satisfying $\cC^*$ also satisfies $\cC$.

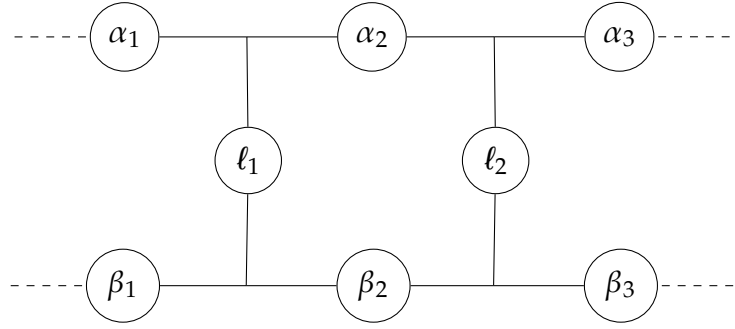
\begin{figure}[!ht]
	\centering
	\begin{tikzpicture}[roundnode/.style={draw,shape=circle,minimum size=0mm}]
	
		% Command
		\newcommand{\hyperedge}[4][180]{
			\draw (#2.#1) ++(#1:1.2)  edge (#2) edge (#3) edge (#4);    
		}
		
		% Vertices
		
		\node[roundnode] (a1)  {$\alpha_1$};
		\node[roundnode] (l1)  [below right = of a1 ]{$\ell_1$};
		\node[roundnode] (a2) [below left = of l1] {$\beta_1$};
		\node[roundnode] (b1) [above right = of l1] {$\alpha_2$};
		\node[roundnode] (b2) [below right = of l1] {$\beta_2$};
		%\node[] (dummy1) [right = of a1] {};
		\node[draw=none] (dummy3) [right = of a2] {};
		\node[roundnode] (l2)  [below right = of b1 ]{$\ell_2$};
		\node[roundnode] (a3)  [above right = of l2 ]{$\alpha_3$};
		\node[roundnode] (a4)  [below right = of l2 ]{$\beta_3$};
		\node[draw=none] (dummy3) [right = of b1] {};
		\node[draw=none] (dummy4) [right = of b2] {};
		
		\node[draw=none] (dummyl) [left = of a1] {};
		\node[draw=none] (dummyll) [left = of a2] {};
		\node[draw=none] (dummyr) [right = of a3] {};
		\node[draw=none] (dummyrr) [right = of a4] {};
			
		% Edges
		
		\hyperedge{b1}{l1}{a1};
		\hyperedge{a3}{l2}{b1};
		\hyperedge{b2}{l1}{a2};
		\hyperedge{a4}{l2}{b2};
		%\draw (a1) -- (dummy1);
		%\draw (b1) -- (dummy1);
		%\draw (l1) -- (dummy1);
		
		\draw[dashed] (dummyl) -- (a1);
		\draw[dashed] (dummyll) -- (a2);
		\draw[dashed] (dummyr) -- (a3);
		\draw[dashed] (dummyrr) -- (a4);
		
		%\node[draw=none] (dummy1) [below left = of T2] {};
		%\node[draw=none] (dummy2) [above left = of T2] {};
		%\node[draw=none] (dummy3) [below right = of T1] {};
		%\node[draw=none] (dummy4) [above right = of T1] {};
		%Lines
	%	\draw[] (T1) -- (T2);
		%\draw[] (T2) -- (dummy1);
	%	\draw[] (T2) -- (dummy2);
	%	\draw[] (T1) -- (dummy3);
	%	\draw[] (T1) -- (dummy4);

		%\draw[bend right]  (v6) edge (v7);
	\end{tikzpicture}
\caption{Representation of part of a hyper-walk satisfying \ref{eq:conditions-hw} for $3$-XOR.}\label{fig:hyper-walk}
\end{figure}

\paragraph{Block non-backtracking hyper walks}
%\Tnote{When it is clear what we need from hypergraphs, add the notation to the preliminaries}
Recall we can encode every term in \cref{eq:nice-hyper-bnbw} and  \cref{eq:annoying-hyper-bnbw} as a sequence $Z$ of multi-indices $(\alpha_1^1,\beta^1_1,\ell_1^1, \alpha_2^2, \beta^1_2,\ell^1_2,\ldots, \ell^{2q}_z, \alpha^{2q}_{z+1}, \beta^{2q}_{z+1})$ where $\alpha^i_j, \beta^i_j$'s have cardinality $k-1$ and $\ell^i_j$'s are indices. We denote the set of sequences over $2q$ blocks of size $z$ satisfying \ref{eq:conditions-hw} as the set of block non-backtracking hyper-walks $Z$ and denote it by $\hbnbw{}{2q,z}$. We denote by $\hbnbw{}{2q,z}(\cC^*)\subseteq\hbnbw{}{2q,z}$ the subset of sequences in $\hbnbw{}{2q,z}$ which also satisfies $\cC^*$. We will interchangeably use the terms hyper-walk and sequence. Notice that each hyper-walk $Z\in \hbnbw{}{2q,z}$ generates a multi-hyper-graph\footnote{That is, a hyper-graph in which hyper-edges may have multiplicity larger than $1$.} $H(Z)$ over vertices in $[n]$.  For simplicity we will say  "hyper-graph" instead of "multi-hyper-graph". We then call $H(Z)$ the underlying hyper-graph of $Z$.We denote with $V(Z)$ its set of vertices (corresponding to the set of distinct indices in the sequence). An hyper-edge $S(\alpha,\alpha',\ell)$ is in $H(Z)$ if $\mathbf{T}_{(\alpha,\alpha',\ell)}$ appears in the product encoded by $Z$. 
%We use $M(H(Z))$ to denote the multi-set of hyper-edges in $H(Z)$ and
We use  $E(H(Z))$  to denote the set of distinct hyper-edges in $H(Z)$. For each $e\in E(Z)$we denote by $m_e(H(Z))$ the multiplicity of $e$ in $H(Z$). We say $H(Z)$ is the underlying multi-hyper-graph of $Z$. For  $Z\in \hbnbw{}{2q,z}$ we can graphically represent each such $H(Z)$ as depicted in \cref{fig:hyper-walk}. 

Furthermore, each $Z\in \hbnbw{}{2q,z}$ can be further split into subsequences $Z_1,\ldots, Z_{2q}$ such that $Z_i=(\alpha_1^i, \beta^i_1, \ell^i_1, \alpha^i_2, \beta^i_2,\ldots,  \ell^i_{z}, \alpha^i_{z+1}, \beta^i_{z+1})$ for all $i\in [n]$. We say $Z_1,\ldots, Z_{2q}$ are non-backtracking hyper-walks and denote the set of such walks as $\hnbw{}{2q,z}$. We use $\hnbw{}{2q,z}(Z)$ to denote the non-backtracking hyper-walks corresponding to $Z$ (picking one arbitrary such partition) and $H(Z_i)$ to denote the underlying hyper-graph of each such sequence. We refer to $Z_1,\ldots, Z_{2q}$ simply as the subsequences of $Z$.

Using our newly introduced notation, we can thus write
\begin{align}
			\E &\sum_{W\in \bnbw{}{2q,z}}\prod_{i=1}^{2q}\prod_{s=1}^{z}\mathbf{A}_{e_s(W_i)}' \nonumber\\
			\leq& O(1)^{2q}\cdot\E \sum_{Z\in \hbnbw{}{2q,z}}\mathbf{T}_{(\alpha^1_1(Z)\alpha^1_2(Z)\ell^1_1(Z))}\cdots \mathbf{T}_{(\beta^{2q}_{z}(Z)\beta^{2q}_{z+1}(Z) \ell^{2q}_{z}(Z))}\nonumber\\
			&+ O(1)^{2q}\cdot\E \sum_{Z\in \hbnbw{}{2q,z}(\cC^*)}\Abs{\mathbf{T}_{(\alpha^1_1(Z)\alpha^1_2(Z)\ell^1_1(Z))}\cdots \mathbf{T}_{(\beta^{2q}_{z}(Z)\beta^{2q}_{z+1}(Z) \ell^{2q}_{z}(Z))}}
			\label{eq:equivalence-walks-hyper-walks}\,.
\end{align}
A useful observation, formalized in the following fact,  is that for any $Z\in \hbnbw{}{2q,z}\setminus\hbnbw{}{2q,z}(\cC^*)$, if the underlying hyper-graph has a hyper-edge with multiplicity $1$ then its expectation is $0$.

\begin{fact}\label{fact:symmetry-hyperwalk}
	Consider the settings of \cref{theorem:bound-preporcessed-A-via-Ihara-Bass}. Let $Z\in \hbnbw{}{2q,z}$ and let $H$ be the underlying hyper-graph of $Z$. If $H$ contains a hyper-edge $e$ of multiplicity $1$, then
	\begin{align*}
		\E \prod_{e\in E(H(Z))} \mathbf{T}_{e}^{m_e(H(Z))} = 0\,.
	\end{align*}
	\begin{proof}
		By independence of hyper-edges and symmetry of their distribution
		\begin{align*}
				\E \prod_{e\in E(H(Z))} \mathbf{T}_{e}^{m_e(H(Z))} &= 	\prod_{e\in E(H(Z))} \E \mathbf{T}_{e}^{m_e(H(Z))} = 0\,.
		\end{align*}
	\end{proof}
\end{fact}

\subsubsection{Encoding tangle-free block non-backtracking hyper walks}\label{section:encoding-tangle-free-hbnbw}
Our goal in this section is to define a meaningful notion of tangle-freeness for hyper-walks in $\hbnbw{}{2q,z}$ and then obtain results along the lines of \cref{lemma:tangle-free-sparse-graph} and \cref{lemma:enumeration-binary-canonical-paths}. For this we need to introduce additional notions.

Recall each $Z$ in $\hbnbw{}{2q,z}$ is a sequence $(\alpha^1_1,\beta^1_1, \ell^1_1,\ldots,  \ell^{2q}_{z}, \alpha^{2q}_{z+1},\beta^{2q}_{z+1})$, which we can decompose it into subsequences $Z_1,\ldots, Z_{2q}\in \hnbw{}{2q,z}$. 
We can imagine each $Z_i$ being revealed through the following discovering process:
\begin{enumerate} 
	\item At time $t_i=0$ we are given multi-indices $\alpha^i_1, \beta^i_1$,
	\item At time $t_i = j >0$ we reveal the multi-indices $\ell_{j}^i\alpha^i_{j+1}\beta^i_{j+1}$.%at positions $3j\,, 3j+1\,, 3j+2$ in $Z_i$.
	%\item At time $t_i = 2j >0$ we reveal the multi-indices $\ell_{2j}^i,\alpha^i_{2j+1},\alpha_{2j+2}^i$.
\end{enumerate}
In other words, having $\alpha^i_1, \beta^i_1$ as our starting indices, we reveal at each time the multi-indices involved in the next two hyper-edges of the underlying hypergraph $H(Z_i)$ of  the subsequence $Z_i$. 
%At each time $j$, the indices in position $3j-2\,, 3j-1$ together with the newly revealed indices in position $3j\,, 3j+1\,, 3j+2$ reveal to two hyper-edges  in the hyper-graph $H(Z_i)$. 
For each time $1\leq j\leq z$, we denote by $R_{ij}\subseteq [n]^3$ the sequence of multi-indices in $Z_i$ revealed at time $j$, we also use $R_\ij$ to denote the corresponding multi-set. We will also refer to $R_\ij$ as a tuple (of size $k$). We denote the two hyper-edges revealed with $R_\ij$ by $E_\ij=\Set{e_{i,j,1}, e_{i,j,2}}$ and by $H(Z_i, j)$ the underlying hyper-graph of $Z_i$ revealed up to time $j$. 
We can now extend this idea to $Z$. We discover $Z$ by revealing in order $Z_1,\ldots, Z_{2q}$ as described above. Notice that due to the block condition in \ref{eq:conditions-hw} we have  $\bigcup_{i\in [2q]\,, j \in [z]}R_\ij = V(Z)$. We define the collection of sets $R_{11},\ldots, R_{2q,z}$ as $\cR(Z)$. We provides two partitions of $\cR(Z)$.
\begin{itemize}
	\item For $0\leq s\leq k$, $R_\ij\in \cG_s$ if exactly $s$ indices in $R_\ij$ did not appear in the sequence $Z$ before. 
	%$R_\ij$ is said to be $i$\textit{-good.} 
	For a subsequence $Z_i$ of $Z$ we write $\cG_s(Z_i)\subseteq\cG_s$ for the subset of tuples in $\cG_s$ corresponding to reveals in $Z_i$.
\end{itemize}
The second partition is:
\begin{itemize}
	\item For $0\leq s\leq 2$, $R_\ij\in \cP_s$ if for the hyper-graph  $\bigoplus_{i'<i} H(Z_i)\oplus H(Z_i,j-1)$ with hyper-edge set $E$ we have $\Card{E\cap E_\ij}=2-s$. That is, $R_\ij\in \cP_s$ if it reveals $s$ new hyper-edges.
	%$R_\ij$ is said to be $s$\textit{-rich}.
\end{itemize}
We are now ready to define $t$-\textit{tangle-free} non-backtracking hyper-walk. Given $Z\in \hbnbw{}{2q, z}$ we say that the subsequence $Z_i\in \hnbw{}{2q,z}$ is $t$-tangle free if
\begin{align}\label{eq:tangle-free}
	\Card{\cT_i:=\Set{R_\ij \in \Paren{\cP_2\cup \cP_1}\cap \cG_0(Z_i) }} \leq t\,.
\end{align}
In words, \cref{eq:tangle-free} is saying that for each $Z_i$ the number of tuples $R_\ij$ that reveal multi-indices containing indices all already seen in $Z_i$, but which reveal at least one new hyper-edge, is at most $t$.
We denote by $\htgf{}{2q,z,t}\subseteq \hnbw{}{2q,z,t}$ the set of $t$-tangle-free non-backtracking hyper-walks. We also use $\hbtgf{}{2q,z,t}\subseteq \hbnbw{}{2q,z}$ to denote the set of block non-backtracking hyper-walks in which each $Z_i$ is in $\htgf{}{2q,z,t}$.
If the block non-backtracking walks in $\bnbw{}{2q,z}$ over $\mathbf{A}'$ only yields $t$-tangle free sequences, we say $\mathbf{A}'$ is $t$-tangle-free.
Using a similar approach as the one shown in the context of sparse graphs in \cref{lemma:tangle-free-sparse-graph},  we will only need to consider walks in $\hbtgf{t}{2q,z,t}$ for $t\leq 100\log\log n$.

\begin{lemma}[Sparse hyper-graphs are tangle-free]\label{lemma:tangle-free-hyper-graph}
	Consider the settings of \cref{theorem:bound-preporcessed-A-via-Ihara-Bass}. Let $n^{-k/2}\leq p\leq n^{-k/2}\log^{10}n$ and $z\leq \frac{\log n}{50}$. Then for  $t\geq 100\log\log n$, $\mathbf{A}'$ is $t$-tangle free with probability $1-o(1)$.
	\begin{proof}
		Let $Z\in \hnbw{}{2q,z}$ be  $t$-tangled and let $v=\card{V(H(Z))}$, $e=\Card{E(H(Z))}$. The probability that $\prod_{e\in E(H(Z))}\mathbf{T}_e\neq 0$ is $p^e$. Now by definition
		\begin{align*}
			e &\geq t + \sum_{1\leq s< (k+1)/2}\Card{\cG_1(Z)}+ 2\sum_{(k+1)/2 \leq s\leq k}\Card{\cG_s(Z)}\,,\\ %2\Card{\cG_3(Z_i)}+2\Card{\cG_2(Z_i)}+\card{\cG_1(Z_i)}\\
			v&\leq  2+\sum_{s\leq k}s\Card{\cG_s(Z)}\,.
		\end{align*}
		Combining the two we get $e\geq t-3+\frac{2}{k}v\,.$
		The number of sequences in $\hnbw{}{2q,z}$ over $v$ indices is at most $n^v\cdot v^{kz+(k-1)}$.
		Thus by union bound%\Tnote{Assuming $k\leq \poly\log n$}
		\begin{align*}
			\bbP \Set{\exists Z\in \hnbw{}{2q,z}\setminus \htgf{}{2q,z,t} \,:\, \prod_{e\in E(H(Z))}\mathbf{T}_e\neq 0}&\leq \sum_{t'\leq t}\sum_{v\leq k(z+1)-1}n^v\cdot v^{k(z+1)-1}\cdot p^{t-3+\frac{2}{k}v}\\
			&\leq \sum_{t'\leq t}\sum_{v\leq k(z+1)-1}\Paren{np^{2/k}}^v\cdot v^{k(z-1)-1}\cdot p^{t'-3}\\
			&\leq\sum_{t'\leq t} \Paren{np^{2/k}v}^{2kz}\cdot p^{t'/2}\\
			&\leq 2\cdot \Paren{k\log n^{11}}^{\frac{k\log n}{11}}\cdot 2^{25k\cdot \log (n)\cdot \log\log n}\\
			&\leq 2^{1+\frac{k\log n}{10}\log\log n-25k\log(n)\cdot \log\log n}\\
			&\leq o(1)\,.
		\end{align*}
	\end{proof}
\end{lemma}
For the remainder of the section we set $t=100\log\log n$.
For simplicity we will say that $t$-tangle free sequences in $\hnbw{}{2q,z}$ are simply tangle-free, we will drop the superscript $t$.

%By \cref{fact:symmetry-hyperwalk} we only need to consider tangle-free block non-backtracking hyper walks in which the underlying hyper-graph does not have hyper-edges of multiplicity $1$. We say such sequences are \textit{interesting}. We denote the corresponding set as $\ihbtgf{}{2q}\subseteq\hbtgf{}{2q,f}$.
Next, we define canonical tangle-free block non-backtracking hyper walks.
We use lexicographic order for the multi-indices: $\alpha\sle \beta$ if at each position $i\leq \min\Set{\card{\alpha},\card{\beta}}$ we have $\alpha(i)\leq \beta(i)$. We say that $Z\in \hbtgf{}{2q,z}$ is \textit{canonical} if $\alpha_1=(1,\ldots, (k-1)/2)$ and for every other multi-index $\beta$ (including $\ell$'s of multiplicity $1$) in the sequence one the following applies:
\begin{enumerate}[(i)]
	\item all the indices in $\beta$ already appeared in the sequence,
	\item indices in $\beta$ are ordered and consecutive (that is for $i\leq\Card{\beta}-1$, $\beta(i)=\beta(i+1)-1$), 
	%for every multi-index $\alpha$ appearing before $\beta$ in the sequence $\alpha\sle \beta$, 
	moreover the index $\beta(0)-1$ already appeared in the sequence.
\end{enumerate}
We use $\cH^{2q,z}$ to denote the set of canonical sequences. 
As in the context of graphs, canonical hyper-walks are convenient as each correspond to a representative for a class of  isomorphic  hyper-walks.
For any canonical sequence over $v$ vertices there are at most %$\binom{n}{v}v!\Brac{(k-1)!}^v\leq n^v\Brac{(k-1)!}^v$
$\binom{n}{v}v!\leq n^v$ isomorphic sequences in $\hbtgf{}{2q,z}$.
We use tuples of the form $\Psi=(v,g_0,\ldots, g_k, r_0, r_1, r_2, d_1, d_2,e^*)$ to encode parameters. Then $\cH^{2q,z}(e, \Psi)$ is the set of canonical sequences $Z$ such that the underlying hyper-graph $H(Z)$ has $e$ distinct hyper-edges, $e^*$ hyper-edges of multiplicity $1$ and:
\begin{align*}
	 \sum_{s\leq k} s\Card\cG_s(Z)=&v\\
	0\leq s\leq k\,,  \Card{\cG_s(Z)}=&g_s\\
	0\leq s \leq 2\,, \Card{\cP_s(Z)}=&r_s\\
	 \Card{\bigcup_{1\leq s\leq 2}\cP_s(Z)\setminus (\bigcup_{1\leq s'\leq k}\cG_{s'}(Z))}=&d_1\\
	 \Card{\bigcup_{1\leq s\leq 2}\cP_s(Z)\setminus\cG_k(Z)} =&d_2\,.
\end{align*} 
The next result, similar in spirit to \cref{lemma:enumeration-binary-canonical-paths}, upper bounds the number of canonical sequences. 

\begin{lemma}[Enumeration of canonical sequences] \label{lemma:enumeration-canonical-sequences} We have
	\begin{align*}
		\Card{\cH^{2q,z}(e, \Psi)}\leq \Brac{\prod_{1\leq s\leq k-1}\Paren{\binom{k}{s}v^{s}}^{g_s}}\cdot v^{kd_1}\cdot k^{4qt}\cdot  z^{16qt(d_2+4)+d_1}
		%\Brac{\prod_{1\leq s\leq 2}\Paren{\binom{3}{s}v^{3-s}}^{g_s}}\cdot v^{3d_1}\cdot z^{2+qt(20+8d_2)}\,.
	\end{align*}
	\begin{proof}
		We can encode each $Z\in \cH^{2q,z}(e,\Psi)$ by encoding each of the tuple $R_{11},\ldots,R_{2q,z}$.
		Notice that since $Z$ is canonical we can encode $Z$ simply by encoding reveals in $\bigcup_{s\leq k-1}\cG_s$ (and their position) as for each $R_\ij \in \cG_s$ we only have one choice for the indices.
		We can encode tuples as follows:
		\begin{itemize}
			\item For $R_\ij\in \cG_s$ for $1\leq s<k$ we have $\binom{k}{s}v^{s}$ choices for the indices that appeared already.  We need not to specify other indices as the sequence is canonical. There are $z$ ways to position such tuples in $Z_i$ so overall $\binom{k}{s}v^{s}z$ possibilities for each such $R_\ij$. In conclusion fixing the cardinalities of $\cG_{1},\ldots, \cG_k$ there are
			$\prod_{1\leq s\leq k-1}\Paren{\binom{k}{s}v^{s}z}^{\Card{\cG_s}}$ choices over $Z$.
			\item For $R_\ij\in \cT_i$, there are $kz$ possible choices for the vertices in $R_\ij$, there are $z$ ways to position such tuples. So overall there are at most $(kz^2)$ choices. By tangle-freeness there are at most $t$ such $R_\ij$'s in each $Z_i$ and thus $\Paren{kz^2}^{2qt}$ possibilities over $Z$.
			\item For $R_\ij\in \Card{\bigcup_{1\leq s\leq 2}\cP_s\setminus (\bigcup_{1\leq s'\leq k}\cG_{s'})}$ we have $v^k$ candidate indices and $z\cdot 2q$ ways to position the tuple. Overall there are $\Paren{v^kz2\cdot q}^{\Card{\bigcup_{1\leq s\leq 2}\cP_s\setminus (\bigcup_{1\leq s'\leq 3}\cG_{s'})}}=\Paren{2v^kzq}^{d_1}$ possibilities over $Z$.
			\item For any subsequence $R_\ij,\ldots, R_{i(j+r)}$ of tuples in $\cP_0$,  we only need to encode the tuples $R_{i(j+r')}$ with $0\leq r'\leq r$ for which at least one of the hyper-edges in $E_{i(j+r')}$ appeared for the first time in the sequence in some $R_{i'j'} \in \bigcup_{1\leq s\leq 2}\cP_s\setminus\cG_k$. In fact we can reconstruct the whole subsequence $R_\ij,\ldots, R_{i(j+r)}$ by the position and the hyper-edges of these tuples. Each of these $R_{i'j'}$ can only appear at most $2t$ times in the subsequence $R_\ij,\ldots, R_{i(j+r)}$ since by assumption $Z_i$ is $t$-tangle-free. For convention if the subsequence ends in $R_{iz}$ we also specify the hyper-edges revealed by this tuple. 
			Each subsequence $R_\ij,\ldots, R_{i(j+r)}$ can be position in $z$ different ways in $Z_i$ and has length at most $z$.
			Since there are $2q$ such $Z_i$'s in $Z$,
			overall we have $z^{8qt\cdot 2\Paren{\Card{\bigcup_{1\leq s\leq 2}\cP_r\setminus\cG_k}+1}}=z^{16qt\Card{\bigcup_{1\leq s\leq 2}\cP_r\setminus\cG_k}+2}=z^{16qt(d_2+2)}$ distinct choices.
		\end{itemize}
		We deduce that for any valid $\Psi=(v,g_0, \ldots, g_k, r_0,r_1,r_2, d_1, d_2,e^*)$ 
		\begin{align*}
			\Card{\cH\Paren{e, \Psi}}&\leq \Brac{\prod_{1\leq s\leq 2}\Paren{\binom{k}{s}v^{s}z}^{g_s}}\cdot (kz^2)^{2qt} \Paren{2v^kzq}^{d_1}z^{16qt(d_2+2)}\\
			%\Card{\cH\Paren{e, \Psi}}&\leq \Brac{\prod_{1\leq s\leq 2}\Paren{\binom{k}{s}v^{s}z}^{g_s}}\cdot \Paren{z}^{20qt}\Paren{v^3z}^{d_1}z^{8qtd_2+2}\\
			&= \Brac{\prod_{1\leq s\leq 2}\Paren{\binom{k}{s}v^{s}}^{g_s}}\cdot v^{kd_1}\cdot k^{4qt}\cdot  z^{16qt(d_2+4)+d_1}\,.
		\end{align*}
	\end{proof}
\end{lemma}

\subsubsection{Putting things together}

We are finally ready to prove \cref{lemma:bound-trace-xor}. 
%We introduce one additional notion.
%We say a tangle-free block non-backtracking hyper walks is $r$-\textit{interesting} if the underlying hyper-graph has $r$ hyper-edges of multiplicity $1.$ We denote the corresponding set as $\ihbtgf{}{2q,k,r}\subseteq\hbtgf{}{2q,k}$. We let $\ihbtgf{}{2q,k}:=\underset{r\leq 4q\cdot k}{\bigcup}\ihbtgf{}{2q,k,r}$.

\begin{proof}[Proof of \cref{lemma:bound-trace-xor}]
	For $Z\hbnbw{}{2q,z}$ we write  $\alpha^i_j$ in place of $\alpha^i_j(Z)$ to denote multi-indices in the sequence $Z$.
	By definition
	\begin{align*}
		\Tr &\Brac{\Paren{\mathbf{B}^{z-1}}\transpose{\Paren{\mathbf{B}^{z-1}}}}^q \\
		&=\sum_{Z\in \hbnbw{}{2q,z}} \prod_{i=1}^{2q} \Brac{\Abs{\sum_{\ell_1^i}\mathbf{T}_{(\alpha_1^i\alpha_2^i\ell_1^i)}\mathbf{T}_{(\beta_1^i\beta^i_2\ell^i_1)}}\Paren{\prod_{\substack{s=2}}^{z-1}\mathbf{T}_{(\alpha^i_s \alpha_{s+1}^i\ell_{s}^i)} \mathbf{T}_{(\beta_{s}^i, \beta_{s+1}^i\ell_{s}^i)}}}\,.
	\end{align*}
	By \cref{lemma:tangle-free-hyper-graph} with probability $1-o(1)$,
	\begin{align*}
		&\sum_{Z\in \hbnbw{}{2q,z}} \prod_{i=1}^{2q} \Brac{\Abs{\sum_{\ell_1^i}\mathbf{T}_{(\alpha_1^i\alpha_2^i\ell_1^i)}\mathbf{T}_{(\beta_1^i\beta^i_2\ell^i_1)}}\Paren{\prod_{\substack{s=2}}^{z-1}\mathbf{T}_{(\alpha^i_s \alpha_{s+1}^i\ell_{s}^i)} \mathbf{T}_{(\beta_{s}^i, \beta_{s+1}^i\ell_{s}^i)}}}\\
		&=\sum_{Z\in \hbtgf{}{2q,z}} \prod_{i=1}^{2q} \Brac{\Abs{\sum_{\ell_1^i}\mathbf{T}_{(\alpha_1^i\alpha_2^i\ell_1^i)}\mathbf{T}_{(\beta_1^i\beta^i_2\ell^i_1)}}\Paren{\prod_{\substack{s=2}}^{z-1}\mathbf{T}_{(\alpha^i_s \alpha_{s+1}^i\ell_{s}^i)} \mathbf{T}_{(\beta_{s}^i, \beta_{s+1}^i\ell_{s}^i)}}}\,.
	\end{align*}
	Now as shown in \cref{eq:equivalence-walks-hyper-walks}
	\begin{align}
		\E \sum_{Z\in \hbtgf{}{2q,z}} &\prod_{i=1}^{2q} \Brac{\Abs{\sum_{\ell_1^i}\mathbf{T}_{(\alpha_1^i\alpha_2^i\ell_1^i)}\mathbf{T}_{(\beta_1^i\beta^i_2\ell^i_1)}}\Paren{\prod_{\substack{s=2}}^{z-1}\mathbf{T}_{(\alpha^i_s \alpha_{s+1}^i\ell_{s}^i)} \mathbf{T}_{(\beta_{s}^i, \beta_{s+1}^i\ell_{s}^i)}}}\nonumber\\
		\leq& O(1)^{2q}\cdot \E \sum_{Z\in \hbtgf{}{2q,z}}\mathbf{T}_{(\alpha^1_1,\alpha^1_2,\ell^1_1)}\cdots \mathbf{T}_{(\beta^{2q}_{z},\beta^{2q}_{z+1}, \ell^{2q}_{z})}\label{eq:3xor-local-tangle-free} \\
		&+ O(1)^{2q}\cdot \E \sum_{Z\in \hbtgf{}{2q,z}(\cC^*)}\Abs{\mathbf{T}_{(\alpha^1_1,\alpha^1_2,\ell^1_1)}\cdots \mathbf{T}_{(\beta^{2q}_{z},\beta^{2q}_{z+1}, \ell^{2q}_{z})}}\,.\label{eq:3xor-local-tangle-free-annoying}
	\end{align}
	We start by studying \cref{eq:3xor-local-tangle-free}. By \cref{fact:symmetry-hyperwalk} we only need to consider hyper-walks  $Z\in\hbtgf{}{2q,z}$ with at most $2qk$ distinct hyper-edges since each hyper-edge must have multiplicity at least $2$ in the underying hyper-graph $H(Z)$.
	%\begin{align*}
	%	\E& \sum_{Z\in \hbtgf{}{2q,k}}\mathbf{T}_{a^1_1(Z),b^1_1(Z),\ell^1_1(Z)}\cdots \mathbf{T}_{a^{2q}_{k+1}(Z),b^{2q}_{k+1}(Z), \ell^{2q}_{k}(Z)} \\
	%	&= \E \sum_{Z\in \ihbtgf{}{2q,k,0}}\mathbf{T}_{a^1_1(Z),b^1_1(Z),\ell^1_1(Z)}\cdots \mathbf{T}_{a^{2q}_{k+1}(Z),b^{2q}_{k+1}(Z), \ell^{2q}_{k}(Z)}
	%\end{align*}
	Thus  we can upper bound \cref{eq:3xor-local-tangle-free} with 
	\begin{align*}
		\underset{\substack{e\leq 2qz\,, v\geq 0\\g_k\,,\ldots\,,g1\geq 0\\r_2\,,r_1\,,r_0\geq 0\\d_1\,,d_2\geq 0}}{\sum} n^v\cdot p^e\cdot \Card{\cH(e, \Psi)}= 	\underset{e\leq 2qz, \Psi}{\sum} n^v\cdot p^e\cdot \Card{\cH(e, \Psi)}\,,
	\end{align*}
	where the right-hand side is a simple rewriting for compactness.
	Since for any $Z\in \hbtgf{}{2q,z}$ we have 
	\begin{align*}
		v&=\sum_{i\leq k}s\Card{\cG_s}\\
		e&=2\sum_{(k+1)/2\leq s\leq k}\Card{\cG_s}+\Card{\bigcup_{1\leq s\leq 2}\cP_s\leq \Paren{\bigcup_{(K+1)/2\leq s'\leq 3}\cG_{s'}}}\,,
	\end{align*}
	it follows using \cref{lemma:enumeration-canonical-sequences}
		\begin{align*}
		\underset{\substack{e\leq 2qz\,, \Psi}}{\sum} &n^v\cdot p^e\cdot \Card{\cH(e,\Psi)}\\
		&\leq 	\sum_{e\leq 2qz\,, \Psi} n^v\cdot p^e\cdot  \Brac{\prod_{1\leq s\leq k-1}\Paren{\binom{k}{s}v^{s}}^{g_s}}\cdot v^{kd_1}\cdot k^{4qt}\cdot  z^{16qt(d_2+4)+d_1}\\
		&\leq 2 \sum_{e\leq 2qz\,, \Psi} \underbrace{
			\prod_{(k+1)/2\leq s\leq k}\Paren{\binom{k}{s}v^sn^sp^2}^{g_s}\cdot
			\prod_{1\leq s<(k+1)/2}\Paren{\binom{k}{s}v^sn^sp}^{g_s}\cdot \Paren{v^kp}^{d_1}\cdot  k^{4qt}\cdot  z^{16qt(d_2+4)+d_1}}_{=:L(e,\Psi)}\,.
	\end{align*}
	It is easy to see that this is a geometric sum.
	Using the assumptions $p\geq n^{-3/2}n\,, z\leq \log n\,, q \leq \frac{\log n}{\Paren{10^3\log\log n}^2}$ and $t=100\log\log n$, as $k\leq O(\log n)$ we have that decreasing $g_s$ in $\Psi$ and increasing $g_{s-1}$ with $(k+1)/2<s\leq k$ to obtain $\Psi'$ we have
	\begin{align*}
		\frac{L(e, \Psi)}{L(e, \Psi')}\geq \frac{n}{v\cdot k\cdot z^{8qt}}\geq \omega(1)\,.
	\end{align*}
	Similarly, decreasing $g_s$ in $\Psi$ and increasing $g_{s-1}$ with $1<s<(k+1)/2$ to obtain $\Psi'$ we have
	\begin{align*}
		\frac{L(e, \Psi)}{L(e, \Psi')}\geq \frac{n}{v\cdot k\cdot z^{8qt}}\geq \omega(1)\,.
	\end{align*}
	Finally, decreasing $g_k$ in $\Psi$ and increasing any $g_{s}$ with $1\leq s<(k+1)/2$ to obtain $\Psi'$ we get%\Tnote{Here tight assumption on $k$: $k^k\leq n^{0.01}$}
	\begin{align*}
		\frac{L(e, \Psi)}{L(e, \Psi')}\geq \frac{n^{k-(k-1)/2}p}{\binom{k}{\frac{k-1}{2}}\cdot v^k\cdot k\cdot z^{8qt}}\geq \frac{\sqrt{n}}{\Paren{2v}^k\cdot z^{8qt}}\geq \omega(1)\,.
	\end{align*}
	Thus as $\sum_{s\leq k}\Card{\cG_k}\leq qz$ we have
	\begin{align*}
		\underset{\substack{e\leq 2qz\,, \Psi}}{\sum} n^v\cdot p^e\cdot \Card{\cH(e,\Psi)}\leq \Paren{n^kp^2}^{qz}\cdot O\Paren{z^{65qt}}\,,
	\end{align*}
	where we also accounted for the $O(1)^{2q}$ multiplicative factor.
	The argument for \cref{eq:3xor-local-tangle-free-annoying} is similar. We can upper bound \cref{eq:3xor-local-tangle-free-annoying} by
	\begin{align*}
		\sum_{e^*\leq 2qz}\sum_{e\leq 2qz+e^*\,, \Psi}n^{v-ke^*}\cdot p^e\cdot \Card{\cH(e,\Psi)}\cdot \Paren{2(k!)qkn}^{e^*}
	\end{align*}
	where we used the crucial fact that for any $Z\in \hbtgf{}{2q,z}(\cC^*)$ any tuple $(\alpha,\beta,\ell)$ appearing an odd number of times must satisfy $\Card{S(\alpha,\beta,\ell)\cap S(\alpha^i_1,\alpha^i_2,\ell^i_1)}\geq k-1$ or $\Card{S(\alpha,\beta,\ell)\cap S(\beta^i_1,\beta^i_2,\ell^i_1)}\geq k-1$ and thus the number of possible choices for those indices is at most $(2(k!)qkn)$. Repeating the analysis above the result follows.
\end{proof}

\section{Strong refutations for random CSPs}\label{section:csp-refutations}
\cref{theorem:main-xor-technical} can also be used to  obtain strong refutations for random CSPs. Similar reductions appeared already in the literature (e.g. \cite{AllenOW15}), however we crucially exploits the sharp novel bound in \cref{theorem:main-xor-technical} to obtain stronger results as in \cref{theorem:main-csp}.
We use the notation introduced in \cref{section:preliminaries-csps}.

\begin{theorem}\label{theorem:main-csp-technical}
	Let $P:\{-1,+1\}^k\rightarrow \{0,1\}$ be a predicate.
	Consider a random instance $\bm \cI\sim  \cspd$ for odd $k\geq 3$. 
	%such that $k\log k \leq o(\log n)$. 
	For $n$ large enough,  there exists a universal constant $C>0$ and a polynomial time algorithm that, if
	\begin{align*}
		p\geq \frac{C\cdot n^{-k/2}}{\eps^2}
	\end{align*}
	certifies with probability $0.99$ that
	\begin{align*}
		\optI{\bm \cI}\leq \E_{\mathbf{z}\overset{u.a.r}{\sim}\Set{\pm 1}^k}\Brac{P(\mathbf{z})} +O(\eps)\,.
	\end{align*}
	%\Tnote{Check if there is a$2^{O(k)}$ factor.}
\end{theorem}
%\begin{remark}
%	The use of the parameter $m$ as opposed to $p$ as in \cref{theorem:main-xor-technical} is mainly due to the way we formulated the random process generating $\bm \cI$. We always have the relation $m=(1\pm o(1))\cdot p\cdot n^{k}$ with high probability.
%\end{remark}
Recall from \cref{section:preliminaries-csps} that  we can represent the predicate $P$ as a multi-linear polynomial of degree $k$,
\begin{align*}
	P(c\circ x^{ \alpha}) =\sum_{d\leq k} P_d(c\circ x^{ \alpha})\,,
\end{align*}
where $P_d$ denotes the degree $d$ part of the predicate. In particular $P_0:=P_0(c\circ x^{ \alpha})$ denotes the constant part of the polynomial, which does not depend on the assignment and the negative pattern.
For a instance $\cI$ with $m$ constraints, we define  
$$
S_{\cI}(x):=\sum_{(c,\alpha)\in \cI} \sum_{1\leq d\leq k} P_d(c\circ x^{\alpha}) \,.
$$
%\E_{\mathbf{z}\overset{u.a.r}{\sim}\Set{\pm 1}^k}\Brac{P(\mathbf{z})} +
Notice that by definition, for any fixed  $x\in \Set{\pm 1}^n$ it holds
\begin{align*}
	S_\cI (x) = m  \cdot \Paren{\valI{\cI}{x}-P_0} = m  \cdot \Paren{\valI{\cI}{x}-\E_{\mathbf{z}\overset{u.a.r}{\sim}\Set{\pm 1}^k}\Brac{P(\mathbf{z})}}\,,
\end{align*} 
where we used the fact that for every $1\leq d\leq k$, $\E_{\mathbf{z}\overset{u.a.r}{\sim}\Set{\pm 1}^k}\Brac{P_d(\mathbf{z})}=0$ by symmetry.
Thus to obtain \cref{theorem:main-csp-technical} it suffices to obtain a tight upper bound on $	\max_{x\in \Set{\pm 1}^n}S_{\bm \cI}(x)$ for $\bm \cI\sim \cspd$.
We further write $S_{\cI,d}(x)$ to denote the degree $d\leq k$ part of $S_{\cI}(x)$. We then have $S_{\cI}(x)=\sum_{d\leq k}S_{\cI,d}(x)$. We can write $S_{\cI,d}(x)$ as a $n^{\lfloor d/2\rfloor }$-by-$n^{\lceil d/2\rceil}$ matrix $M_{\cI,d}$ such that $S_{\cI,d}(x)=\iprod{\tensorpower{x}{\lfloor d/2\rfloor }, M_{\cI, d}\tensorpower{x}{\lceil d/2\rceil}}$ for any $x\in \R^n$.
The next lemma, provides a rough bound on the spectral norm of each $M_{\bm \cI, d}$ when $\bm \cI\sim \cspd$. We remark that these bounds on $S_{\bm \cI,d}(x)$ are significantly less sharp than \cref{theorem:main-xor-technical}, nevertheless they will be good enough for our needs.

\begin{lemma}\label{lemma:csps-rough-matrix-bounds}
	Consider the settings of \cref{theorem:main-csp-technical}. % and let $m$ be the number of clauses in $\bm \cI$.
	For some $d<k$ let $M_{\bm \cI,d}$ be the $n^{\lfloor d/2\rfloor }$-by-$n^{\lceil d/2\rceil}$  matrix representing $S_{\bm \cI, d}$. 
	Then with probability $1-n^{\Omega(1)}$,
	\begin{align*}
		\Norm{M_{\bm \cI,d}}\leq O\Paren{\sqrt{p\cdot 2^k\cdot n^{k-\lfloor d/2\rfloor}}+1}\sqrt{\log n},.
		%O\Paren{1+\sqrt{\frac{m}{n^{\lfloor d/2\rfloor}}}}\cdot \sqrt{\log (n)}\,.
	\end{align*}
	\begin{proof}
		%Recall that  $P(\mathbf{c}\circ x^{\bm \alpha})$ can be written as a polynomial of degree $k$, we use the notation
		%\begin{align*}
		%	P(\mathbf{c}\circ x^{\bm \alpha}) =\sum_{d\leq k} P_d(\mathbf{c}\circ x^{\bm \alpha})\,,
		%\end{align*}
		%where $P_d$ denotes the degree $d$ part of the predicate.
		Let $0\leq\tau\leq O(1)$ be the largest coefficient in absolute value in the polynomial $P(z)$, for any $z\in \Set{\pm 1}^k$.
		For each $\alpha \in [n]^k$ and $c\in \Set{\pm 1}^k$ let $\mathbf{M}_{(c, \alpha)}$ be the $n^{\lfloor d/2\rfloor}\times n^{\lceil d/2\rceil}$ matrix flattening  of $P_d(c\circ x^{\alpha})$ so that
		\begin{align*}
			P_d(c\circ x^{\alpha})= \iprod{\tensorpower{x}{\lfloor d/2\rfloor},\mathbf{M}_{(c, \alpha)} \tensorpower{x}{\lceil d/2\rceil}}\,,
		\end{align*}
		for any $x\in \R^{n}\,.$
		We use the decomposition 
		\begin{align*}
			M_{\bm \cI,d} = \sum_{(c,\alpha)\in \Set{\pm 1}^k\times [n]^k} \mathbf{M}_{(c, \alpha)}\,.
		\end{align*}
		Now each $\mathbf{M}_{(c, \alpha)}$ satisfies, for any $\alpha_1,\alpha_1'\in [n]^{\lfloor d/2 \rfloor}$ and $\alpha_2,\alpha_2'\in [n]^{\lceil d/2\rceil}$
		\begin{align*}
			\E \Brac{\Paren{\mathbf{M}_{(c, \alpha)}}_{\alpha_1\alpha_2}}&=0\\
			\text{if }(\alpha_1,\alpha_2)\neq \alpha\,,\quad\bbP \Brac{\Paren{\mathbf{M}_{(c, \alpha)}}_{\alpha_1\alpha_2}\neq 0}&=0\\
			\text{if }(\alpha_1,\alpha_2)= \alpha\,,\quad\bbP \Brac{\Paren{\mathbf{M}_{(c, \alpha)}}_{\alpha_1\alpha_2}\neq 0}&\leq p\\
			\Abs{\Paren{\mathbf{M}_{(c, \alpha)}}_{\alpha_1\alpha_2}}&\leq \tau\\
			\text{for }\alpha_1\neq \alpha_1' \text{ and }\alpha_2\neq \alpha_2'\,,\quad \E \brac{\Paren{\mathbf{M}_{(c, \alpha)}}_{\alpha_1\alpha_2}\Paren{\mathbf{M}_{(c, \alpha)}}_{\alpha_1'\alpha_2'}}&=0\,.
		\end{align*}
		We can thus apply Bernstein's inequality for matrices as in  \cref{theorem:matrix-bernstein}.
		We have
		\begin{align*}
			\sigma^2:=& \max\Set{\Norm{\sum_{(c,\alpha)}\E \mathbf{M}_{(c,\alpha)}\transpose{\mathbf{M}}_{(c,\alpha)}}\,, \Norm{\sum_{(c,\alpha)}\E \transpose{\mathbf{M}}_{(c,\alpha)}\mathbf{M}_{(c,\alpha)}}}\\
			\leq& p\cdot\tau^2\cdot 2^k\cdot n^{k-\lfloor d/2\rfloor}\,.
		\end{align*}
		Thus choosing $t=O\Paren{\sqrt{p\cdot\tau^2\cdot 2^k\cdot n^{k-\lfloor d/2\rfloor}}+\tau}\sqrt{\log n}$ the result follows.
		%Let us simply write $\mathbf{M}$ in place of $M_{\bm \cI,d}$.  
		%We may decompose $\mathbf{M}$ into $m$ $n^{\lfloor d/2\rfloor }$-by-$n^{\lceil d/2\rceil}$ independent matrices $\mathbf{M}_{\bm \alpha_1},\ldots,\mathbf{M}_{\bm \alpha_m}$ each corresponding to the flattening of the degree $d$ part of a constraint $P(\mathbf{c}\circ x^{\bm \alpha_i})$ in $\bm \cI$. %Then for $\ell\in [m]$, $\mathbf{M}_\ell$ satisfies
		%\begin{align*}
		%	\E \Brac{\Paren{\mathbf{M}_\ell}_\ij} &= 0\\
		%	\bbP \Brac{\Paren{\mathbf{M}_\ell}_\ij\neq 0}&\leq \frac{\binom{k}{d}}{n^d}\\
		%	\Abs{\Paren{\mathbf{M}_\ell}_\ij}&\leq \tau\\
		%	\text{for }(i,j)\neq ('i,j')\quad \E %\brac{\Paren{\mathbf{M}_\ell}_\ij\Paren{\mathbf{M}_\ell}_{i'j'}}&=0\,,
		%\end{align*}
		%where the last identity follows by randomness of the negation pattern $\bm c\in %\Set{\pm 1}^k$ of the $\ell$-th predicate.
		%Thus applying the concentration inequality \cref{lemma:matrix-concentration} we get
		%\begin{align*}
		%	\Norm{M_{\bm \cI,d}}&\leq O\Paren{1+\sqrt{m\cdot \Paren{\frac{k}{n}}^d\cdot %n^{\lceil d/2\rceil}}}\cdot \tau\cdot \sqrt{d\log (n)}\\
		%	&\leq O\Paren{1+\sqrt{\frac{m\cdot k^d}{n^{\lfloor d/2\rfloor}}}}\cdot \tau\cdot %\sqrt{d\log (n)}\,.
		%\end{align*}
	\end{proof}
\end{lemma}

We can now prove the main theorem of the section.

\begin{proof}[Proof of \cref{theorem:main-csp-technical}]
	As for \cref{theorem:refutation-k-xor}, we focus on the case $p\leq n^{-k/2}\polylog(n)$ as for larger $p$ results are known already \cite{AllenOW15}.
	Moreover with high probability  the number of clauses in  $\bm \cI$ is $m=(1+o(1))pn^{k}$.
	For any $x\in \Set{\pm 1}^n$ we have
	\begin{align*}
		\valI{\bm \cI}{x} &=  \E_{\mathbf{z}\overset{u.a.r}{\sim}\Set{\pm 1}^k}\Brac{P(\mathbf{z})}+ \frac{1}{m}\sum_{1\leq d\leq k} S_{\bm \cI, d}(x)\,.
	\end{align*}
	By \cref{lemma:csps-rough-matrix-bounds}, with probability at least  $1-kn^{-\Omega(1)}=1-o(1)$,
	\begin{align*}
		\max_{x\in \Set{\pm 1}^n}\frac{1}{m}\sum_{1\leq d< k} S_{\bm \cI, d}(x)&\leq 	\frac{1}{m}\sum_{1\leq d<k}n^{d/2}\cdot 	\Norm{M_{\bm \cI,d}}\\
		&\leq \frac{1}{m}\sum_{1\leq d<k}n^{d/2}\cdot O\Paren{\sqrt{p\cdot 2^k\cdot n^{k-\lfloor d/2\rfloor}}+1}\sqrt{\log n}\\
		&\leq \sum_{1\leq d<k}n^{-k/2+d/2-k/4+k/2-\frac{\lfloor d/2\rfloor}{2}}\cdot 2^{k/2}\cdot \polylog(n)\\
		%&\leq 	\frac{1}{m}\sum_{d<k}n^{d/2}\cdot O\Paren{\sqrt{\frac{m}{n^{\lfloor d/2\rfloor}}}}\cdot\log (n)\\
		&\leq O\Paren{n^{-\frac{k}{4}+\frac{\lceil d/2\rceil}{2}}}\polylog(n)\\
		&\leq o(1)\,.
	\end{align*}
	Applying \cref{theorem:main-xor-technical} to $\max_{x\in \Set{\pm 1}^n}S_{\bm \cI, k}(x)$ the desired bound on $\optI{\bm \cI}$ follows.
	Finally, we also immediately obtain a polynomial time applying any efficient method to certify the spectral norm of $\sum_{d<k}M_{\bm \cI,d}$.
\end{proof}

%\section{Algorithm for semi-random k-CSPs}\label{section:algorithm}
%In this section, we exploit the bounds obtained in \cref{section:random-3xor} to design a polynomial time algorithm that, given a semi-random $k$-CSP instance $\cI$ with $n$ variables and \textit{arbitrary} (possibly adversarial) signs, achieves a $(1-\epsilon)$-approximation in time $n^{O(k/\epsilon^2)}$.

%\subsection{Algorithm for semi-random k-XOR}\label{section:algorithm-semi-random-k-xor}

\section{Algorithm for k-XOR with adversarial signs}\label{section:algorithm}
In this section, we exploit the bounds obtained in \cref{section:random-3xor} to design a polynomial time algorithm that, given a semi-random $k$-XOR instance $\cI$ with $n$ variables and \textit{arbitrary} (possibly adversarial) signs, achieves a $(1-\epsilon)$-approximation in time $n^{O(k/\epsilon^2)}$.
We start by defining the semi-random model of interest.

\begin{definition}[Semi-random $k$-XOR]\label{definition:semi-random-xor}
A semi-random $k$-XOR instance $\cI$ with $n$ variables and $m:=p\binom{n}{k}(1\pm o(1))$ clauses can be generated through the following process:
\begin{enumerate}
    \item[(i)]  Pick a
random symmetric tensor $\mathbf{T}'$, with independent entries, such that $\mathbf{T}'_{\alpha}=0$ if the indices in the multi-index $\alpha\in [n]^{k}$ are not distinct and otherwise:%
\begin{align*}
	\mathbf{T}'_{\alpha} = \begin{cases}
		0& \text{with probability }1-p\,,\\
		+1&\text{with probability }p/2\,,\\
		-1&\text{with probability }p/2\,.
	\end{cases}
\end{align*}
\item[(ii)] Given $\mathbf{T}'$, pick an arbitrary binary function (possibly chosen adversarially from $\mathbf{T}'$) $\sigma:[n]^k\rightarrow \Set{\pm 1}$ and let $T$ be the tensor with entries $T_\alpha:=\sigma(\alpha)\mathbf{T}'_\alpha\,.$
\end{enumerate}
$\cI$ consists of the $k$-XOR predicates $k\text{-XOR}(\alpha)=\frac{1-x^{\alpha}(-T)_\alpha}{2}=\frac{1-\sigma(\alpha)\cdot x^{\alpha}(-\mathbf{T}')_\alpha}{2}$ where $\mathbf{T}'_{\alpha}$ is non-zero.
\end{definition}

For a semi-random instance $\cI$, we denote by $m(\cI)$ the exact number of clauses in the instance.
For convenience, we  denote by $\mathbf{T}'_\cI$ and $\sigma_\cI$ respectively the random tensor and the  adversarial binary function associated to $\cI$. We also use $T_\cI$ to denote the tensor with entries $(T_\cI)_\alpha:=\sigma(\alpha)(\mathbf{T}'_\cI)_\alpha\,.$ When the context is clear we drop the subscripts.
The max $k$-XOR problem is that of finding an assignment with value
\begin{align*}
    \optI{\cI} :=\frac{1}{m(\cI)}\max_{x\in \Set{\pm 1}^n}\sum_{\alpha\in [n]^k} T_\alpha x^{\alpha}=\frac{1}{m(\cI)}\max_{x\in \Set{\pm 1}^n}\sum_{\alpha\in [n]^k}  \sigma(\alpha)\cdot \mathbf{T}'_\alpha x^{\alpha}\,.
\end{align*}

With the above objective and  adversarial model and in mind, we prove the following theorem, which implies \cref{theorem:main-algorithm-xor}.

\begin{theorem}\label{theorem:algorithm-technical-xor}
Let $n\,,k$ be positive integers, $\epsilon > 0\,, n$ and $n^{-k/2}/\epsilon^2 < p(n):=p < 1$.
Let $\cI$ be a semi-random $k$-XOR instance with parameters $n,p$ as in \cref{definition:semi-random-xor}. There exists a randomized algorithm (\cref{algorithm:k-xor}), running in time $n^{O(k/\epsilon^2)}$, that returns an assignment $\hat{\mathbf{x}}$ satisfying
\begin{align*}
    \valI{\cI}{\hat{\mathbf{x}}}\geq \optI{\cI}-O(\epsilon)\,,
\end{align*}
with probability at least $0.99$.
\end{theorem}

Before proving the result, we introduce some additional notation. We focus on the settings with odd $k$, since for even $k$ \cref{theorem:algorithm-technical-xor} is implied by \cite{Alev2019approximating}.
We denote by $\cQ_t$ the set of degree-$t$ pseudo-distributions in indeterminates $x_1,\ldots, x_n$ satisfying 
\begin{align*}
    \Set{x_i^2 = 1\,, \forall i \in [n]}\,.
\end{align*}
We write $x=(x_1,\ldots, x_n)$.
Let $\cI$ be a $k$-XOR instance for odd $k\geq 3$ and let $T_{\cI}\in \R^{\tensorpower{n}{k}}$ be the associated tensor.  As usual, when the context is clear we drop the subscript.
Throughout the rest of the section we also use the symbols $\alpha, \beta$ to denote multi-indices in $[n]^{q}$, for some $q\in \N$.
%We denote the set of multi-linear multi-indices in $[n]^{q}$ by $\cS_{q}$ for any $q\geq 1$.
For an index $\ell \in [n]$ and multi-indices $\alpha,\alpha'\in [n]^{\frac{k-1}{2}}$, we write $(\bm \alpha, \bm \alpha', \bm \ell)\sim T$ to denote an entry picked uniformly at random among the \textit{non-zero} entries of $T$. For an assignment $x\in \Set{\pm 1}^n$,  $\E_{(\bm \alpha, \bm \alpha', \bm \ell)\sim T}x^{(\bm \alpha. \bm\alpha', \bm \ell)}$ is the expectation of the monomial $x^{(\bm \alpha. \bm\alpha', \bm \ell)}$ for the uniform distribution over all \textit{non-zero} entries of $T\,.$
For a fixed $\ell\in [n]$, we write $(\bm \alpha, \bm \alpha') \sim T_\ell$ to denote a non-zero entry in $T$ picked uniformly at random among those containing index $\ell$.
Furthermore, we denote by $D(T)$ the distribution over $[n]$ such that, for each $\ell\in [n]$, its probability is proportional to the number of non-zero entries in $T$ with index $\ell$.
Therefore, we have
\begin{align*}
    \E_{(\bm \alpha, \bm \alpha', \bm \ell) \sim T}x^{(\bm \alpha, \bm \alpha',\bm \ell)} = \E_{\bm \ell \sim D(T)}\E_{(\bm \alpha, \bm \alpha')\sim T_{\bm \ell}}x^{(\bm \alpha, \bm \alpha',\bm \ell)}\,.
\end{align*}
%We can further generalize the above notation and write, for $\ell\in [n]$, $\tilde{\E}_{\bm \alpha\in E(H, \ell, -)}x^{\bm \alpha}$ to denote the expectation of $x^{\bm \alpha}$ of the uniform distribution over all edges in $H$ containing the indices in $(\bm \alpha, \ell)$.
Finally, we introduce the following crucial definitions.

\begin{definition}[Local correlation]\label{definition:local-correlation}
Let $t\geq 2$.
Let $T\in \R^{n^{\otimes k}}$ be a symmetric tensor. Let $\mu \in \cQ_t$, we define the local correlation of $\mu$ on $T$ to be
\begin{align*}
    \LC{T}{\mu} :=  \E_{\bm \ell \sim  D(T)}\E_{\substack{(\bm \alpha, \bm \alpha')\sim T_{\bm \ell}\\
    (\bm \beta, \bm \beta')\sim T_{\bm \ell}}}\tilde{\E}\Paren{x^{(\bm \alpha, \bm \alpha',\bm \beta, \bm \beta')} - 2x^{(\bm \alpha, \bm \alpha')}\tilde{\E}x^{\bm \beta}\tilde{\E}x^{\bm \beta'} + \tilde{\E}x^{\bm \alpha}\tilde{\E}x^{\bm \alpha'}\tilde{\E}x^{\bm \beta}\tilde{\E}x^{\bm \beta'}}
\end{align*}
\end{definition}

For an instance $\cI$ as in \cref{definition:semi-random-xor} we interchangeably use  $\LC{\cI}{\mu}$, $\LC{\mathbf{T}'}{\mu}$ and $\LC{T}{\mu}$.
 
\begin{definition}[Global correlation]\label{definition:global-correlation}
Let $t\geq 2$. Let $\mu \in \cQ_t$, we define the global correlation of $\mu$ to be
    \begin{align*}
        \GC{\mu} :=  \E_{\bm \ell \sim [n]}\E_{\substack{(\bm \alpha, \bm \alpha')\sim [n]^{k-1}\\
        (\bm \beta, \bm \beta')\in [n]^{k-1}}}\Abs{\Cov{\mu}{x^{(\bm \alpha, \bm \alpha')}, x^{(\bm \beta, \bm \beta')}}} + \Abs{\Cov{\mu}{x^{\bm \alpha}, x^{\bm \beta}}} +  \Abs{\Cov{\mu}{x^{\bm \alpha'}, x^{\bm \beta'}}}\,.
    \end{align*}
\end{definition}

Our proof of \cref{theorem:algorithm-technical-xor} will be inspired by the \textit{local correlation to global correlation} approach of \cite{Barak2011rounding}.
First, we show that if we can find a pseudo-distribution in $\cQ_t$ with local correlation at most $\epsilon^2$, then we can obtain the desired approximation.
Second, we argue that we can always find a pseudo-distribution in $\cQ_t$ with low global correlation. Finally, using \cref{theorem:main-xor-technical}, we show that high local correlation implies high global correlation. This will yield the desired result.

We start by present the algorithm behind \cref{theorem:algorithm-technical-xor}. Its correctness is then analyzed in the subsequent sections.

\begin{algorithmbox}[Algorithm for semi-random $k$-XOR] \label{algorithm:k-xor}
	\mbox{}\\
	\textbf{Input:} A $k$-XOR instance $\cI$ as in \cref{definition:semi-random-xor}, $ \epsilon>0\,, t\geq \Omega(k/\epsilon^2)$.
  
  \noindent
  \textbf{Output:} assignment $\hat{\mathbf{x}}\in\Set{\pm 1}^n$

  \noindent
	\textbf{Operations:}
	\begin{enumerate}
    \item Find a pseudo-distribution $\mu \in \cQ_t$ maximizing $\E_{(\bm \alpha, \bm \alpha',\bm\ell)\sim T}\tilde{\E}_\mu x^{(\bm \alpha, \bm\alpha',\bm \ell)}$, for large enough $t\geq \Omega(k/\epsilon^2)\,.$
    \item If $\GC{\mu}> \epsilon^2$, let $\bm \mu'$ be the pseudo-distribution returned by \cref{algorithm:driving-global-correlation} on input $\mu$.
    \item For each $i\in [n]$, set $\hat{\mathbf{x}}_i = 1$ with probability $\frac{1+\tilde{\E}_{\bm \mu'} x_i}{2}$ and $-1$ otherwise. Return $\hat{\mathbf{x}}\,.$
  \end{enumerate}
\end{algorithmbox}

\begin{remark}[Running time]
    Finding a pseudo-distribution in $\cQ_t$ requires time $n^{O(k/\epsilon^2)}\,.$
    Step $2$ is repeated at most $O(1)$ times. The first part can be checked in time $n^{O(k/\epsilon^2)}$  and the second depends on the running time $R$ of \cref{algorithm:driving-global-correlation}.
    Finally, step $3$ requires time $O(n).$ All in all the algorithm takes time $n^{O(k/\epsilon^2)}+R\,.$
\end{remark}

A crucial building block of \cref{algorithm:k-xor} is the  subroutine below, used to find a pseudo-distribution in $\cQ_t$ with low global correlation. The idea behind its approach is that, conditioning our pseudo-distribution, we can obtain a new pseudo-distribution "significantly closer" to a product distribution. We analyze the guarantees of \cref{algorithm:driving-global-correlation} in \cref{section:driving-down-global-correlation}.

\begin{algorithmbox}[Driving down global correlation] \label{algorithm:driving-global-correlation}
	\mbox{}\\
	\textbf{Input:} pseudo-distribution $\mu\in \cQ_t$
  
  \noindent
  \textbf{Output:} pseudo-distribution $\bm \mu'\in \cQ_{t-k/\epsilon^2}$

  \noindent
	\textbf{Operations:}
	\begin{enumerate}
    \item[0.] Set $\mu'=\mu\,.$ Sequentially repeat for both  $c\in\Set{k-1, \frac{k-1}{2}}$ on input $\mu'$. 
    \begin{enumerate}
        \item Pick uniformly at random  $\bm \alpha_{1},\ldots,\bm \alpha_{C/\epsilon^2}\in [n]^{c}$, for a large enough constant $C$.
    \item Sequentially set 
    \begin{align*}
        \mathbf{x}^{\bm \alpha_{i}} =
        \begin{cases}
            1\,, &\textnormal{ with prob. } \frac{1+\tilde{\E}_\mu\Brac{x^{\bm \alpha_{i}}\given \mathbf{x}^{\bm \alpha_{1}},\ldots,\mathbf{x}^{ \bm \alpha_{i-1}} }}{2}\\
            -1\,,&\textnormal{ otherwise.}
        \end{cases}
    \end{align*}
    \item For $i\in [C/\epsilon^2]$, let $\bm \mu_i$ be the pseudo-distribution obtained from $\bm \mu'$ conditioning on the sampled values of $\mathbf{x}^{\bm \alpha_{1}},\ldots,\mathbf{x}^{ \bm \alpha_{i-1}}$.
    \item Find $\bm \mu'$ among $\set{\bm\mu_i}_{i\in[C/\eps^2]}$ minimizing
    \begin{align*}
        \E_{\bm \alpha, \bm \alpha'\in [n]^{\mathbf{c}}} \Brac{\Cov{\bm \mu'}{x^{\bm \alpha}, x^{\bm \alpha'}}^2}\,.
    \end{align*}
    \end{enumerate}
    \item Return $\bm \mu'\,.$
  \end{enumerate}
\end{algorithmbox}

\begin{remark}[Running time]
    It suffices to consider steps $(a)$-$(d)$.
    The first two steps of the algorithm require time $O(n^{O(k)}C/\eps^2)$. In the third step we can compute each $\bm \mu_i$ in time $n^{O(k/\epsilon^2)}$ and there are $O(C/\epsilon^2)$ of them. The last step also takes time $n^{O(k/\epsilon^2)}$.
\end{remark}

\subsection{Rounding with low local correlation}\label{section:local-correlation-rounding}

We show here that, given a pseudo-distribution with low local correlation, the rounding step provides nearly optimal guarantees.

\begin{lemma}[Low local correlation rounding]\label{lemma:local-correlation-rounding}
Consider the settings of \cref{theorem:algorithm-technical-xor}. Let $t\geq 2k$ and $\mu\in \cQ_t$.
Let $\cI$ be a semi-random $k$-XOR instance with parameters $n,p$ as in \cref{definition:semi-random-xor}. Let $T$ be the associated tensor. Suppose $\LC{T}{\mu}\leq \epsilon^2\,.$ Then the last step in
\cref{algorithm:k-xor} outputs an assignment $\hat{\mathbf{x}}$ satisfying
\begin{align*}
    \valI{\cI}{\hat{\mathbf{x}}}\geq \optI{\cI}-O(\epsilon)\,,
\end{align*}
with probability at least $0.99$.
\begin{proof}
By Markov's inequality, it will suffices to show that
\begin{align*}
    \E_{(\bm \alpha, \bm \alpha', \bm \ell)\sim T}\sigma(\bm \alpha, \bm \alpha', \bm \ell)\Paren{\tilde{\E}x^{(\bm \alpha, \bm \alpha', \bm \ell)}- \tilde{\E}x^{\bm \alpha}\tilde{\E}x^{\bm \alpha'}\tilde{\E}x^{\bm \ell}}\leq O(\epsilon)\,.
\end{align*}
So applying \cref{fact:cauchy-schwarz}, since $t\geq 2k$ and for any $0\leq c\leq 1$ we have $c^2\leq c\,,$
\begin{align*}
    \E_{(\bm \alpha, \bm \alpha', \bm \ell)\sim T}&\sigma(\bm \alpha, \bm \alpha', \bm \ell)\Paren{\tilde{\E}x^{(\bm \alpha, \bm \alpha', \bm \ell)}- \tilde{\E}x^{\bm \alpha}\tilde{\E}x^{\bm \alpha'}\tilde{\E}x^{\bm \ell}}\\
    &= \tilde{\E}\E_{\bm \ell\sim  D(T)} \E_{(\bm \alpha, \bm \alpha')\sim T_{\bm \ell}}\sigma(\bm \alpha, \bm \alpha', \bm \ell)\Paren{x^{(\bm \alpha, \bm \alpha', \bm \ell)}- \tilde{\E}x^{\bm \alpha}\tilde{\E}x^{\bm \alpha'}\tilde{\E}x^{\bm \ell}}\\
    &= \tilde{\E}\E_{\bm \ell\sim  D(T)} \E_{(\bm \alpha, \bm \alpha')\sim T_{\bm \ell}}\sigma(\bm \alpha, \bm \alpha', \bm \ell)x^{\bm \ell}\Paren{x^{(\bm \alpha, \bm \alpha')}- \tilde{\E}x^{\bm \alpha}\tilde{\E}x^{\bm \alpha'}}\\
    &\leq \sqrt{ \tilde{\E}\E_{\bm \ell\sim  D(T)} \Paren{\E_{(\bm \alpha, \bm \alpha')\sim T_{\bm \ell}}\sigma(\bm \alpha, \bm \alpha', \bm \ell)x^{\bm \ell}}^2}\sqrt{ \tilde{\E}\E_{\bm \ell\sim [n]}\Paren{\E_{(\bm \alpha, \bm \alpha')\sim T_{\bm \ell}}x^{(\bm \alpha, \bm \alpha')}- \tilde{\E}x^{\bm \alpha}\tilde{\E}x^{\bm \alpha'} }^2}\\
    &\leq \sqrt{ \tilde{\E}\E_{\bm \ell\sim  D(T)}\Paren{\E_{(\bm \alpha, \bm \alpha')\sim T_{\bm \ell}}x^{(\bm \alpha, \bm \alpha')}- \tilde{\E}x^{\bm \alpha}\tilde{\E}x^{\bm \alpha'} }^2}\\
    &= \LC{T}{\mu}^{1/2}\,.
\end{align*}
\end{proof}
\end{lemma}

\subsection{Driving down global correlation}\label{section:driving-down-global-correlation}

Here we show that, via \cref{algorithm:driving-global-correlation}, we can always efficiently obtain a pseudo-distribution in $\cQ_t$ with low global correlation. 
Concretely, we prove the following statement.

\begin{lemma}[Driving down global correlation]\label{lemma:driving-down-global-correlation}
Consider the settings of \cref{theorem:algorithm-technical-xor}. Let $t\geq C\cdot k/\epsilon^2$, for a large enough constant $C.$ Let $\mu\in \cQ_t$ be the pseudo-distribution in input for Step 2 of \cref{algorithm:k-xor} and let $\bm \mu'\in \cQ_{k}$ be its output. Then with probability at least $0.998$, it holds that $\GC{\bm \mu'}\leq \epsilon^2\,.$
\end{lemma}

We need an intermediate observation to prove the lemma: a simple, yet crucial, statement  about the covariance of (pseudo)distributions after conditioning \cite{Barak2011rounding, raghavendra2012approximating, manurangsi2016birthday}.

\begin{lemma}[See \cite{tselil-notes}]\label{lemma:low-correlation-after-conditioning}
    Let $0\leq q < t-2$ and let $0< c < (t-q)/2\,.$
    Let $\mu$ be a pseudodistribution in $\cQ_t$.
    %Let $\nu$ be a distribution over $\mathbf{x}_1,\ldots, \mathbf{x}_n$ be random variables in $\Set{\pm 1}\,.$ 
    Let $\mathbf{i}_1,\ldots,\mathbf{i}_\ell$ be indices sampled uniformly at random from $[n]$ without replacement. Suppose we sequentially set, for all $q\leq \ell$,
    \begin{align*}
        \mathbf{x}_{\mathbf{i}_q} = 
        \begin{cases}
            1\,, & \textnormal{ with prob. } \frac{1 + \tilde{\E}_\mu\Brac{\mathbf{x}_{\mathbf{i}_q}\given \mathbf{x}_{\mathbf{i}_1},\ldots,\mathbf{x}_{\mathbf{i}_{q-1}}}}{2}\\
            -1\,, &\textnormal{ otherwise.}
        \end{cases}
    \end{align*}
    Then there exists some $q\leq \ell$ such that
    \begin{align*}
        \E_{\mathbf{x}_{\mathbf{i}_1},\ldots, \mathbf{x}_{\mathbf{i}_q}}\E_{\bm \alpha,\bm \alpha' \in [n]^c} \Paren{
        \tilde{\E}_\mu \Brac{ \mathbf{x}^{\bm \alpha} \mathbf{x}^{\bm \alpha'} - \tilde{\E} \mathbf{x}^{\bm \alpha} \tilde{\E} \mathbf{x}^{\bm \alpha'} \given \mathbf{x}_{\mathbf{i}_1},\ldots\mathbf{x}_{\mathbf{i}_q}}^2}\leq \frac{2\log 2}{\ell}\,.
    \end{align*}
\end{lemma}

%It is important to notice that, since for any multi-index $\alpha$ over $[n]$, the monomial $x^{\alpha}$ takes values in $\Set{\pm 1}$, the statement holds for multi-indices as well.
We can use this result to argue that \cref{algorithm:driving-global-correlation} decreases global correlation. Thus proving \cref{lemma:driving-down-global-correlation}.

\begin{proof}[Proof of \cref{lemma:driving-down-global-correlation}]
Consider without loss of generality the  iteration  of \cref{algorithm:driving-global-correlation} with $c=k-1$. The argument for $c=\frac{k-1}{2}$ is analogous.
Let $\mu\in \cQ_t$ be the pseudo-distribution in input and let $\bm \mu'\in \cQ_{t-k/\epsilon^2}$ be its output.
It suffices to show that for a large enough constant $C> 1$, chosen by \cref{algorithm:driving-global-correlation}, with probability at least $0.999$, it holds:
\begin{align*}
    \E_{\bm \alpha, \bm \alpha'\sim [n]^{k-1}} \Brac{\Cov{\bm \mu'}{x^{\bm \alpha}, x^{\bm \alpha'}}^2}&\leq \frac{1}{10\epsilon^2}\,.
\end{align*}
This follows immediately by \cref{lemma:low-correlation-after-conditioning} and Markov's inequality.% hence $\bm \mu'$ satisfies the first inequality with probability at least $0.999$. 
By union bound over the two iterations of the algorithm, the result follows.
\end{proof}

\subsection{From local correlation to global correlation}\label{section:local-to-global}
In this section we obtain \cref{theorem:algorithm-technical-xor}.
We start introducing some additional notation.
Let $q\in \N$, %Let $\cS_{q}\subseteq [n]^q\times [n]^q$ be the set of pairs of disjoint multi-linear multi-indices in $[n]^q$.
let $\mu\in \cQ_t$ for some $t\geq 2q$. For each $\alpha, \alpha'\in [n]^q$, let $y^{(\alpha, \alpha')}= x^{(\alpha, \alpha')}-\tilde{\E}_\mu x^{\alpha}\tilde{\E}_\mu x^{\alpha'}$. Furthermore, recall that we may specify any semi-random instance $\cI$ by the pair $(\sigma, \mathbf{T}')$ where $\sigma:[n]^k\rightarrow \Set{\pm 1}$ and $\mathbf{T}'$ is a random symmetric tensor in $\R^{n^{\otimes q}}$, as specified in \cref{definition:semi-random-xor}. 
Then we may write 
\begin{align}
    \LC{\mathbf{T}'}{\mu} =& \E_{\bm \ell \in  D(\mathbf{T}')}\E_{\substack{(\bm \alpha, \bm \alpha')\sim \mathbf{T}'_{\bm \ell}\\
    (\bm \beta, \bm \beta')\sim \mathbf{T}'_{\bm \ell}}}\tilde{\E}\Paren{x^{(\bm \alpha, \bm \alpha',\bm \beta, \bm \beta')} - 2x^{(\bm \alpha, \bm \alpha')}\tilde{\E}x^{\bm \beta}\tilde{\E}x^{\bm \beta'} + \tilde{\E}x^{\bm \alpha}\tilde{\E}x^{\bm \alpha'}\tilde{\E}x^{\bm \beta}\tilde{\E}x^{\bm \beta'}}\nonumber\\
    = & \E_{\ell\sim D(\mathbf{T}')}\tilde{\E}\Paren{\E_{(\bm \alpha, \bm \alpha')\sim \mathbf{T}'_\ell}y^{(\bm \alpha,\bm \alpha')}}^2 \nonumber\\
    =&\E_{\ell\sim D(\mathbf{T}')}\E_{\substack{(\bm \alpha, \bm \alpha')\sim \mathbf{T}'_{\bm \ell}\\(\bm \beta, \bm \beta')\sim \mathbf{T}'_{\bm \ell}}}\tilde{\E}y^{(\bm \alpha,\bm \alpha')}y^{(\bm\beta,\bm\beta')}\,.\label{eq:local-correlation-rewriting}
\end{align}
Now, \cref{eq:local-correlation-rewriting} can be rewritten as the matrix product $\iprod{\mathbf{A}, X}$ where $X\in \R^{n^{k-1}\times n^{k-1}}$ is the positive semidefinite matrix with entries $X_{(\alpha, \beta),(\alpha',\beta')}:=\tilde{\E}y^{(\alpha,\alpha')}y^{(\beta,\beta')}$ and 
$\mathbf{A}$ is a random matrix with entries
\begin{align}
    \mathbf{A}_{(\alpha,\beta), (\alpha',\beta')} \propto \sum_\ell \mathbf{T}'_{(\alpha,\alpha',\ell)}\mathbf{T}'_{(\beta,\beta',\ell)}\,.\label{eq:algorithm-xor-A}
\end{align}
Similarly, we may rewrite
\begin{align*}
    \E_{\bm \ell \sim [n]}\E_{\substack{(\bm \alpha, \bm \alpha')\sim [n]^{k-1}\\
    (\bm \beta, \bm \beta')\sim [n]^{k-1}}}\tilde{\E}\Paren{x^{(\bm \alpha, \bm \alpha',\bm \beta, \bm \beta')} - 2x^{(\bm \alpha, \bm \alpha')}\tilde{\E}x^{\bm \beta}\tilde{\E}x^{\bm \beta'} + \tilde{\E}x^{\bm \alpha}\tilde{\E}x^{\bm \alpha'}\tilde{\E}x^{\bm \beta}\tilde{\E}x^{\bm \beta'}} = \iprod{\bar{J}, X}\,,
\end{align*}
where $\bar{J}$ is the normalization of  the all-ones matrix $J$.
We can now use this notation to prove the following two statements, which will allow us to relate local and global correlation.

\begin{lemma}\label{lemma:local-implies-global}
Consider the settings of \cref{theorem:algorithm-technical-xor}.
Let $X\in \R^{n^{k-1}\times n^{k-1}}$ be a positive semi-definite matrix satisfying $\Norm{X}_{\textnormal{max}}\leq O(1)\,.$ For a semi-random instance $\cI$ as in \cref{definition:semi-random-xor}, let $\mathbf{A}$ be the associate matrix as defined in \cref{eq:algorithm-xor-A}. Then, with probability at least $0.999$
\begin{align*}
    \iprod{\mathbf{A}-\bar{J}, X}\leq O(\epsilon^2)\,.
\end{align*}
    \begin{proof}
        Recall, for any matrix $M$ we used the notation $\Norm{M}_{\textnormal{Gr}}=\max\Set{\iprod{M, X}\given X\sge 0, X_{ii}\leq 1\,, \forall i \in [n]}\,.$ Observe that by Cauchy-Schwarz and Grothendieck's inequality
        \begin{align*}
             \iprod{\mathbf{A}-\bar{J}, X}\leq \Norm{X}_{\textnormal{max}}\cdot \Norm{\mathbf{A}-\bar{J}}_{\textnormal{Gr}}
             \leq O(\Normio{\mathbf{A}-\bar{J}})\,.
        \end{align*}
        Thus it remains to bound $\Normio{\mathbf{A}-\bar{J}}$.
        Define $\mathbf{A}^*$ to the be matrix with entries
        \begin{align*}
            \mathbf{A}^*_{(\alpha,\beta),(\alpha',\beta')} = 
            \begin{cases}
                (\mathbf{A}-\bar{J})_{(\alpha,\beta),(\alpha,\beta')}& \text{ if }\card{S(\alpha, \alpha')\cap S(\beta,\beta')}=\emptyset\,\\
                0& \text{ otherwise.}
            \end{cases}
        \end{align*}
        By the triangle inequality
        \begin{align*}
            \Normio{\mathbf{A}-\bar{J}}&\leq \Normio{\mathbf{A}^*} + \Normio{\mathbf{A} -\bar{J}-\mathbf{A}^*}\,.
        \end{align*}
        Now 
        \begin{align*}
            \Normio{\mathbf{A} -\bar{J}-\mathbf{A}^*}&\leq \sum_{\alpha,\beta,\alpha',\beta} \Abs{\mathbf{A} -\bar{J}-\mathbf{A}^*}_{(\alpha,\beta),(\alpha',\beta')}\\
            &\leq \sum_{\substack{\alpha,\beta,\alpha',\beta\\\textnormal{s.t. } S(\alpha,\alpha')\cap S(\beta,\beta')\neq \emptyset}} \Abs{\mathbf{A}_{(\alpha,\beta),(\alpha',\beta')}}+\Abs{\bar{J}_{(\alpha,\beta),(\alpha',\beta')}}\,.
        \end{align*}
        The first term can be shown to be $o(1)$ repeating the argument of \cref{lemma:cut-norm-residual-A}. The second term is a sum of at most $n^{2k-3}$ elements of value at most $O(n^{-2k+2})$ and thus also $o(1)$.
        It remains to bound $\Normio{\mathbf{A}^*}$. Here $\mathbf{A}^*$ satisfies the premises of \cref{lemma:cut-norm-A} and thus for $p\geq n^{-k/2}/\epsilon^2$ we immediately get
        \begin{align*}
            \Normio{\mathbf{A} -\bar{J}-\mathbf{A}^*}\leq O(\epsilon^2)+o(1)
        \end{align*}
        with probability at least $0.999$ over the draw of $\mathbf{T}$.
    \end{proof}
\end{lemma}

Next we show that $\iprod{\bar{J}, X}$ is a lower bound to the global correlation.

\begin{lemma}\label{lemma:lower-bound-global-correlation}
    For a semi-random instance $\cI$ as in \cref{definition:semi-random-xor}, let $\mathbf{A}$ be the associate matrix as defined in \cref{eq:algorithm-xor-A}. Let $\mu$ be a pseudo-distribution of degree at least $2k$, and let $X\in \R^{n^{k-1}\times n^{k-1}}$ be the positive semi-definite matrix satisfying 
    \begin{align*}
        \LC{\mathbf{T}}{\mu} = \iprod{\mathbf{A}, X}\,.
    \end{align*}
    Then 
    \begin{align*}
        \iprod{\bar{J}, X} = \E_{\bm \ell \sim [n]}\E_{\substack{(\bm \alpha, \bm \alpha')\sim [n]^{k-1}\\
    (\bm \beta, \bm \beta')\sim [n]^{k-1}}}\tilde{\E}\Paren{x^{(\bm \alpha, \bm \alpha',\bm \beta, \bm \beta')} - 2x^{(\bm \alpha, \bm \alpha')}\tilde{\E}x^{\bm \beta}\tilde{\E}x^{\bm \beta'} + \tilde{\E}x^{\bm \alpha}\tilde{\E}x^{\bm \alpha'}\tilde{\E}x^{\bm \beta}\tilde{\E}x^{\bm \beta'}}\leq \GC{\mu}\,.
    \end{align*}
    \begin{proof}
        For fixed $\ell, \alpha,\alpha',\beta,\beta'$ we may rewrite
        \begin{align*}
        \tilde{\E}&\Paren{x^{(\alpha, \alpha',\beta, \beta')} - 2x^{(\alpha, \alpha')}\tilde{\E}x^{\beta}\tilde{\E}x^{\beta'} + \tilde{\E}x^{\alpha}\tilde{\E}x^{\alpha'}\tilde{\E}x^{\beta}\tilde{\E}x^{\beta'}} \\
        &\leq\tilde{\E} \Paren{x^{(\alpha,\alpha',\beta,\beta')}-x^{(\alpha,\alpha')}\tilde{\E}x^{\beta}\tilde{\E}x^{\beta}} - \Cov{\mu}{x^\alpha,x^{\alpha'}}\\
        &\leq \tilde{\E} \Paren{x^{(\alpha,\alpha',\beta,\beta')}-x^{(\alpha,\alpha')}\tilde{\E}x^{\beta}\tilde{\E}x^{\beta}} + \Abs{\Cov{\mu}{x^\alpha,x^{\alpha'}}}\,.
        \end{align*}    
        Furthermore 
        \begin{align*}
            \tilde{\E} &\Paren{x^{(\alpha,\alpha',\beta,\beta')}-x^{(\alpha,\alpha')}\tilde{\E}x^{\beta}\tilde{\E}x^{\beta}} \\
            &= \tilde{\E} \Paren{x^{(\alpha,\alpha',\beta,\beta')}-x^{(\alpha,\alpha')}\tilde{\E}x^{\beta}\tilde{\E}x^{\beta}} + \tilde{\E}x^{(\alpha,\alpha')}\tilde{\E}x^{(\beta,\beta')} - \tilde{\E}x^{(\alpha,\alpha')}\tilde{\E}x^{(\beta,\beta')}\\
            &=\Cov{\mu}{x^{(\alpha,\alpha'),x^{(\beta,\beta')}}} - \tilde{\E}x^{(\alpha,\alpha')}\Cov{\mu}{x^{\beta},x^{\beta'}}\\
            &\leq\Abs{\Cov{\mu}{x^{(\alpha,\alpha'),x^{(\beta,\beta')}}}} + \Abs{\Cov{\mu}{x^{\beta},x^{\beta'}}}\,.
        \end{align*}
        The result follows.
    \end{proof}
\end{lemma}

We are finally ready to prove the theorem.

\begin{proof}[Proof of \cref{theorem:algorithm-technical-xor}]
Let $\bm \mu$ be the pseudo-distribution used by \cref{algorithm:k-xor} in the last step.
By \cref{lemma:driving-down-global-correlation}, with probability at least $0.999$, it satisfies $\GC{\bm \mu}\leq \epsilon^2$. Combining \cref{lemma:local-implies-global} and \cref{lemma:lower-bound-global-correlation} it follows that $\LC{\cI}{\bm \mu}\leq O(\epsilon^2)$. We obtain the desired result applying \cref{lemma:local-correlation-rounding}.
\end{proof}

\section{Algorithm for CSPs with adversarial signs patterns}

We \textit{sketch} here how to extend \cref{theorem:algorithm-technical-xor} to arbitrary predicates --with adversarial sign patterns--  on $k$ Boolean variables.
We start by introducing the model.

\begin{definition}\label{definition:semi-random-csp}
Let $P:\Set{-1,1}^k\rightarrow \Set{0,1}$. 
A semi-random $k$-CSP instance $\cI$ with $n$ variables and $m:=p\cdot 2^k\cdot n^k(1\pm o(1))$ constraints can be generated as follows. 
\begin{enumerate}[(i)]
    \item Pick independently with probability $p$ each  pair $(\mathbf{c}', \bm \alpha)$ where $\mathbf{c}'$ is a random negation pattern from $\Set{\pm 1}^k$ and $\bm \alpha$ is a multi-index from $[n]^k$,
    \item Given the $m$ pairs, replace each such $\mathbf{c}'$ with an arbitrary, possibly  adversarially chosen, negation pattern $c$.
    \item For each pair $(c, \bm \alpha)$ add the constraint $P(c\circ x^{\bm \alpha})=1$ to $\cI$.
\end{enumerate}
\end{definition}

We prove the following theorem, which implies \cref{theorem:main-algorithm-csp}.
\begin{theorem}\label{theorem:algorithm-technical-csp}
Let $n\,,k$ be positive integers, $\epsilon > 0\,, n$ and $n^{-k/2}/\epsilon^2 < p(n):=p < 1$.
Let $P:\Set{\pm 1}^k\rightarrow \Set{0,1}$ be a predicate.
Let $\cI$ be a  $CSP(P)$ instance with parameters $n,p$ as in \cref{definition:semi-random-csp}. There exists a randomized algorithm (\cref{algorithm:k-csp}), running in time $n^{O(k^2/\epsilon^2)}$, that returns an assignment $\hat{\mathbf{x}}$ satisfying
\begin{align*}
    \valI{\cI}{\hat{\mathbf{x}}}\geq \optI{\cI}-O(\epsilon)\,,
\end{align*}
with probability at least $0.99$.
\end{theorem}

%We denote by $\bm \cI'$ the random k-CSP instance obtained setting $c=\mathbf{c}'$ in step (ii) for all negation patterns.
Given a predicate $P:\Set{\pm 1}^k\rightarrow {0,1}$, for each $c\in \Set{\pm 1}^k$ and $\alpha \in [n]^k$ we may rewrite
\begin{align*}
    P(c\circ \alpha) = \sum_{\beta \subseteq \alpha} q_\beta \cdot \sigma(\beta) \cdot x^\beta\,,
\end{align*}
where $q_\beta$ is a constant coefficient and $\sigma(\beta)\in \Set{\pm 1}\,.$
For an instance $\cI$, and $d\leq k$, %let $T(d, \cI)\in \R^{(2n)^{\otimes d}}$ be tensor with entries $T_{(c,\beta)}(d,\cI)=1$ if and only if there exists a pair $(c, \alpha)\in \cI$ such that $\beta \subseteq \alpha$ and the corresponding coefficient $q_\beta$ in $P(c\circ \alpha)$ is non-zero.
%Similarly, 
let $M(d, \cI)\in \R^{n^{\otimes d}}$ be the tensor with entries $M_{\beta}(d,\cI)=q$ where $q$ is the \textit{number} of pairs such that $(c, \alpha)\in \cI$ with $\beta \subseteq \alpha$ and the corresponding coefficient $q_\beta$ in $P(c\circ \alpha)$ is non-zero.
Furthermore, let $D(\cI, d)$ be the distribution over indices in $[n]$ such that the probability of $\ell\in [n]$ is proportional to the fraction of pairs $(c, \alpha)\in \cI$ with $\ell\subseteq \alpha.$ Then we write $T(d, \cI)_\ell$ for the set of multi-indeces $(\alpha, \ell)$ corresponding to non-zero in $T(d, \cI)\,.$
We can now introduce new notions of local and global correlation. Let $\mu\in \cQ_t$ for some $t\geq 2k$. For even $d\leq k$ let 
\begin{align*}
    \LC{\cI}{\mu, d}&:=\Paren{\E_{ (\bm \alpha, \bm \alpha') \sim T(d, \cI)} \Abs{\Cov{\mu}{x^{\bm \alpha}, x^{\bm \alpha'}}}}^2\\
    \GC{\mu, d}&:=\Paren{\E_{\bm \alpha, \bm \alpha' \sim[n]^{d/2}} \Abs{\Cov{\mu}{x^{\bm \alpha}, x^{\bm \alpha'}}}}^2\,.
\end{align*}
For odd $d\leq k$ let
\begin{align*}
    \LC{\cI}{\mu, d}&:=\E_{\bm \ell \sim D(\cI, d)} \E_{\substack{(\bm \alpha, \bm \alpha')\sim T(d, \cI)_\ell\\(\bm \beta, \bm \beta')\sim T(d, \cI)_\ell}} \tilde{\E}\Paren{x^{(\bm \alpha, \bm \alpha', \bm \beta, \bm \beta')}-2x^{(\bm \alpha, \bm \alpha')}\tilde{\E}x^{(\bm \beta, \bm \beta')} + \tilde{\E}x^{\bm \alpha}\tilde{\E}x^{\bm \alpha'}\tilde{\E}x^{\bm \beta}\tilde{\E}x^{\bm \beta'}}\\
    \GC{\mu d}&:=\E_{\bm \ell \in [n]}\E_{\substack{(\bm \alpha, \bm \alpha') \sim[n]^{d-1}\\
    (\bm \beta, \bm \beta')\sim [n]^{d-1}}} \Abs{\Cov{\mu}{x^{\bm \alpha}, x^{\bm \alpha'}}}+\Abs{\Cov{\mu}{x^{\bm \beta}, x^{\bm \beta'}}} +\Abs{\Cov{\mu}{x^{(\bm \alpha, \bm \alpha')}, x^{(\bm \beta, \bm \beta')}}}\,.
\end{align*}

Both the algorithm and the proof structure closely resemble the ones used for \cref{theorem:algorithm-technical-xor}, so we only discuss the steps that differ. We start by presenting the algorithm.

\begin{algorithmbox}[Algorithm for semi-random $k$-XOR] \label{algorithm:k-csp}
	\mbox{}\\
	\textbf{Input:} A $k$-CSP instance $\cI$ as in \cref{definition:semi-random-csp}, $ \epsilon$.
  
  \noindent
  \textbf{Output:} assignment $\hat{\mathbf{x}}\in\Set{\pm 1}^n$

  \noindent
	\textbf{Operations:}
	\begin{enumerate}
    \item Find a pseudo-distribution $\mu \in \cQ_t$  maximizing $\E_{(c,\bm \alpha)\sim \bm\cI}\tilde{\E}_\mu P(c\circ\bm \alpha)$, for large enough $t\geq \Omega(k/\epsilon^2)\,.$
    \item If $\GC{\mu}> \epsilon^2$, let $\bm \mu'$ be the pseudo-distribution returned by \cref{algorithm:driving-global-correlation-csp} on input $\mu$.
    \item For each $i\in [n]$, set $\hat{\mathbf{x}}_i = 1$ with probability $\frac{1+\tilde{\E}_{\bm \mu'} x_i}{2}$ and $-1$ otherwise. Return $\hat{\mathbf{x}}\,.$
  \end{enumerate}
\end{algorithmbox}

\begin{algorithmbox}[Driving down global correlation] \label{algorithm:driving-global-correlation-csp}
	\mbox{}\\
	\textbf{Input:} pseudo-distribution $\mu\in \cQ_t$
  
  \noindent
  \textbf{Output:} pseudo-distribution $\bm \mu'\in \cQ_{t-k/\epsilon^2}$

  \noindent
	\textbf{Operations:}
	\begin{enumerate}
    \item[0.] Set $\mu'=\mu\,.$ Sequentially repeat for  $d\in [k]$ on input $\mu'$. 
    \begin{enumerate}
        \item Let $C$ be a large enough constant. Let $q=C\cdot k/\epsilon^2$. Pick uniformly at random  $\bm \alpha_{1},\ldots,\bm \alpha_{q}\in [n]^{d}$.
    \item Sequentially set 
    \begin{align*}
        \mathbf{x}^{\bm \alpha_{i}} =
        \begin{cases}
            1\,, &\textnormal{ with prob. } \frac{1+\tilde{\E}_\mu\Brac{x^{\bm \alpha_{i}}\given \mathbf{x}^{\bm \alpha_{1}},\ldots,\mathbf{x}^{ \bm \alpha_{i-1}} }}{2}\\
            -1\,,&\textnormal{ otherwise.}
        \end{cases}
    \end{align*}
    \item For each $i\in [q]$, let $\bm \mu_i$ be the pseudo-distribution obtained from $\bm \mu'$ conditioning on the sampled values of $\mathbf{x}^{\bm \alpha_{1}},\ldots,\mathbf{x}^{ \bm \alpha_{i-1}}$.
    \item Find $\bm \mu'$ among $\Set{\bm \mu_i}_{i\in [q]}$ minimizing
    \begin{align*}
        \E_{\bm \alpha, \bm \alpha'\in [n]^{d}} \Brac{\Cov{\bm \mu'}{x^{\bm \alpha}, x^{\bm \alpha'}}^2}\,.
    \end{align*}
    \end{enumerate}
    \item Return $\bm \mu'\,.$
  \end{enumerate}
\end{algorithmbox}
\begin{remark}[Running time]
    The running time of steps (a)-(d) is at most $O(n^{k^2/\epsilon^2})$. The steps are called $k$ times so overall the running time is $O(n^{k^2/\epsilon^2})$.
\end{remark}

\paragraph{Rounding with low local correlation}
The next result is the CSP version of \cref{lemma:local-correlation-rounding}.

\begin{lemma}[Low local correlation rounding]\label{lemma:local-correlation-rounding-csp}
Consider the settings of \cref{theorem:algorithm-technical-csp}. Let $t\geq 2k$ and $\mu\in \cQ_t$.
Let $\cI$ be a semi-random $k$-CSP instance with parameters $n,p$ as in \cref{definition:semi-random-csp}. Let $T$ be the associated tensor. 
Suppose $\sum_{d\leq k}\LC{\cI}{\mu, d}\leq \epsilon^2$. Then the last step in
\cref{algorithm:k-csp} outputs an assignment $\hat{\mathbf{x}}$ satisfying
\begin{align*}
    \valI{\cI}{\hat{\mathbf{x}}}\geq \optI{\cI}-O(\epsilon)\,,
\end{align*}
with probability at least $0.99$.
\begin{proof}
%For a multi-linear monomial in indeterminate $x$, $s(x):=x^{\alpha}$, let $s(\tilde{\E}x):= \prod_{i \in \alpha} \tilde{\E}x_i\,.$ Similarly, 
By Markov's inequality it suffices to show that
\begin{align*}
    \E_{(\mathbf{c}, \bm \alpha)\sim \cI} \Abs{\tilde{\E}_\mu P\Paren{\mathbf{c}\circ \bm \alpha} - P\Paren{\mathbf{c}\circ \bm \alpha, \mu}} \leq O\Paren{\sum_{d \leq k} \sqrt{\LC{\cI}{\mu, d}}}\leq O(\epsilon)\,.
\end{align*}
To do so we may rewrite
\begin{align*}
     \E_{(\mathbf{c}, \bm \alpha)\sim \cI} \Abs{\tilde{\E}_\mu P\Paren{\mathbf{c}\circ \bm \alpha} - P\Paren{\mathbf{c}\circ \bm \alpha, \mu}} \leq  \E_{(\mathbf{c}, \bm \alpha)\sim \cI}\sum_{\beta\subseteq \bm \alpha} O\Paren{1}\Abs{\tilde{\E}x^{\beta}-\prod_{b\in \beta}\tilde{\E}x_b}\,.
\end{align*}
Now, using a derivation as in \cref{lemma:local-correlation-rounding} the result follows.
%Given a predicate $P:\Set{\pm 1}^k\rightarrow \Set{0,1}$ and $\alpha\in [n]^k$ we may write $P(c\circ \alpha, \mu)$ to denote the value obtained by replacing each $x_i$ in each monomial in $P(c \circ \alpha)$ with its pseudo-expectation over $\mu$.
%Then, by Markov's inequality, the result follows observing that
%\begin{align*}
%    \E_{(\mathbf{c}, \bm \alpha)\sim \cI} \Abs{\tilde{\E}_\mu P\Paren{\mathbf{c}\circ \bm \alpha} - P\Paren{\mathbf{c}\circ \bm \alpha, \mu}} \leq O\Paren{\sum_{d < k} \LC{\cI, d}{\mu} + \sqrt{\LC{\cI, k}{\mu}}}\leq O(\epsilon)\,,
%\end{align*}
%where for the degree-$k$ terms we used the derivation in \cref{lemma:local-correlation-rounding}.
\end{proof}
\end{lemma}

\paragraph{Driving down global correlation}

Next we analyze the guarantees of \cref{algorithm:driving-global-correlation-csp} and obtain a statement resembling \cref{lemma:driving-down-global-correlation}.

\begin{lemma}[Driving down global correlation]\label{lemma:driving-down-global-correlation-csp}
Consider the settings of \cref{theorem:algorithm-technical-csp}. Let $t\geq C\cdot k^2/\epsilon^2$, for a large enough constant $C$. Let $\mu\in \cQ_t$ be the pseudo-distribution in input for Step 2 of \cref{algorithm:k-csp} and let $\bm \mu'\in \cQ_{k}$ be its output. Then with probability at least $0.998$, it holds that $\sum_{d\leq k}\GC{\bm \mu', d}\leq \epsilon^2\,.$
\begin{proof}
For the iteration with $d=k$, we know by the proof of \cref{lemma:driving-down-global-correlation} that with probability at least $0.999$ it holds:
\begin{align*}
    \E_{\bm \alpha, \bm \alpha'\sim [n]^{k-1}} \Brac{\Cov{\bm \mu'}{x^{\bm \alpha}, x^{\bm \alpha'}}^2}&\leq \frac{1}{10\epsilon^2}\,.
\end{align*}
So consider now the other iterations. Repeating the analysis as in the previous case, we have that in expectation, for each $d< k$
\begin{align*}
    \E_{\bm \alpha, \bm \alpha'\sim [n]^{d}} \Brac{\Cov{\bm \mu'}{x^{\bm \alpha}, x^{\bm \alpha'}}^2}&\leq \frac{1}{10000k\epsilon^2}\,.
\end{align*}
By linearity of expectations, applying Markov's inequality we get
\begin{align*}
    \sum_{d\leq  k}\E_{\bm \alpha, \bm \alpha'\sim [n]^{d}} \Brac{\Cov{\bm \mu'}{x^{\bm \alpha}, x^{\bm \alpha'}}^2}&\leq \frac{1}{10\epsilon^2}\,,
\end{align*}
with probability at least $0.998$.
\end{proof}
\end{lemma}

\paragraph{From local correlation to global correlation}
As in the case of k-XOR, for odd $d\leq k$ we may rewrite 
\begin{align*}
    \LC{\cI,d}{\mu}=\iprod{\mathbf{A}_d, X_d}
\end{align*}
for a positive semidefinite matrix $X_d$ and a matrix $\mathbf{A}$ with entries
\begin{align*}
    \mathbf{A}_{(\alpha,\beta), (\alpha',\beta')} \propto \sum_\ell M_{(\alpha,\alpha,\ell)}(d, \bm \cI')\cdot M_{(\beta,\beta',\ell)}(d, \bm \cI')\,.
\end{align*}
Similarly we may write
\begin{align*}
    \E_{\bm \ell \sim [n]} \E_{\substack{(\bm \alpha, \bm \alpha')\sim [n]^{k-1}\\(\bm \beta, \bm \beta')\sim [n]^{k-1}}} \tilde{\E}\Paren{x^{(\bm \alpha, \bm \alpha', \bm \beta, \bm \beta')}-2x^{(\bm \alpha, \bm \alpha')}\tilde{\E}x^{(\bm \beta, \bm \beta')} + \tilde{\E}x^{\bm \alpha}\tilde{\E}x^{\bm \alpha'}\tilde{\E}x^{\bm \beta}\tilde{\E}x^{\bm \beta'}}=\iprod{\bar{J}, X_d}\,.
\end{align*}
Moreover, by the analysis in \cref{lemma:lower-bound-global-correlation}, we have for all odd $d\leq k$
\begin{align*}
    \iprod{\bar{J}, X_d} \leq \GC{\mu, d}\,.
\end{align*}
It remains to prove an analogue of \cref{lemma:local-implies-global}.

\begin{lemma}\label{lemma:local-implies-global-csp}
Consider the settings of \cref{theorem:algorithm-technical-xor}.
Let $\mu$ be a pseudo-distribution in $\cQ_t$ for some $t\geq 2k\,.$
Suppose $\GC{\mu, d}\leq \epsilon^2$ for all $d\leq k$.
Then, with probability at least $0.998$ over the randomness of the instance $\cI$, for all $d\leq k$,
\begin{align}\label{eq:bounded-local-correlation}
    \LC{\cI}{\mu, d}\leq O(\epsilon^2)\,.
\end{align}
    \begin{proof}
        By \cref{lemma:local-implies-global-csp},  \cref{eq:bounded-local-correlation} holds with probability $0.999$ for $d=k\,.$
        Now, for  odd $d< k$, a similar analysis combined with the bounds of \cref{lemma:csps-rough-matrix-bounds} implies \cref{eq:bounded-local-correlation} with probability $1-o(1)$.
        For even $d$ we instead combine \cref{eq:bounded-local-correlation} with the standard local-to-global correlation result (e.g. see Lemma 4.1 in \cite{Barak2011rounding}).
        Taking a union bound over all $d\leq k$ the result follows.
    \end{proof}
\end{lemma}

Finally, \cref{theorem:algorithm-technical-csp} immediately follows combining \cref{lemma:local-correlation-rounding-csp}, \cref{lemma:driving-down-global-correlation-csp} and \cref{lemma:local-implies-global-csp}.

%%%% BIBLIOGRAPHY
% assumes hyperref
\phantomsection
\addcontentsline{toc}{section}{Bibliography}
\bibliographystyle{amsalpha}
\bibliography{refuting-3xor/bib/custom, refuting-3xor/bib/mathreview,refuting-3xor/bib/scholar}

\newcommand{\etalchar}[1]{$^{#1}$}
\providecommand{\bysame}{\leavevmode\hbox to3em{\hrulefill}\thinspace}
\providecommand{\MR}{\relax\ifhmode\unskip\space\fi MR }
\providecommand{\MRhref}[2]{%
  \href{http://www.ams.org/mathscinet-getitem?mr=#1}{#2}
}
\providecommand{\href}[2]{#2}
\begin{thebibliography}{BBH{\etalchar{+}}12}

\bibitem[AJT19]{Alev2019approximating}
Vedat~Levi Alev, Fernando~Granha Jeronimo, and Madhur Tulsiani,
  \emph{Approximating constraint satisfaction problems on high-dimensional
  expanders}, 2019 IEEE 60th Annual Symposium on Foundations of Computer
  Science (FOCS), IEEE, 2019, pp.~180--201.

\bibitem[AN04]{AlonN04}
Noga Alon and Assaf Naor, \emph{Approximating the cut-norm via grothendieck's
  inequality}, Proceedings of the 36th Annual {ACM} Symposium on Theory of
  Computing, Chicago, IL, USA, June 13-16, 2004, 2004, pp.~72--80.

\bibitem[AOW15]{AllenOW15}
Sarah~R. Allen, Ryan O'Donnell, and David Witmer, \emph{How to refute a random
  {CSP}}, {IEEE} 56th Annual Symposium on Foundations of Computer Science,
  {FOCS} 2015, Berkeley, CA, USA, 17-20 October, 2015, 2015, pp.~689--708.

\bibitem[Bas92]{bass1992ihara}
Hyman Bass, \emph{The ihara-selberg zeta function of a tree lattice},
  International Journal of Mathematics \textbf{3} (1992), no.~06, 717--797.

\bibitem[BBH{\etalchar{+}}12]{Barak2012hypercontractivity}
Boaz Barak, Fernando~GSL Brandao, Aram~W Harrow, Jonathan Kelner, David
  Steurer, and Yuan Zhou, \emph{Hypercontractivity, sum-of-squares proofs, and
  their applications}, Proceedings of the forty-fourth annual ACM symposium on
  Theory of computing, 2012, pp.~307--326.

\bibitem[BLM15]{BordenaveLM15}
Charles Bordenave, Marc Lelarge, and Laurent Massouli{\'{e}},
  \emph{Non-backtracking spectrum of random graphs: Community detection and
  non-regular ramanujan graphs}, {IEEE} 56th Annual Symposium on Foundations of
  Computer Science, {FOCS} 2015, Berkeley, CA, USA, 17-20 October, 2015, 2015,
  pp.~1347--1357.

\bibitem[BRS11]{Barak2011rounding}
Boaz Barak, Prasad Raghavendra, and David Steurer, \emph{Rounding semidefinite
  programming hierarchies via global correlation}, 2011 ieee 52nd annual
  symposium on foundations of computer science, IEEE, 2011, pp.~472--481.

\bibitem[BS16]{BoazNotes}
Boaz Barak and David Steurer, \emph{Proofs, beliefs, and algorithms through the
  lens of sum- of-squares}, Course notes: www.sumofsquares.org (2016).

\bibitem[Cha16]{Chan2016approximation}
Siu~On Chan, \emph{Approximation resistance from pairwise-independent
  subgroups}, Journal of the ACM (JACM) \textbf{63} (2016), no.~3, 1--32.

\bibitem[Fei02]{Feige02}
Uriel Feige, \emph{Relations between average case complexity and approximation
  complexity}, 2002, pp.~534--543.

\bibitem[Fei07]{Feige2007refuting}
Uriel Feige, \emph{Refuting smoothed 3cnf formulas}, 48th Annual IEEE Symposium
  on Foundations of Computer Science (FOCS'07), IEEE, 2007, pp.~407--417.

\bibitem[FG01]{FriedmanG01}
Joel Friedman and Andreas Goerdt, \emph{Recognizing more unsatisfiable random
  3-{SAT} instances efficiently}, Automata, Languages and Programming, 28th
  International Colloquium, {ICALP} 2001, Crete, Greece, July 8-12, 2001,
  Proceedings, Lecture Notes in Computer Science, vol. 2076, Springer, 2001,
  pp.~310--321.

\bibitem[FLP15]{Fotakis2015sub}
Dimitris Fotakis, Michael Lampis, and Vangelis~Th Paschos,
  \emph{Sub-exponential approximation schemes for csps: From dense to almost
  sparse}, arXiv preprint arXiv:1507.04391 (2015).

\bibitem[FM17]{FanM17}
Zhou Fan and Andrea Montanari, \emph{How well do local algorithms solve
  semidefinite programs?}, Proceedings of the 49th Annual {ACM} {SIGACT}
  Symposium on Theory of Computing, {STOC} 2017, Montreal, QC, Canada, June
  19-23, 2017, {ACM}, 2017, pp.~604--614.

\bibitem[FO05]{FeigeO05}
Uriel Feige and Eran Ofek, \emph{Spectral techniques applied to sparse random
  graphs}, Random Struct. Algorithms \textbf{27} (2005), no.~2, 251--275.

\bibitem[GK01]{GoerdtK01}
Andreas Goerdt and Michael Krivelevich, \emph{Efficient recognition of random
  unsatisfiable k-{SAT} instances by spectral methods}, {STACS} 2001, 18th
  Annual Symposium on Theoretical Aspects of Computer Science, Dresden,
  Germany, February 15-17, 2001, Proceedings, Lecture Notes in Computer
  Science, vol. 2010, Springer, 2001, pp.~294--304.

\bibitem[GKM21]{GKM22}
Venkatesan Guruswami, Pravesh~K. Kothari, and Peter Manohar, \emph{Algorithms
  and certificates for boolean {CSP} refutation: "smoothed is no harder than
  random"}, arXiv \textbf{2109.04415} (2021).

\bibitem[GKM22]{Guruswami2022algorithms}
Venkatesan Guruswami, Pravesh~K Kothari, and Peter Manohar, \emph{Algorithms
  and certificates for boolean csp refutation: smoothed is no harder than
  random}, Proceedings of the 54th Annual ACM SIGACT Symposium on Theory of
  Computing, 2022, pp.~678--689.

\bibitem[GLS81]{MR625550-Grotschel81}
M.~Gr\"otschel, L.~Lov\'asz, and A.~Schrijver, \emph{The ellipsoid method and
  its consequences in combinatorial optimization}, Combinatorica \textbf{1}
  (1981), no.~2, 169--197. \MR{625550}

\bibitem[HST06]{horton2006zeta}
Matthew~D Horton, HM~Stark, and Audrey~A Terras, \emph{What are zeta functions
  of graphs and what are they good for?}, Contemporary Mathematics \textbf{415}
  (2006), 173--190.

\bibitem[Kot22]{Kothari-personal}
Pravesh~K. Kothari, Personal communication (2022).

\bibitem[Las01]{MR1846160-Lasserre01}
Jean~B. Lasserre, \emph{New positive semidefinite relaxations for nonconvex
  quadratic programs}, Advances in convex analysis and global optimization
  ({P}ythagorion, 2000), Nonconvex Optim. Appl., vol.~54, Kluwer Acad. Publ.,
  Dordrecht, 2001, pp.~319--331. \MR{1846160}

\bibitem[MR08]{Moshkovitz2008two}
Dana Moshkovitz and Ran Raz, \emph{Two-query pcp with subconstant error},
  Journal of the ACM (JACM) \textbf{57} (2008), no.~5, 1--29.

\bibitem[MR16]{manurangsi2016birthday}
Pasin Manurangsi and Prasad Raghavendra, \emph{A birthday repetition theorem
  and complexity of approximating dense csps}, arXiv preprint arXiv:1607.02986
  (2016).

\bibitem[Nes00]{MR1748764-Nesterov00}
Yurii Nesterov, \emph{Squared functional systems and optimization problems},
  High performance optimization, Appl. Optim., vol.~33, Kluwer Acad. Publ.,
  Dordrecht, 2000, pp.~405--440. \MR{1748764}

\bibitem[O'D17]{o2017sos}
Ryan O'Donnell, \emph{Sos is not obviously automatizable, even approximately},
  8th Innovations in Theoretical Computer Science Conference (ITCS 2017),
  Schloss Dagstuhl-Leibniz-Zentrum fuer Informatik, 2017.

\bibitem[Par00]{parrilo2000structured}
Pablo~A Parrilo, \emph{Structured semidefinite programs and semialgebraic
  geometry methods in robustness and optimization}, Ph.D. thesis, California
  Institute of Technology, 2000.

\bibitem[RRS17]{RaghavendraRS17}
Prasad Raghavendra, Satish Rao, and Tselil Schramm, \emph{Strongly refuting
  random {CSP}s below the spectral threshold}, Proceedings of the 49th Annual
  {ACM} {SIGACT} Symposium on Theory of Computing, {STOC} 2017, Montreal, QC,
  Canada, June 19-23, 2017 (Hamed Hatami, Pierre McKenzie, and Valerie King,
  eds.), {ACM}, 2017, pp.~121--131.

\bibitem[RT12]{raghavendra2012approximating}
Prasad Raghavendra and Ning Tan, \emph{Approximating csps with global
  cardinality constraints using sdp hierarchies}, Proceedings of the
  twenty-third annual ACM-SIAM symposium on Discrete Algorithms, SIAM, 2012,
  pp.~373--387.

\bibitem[Sch22]{tselil-notes}
Tselil Schramm, \emph{The sum-of-squares algorithmic paradigm in statistics},
  Lecture notes (2022).

\bibitem[Sho87]{MR939596-Shor87}
N.~Z. Shor, \emph{Quadratic optimization problems}, Izv. Akad. Nauk SSSR Tekhn.
  Kibernet. (1987), no.~1, 128--139, 222. \MR{939596}

\bibitem[Tro12]{Tropp12}
Joel~A. Tropp, \emph{User-friendly tail bounds for sums of random matrices},
  Found. Comput. Math. \textbf{12} (2012), no.~4, 389--434.

\bibitem[WAM19]{WeinAM19}
Alexander~S. Wein, Ahmed~El Alaoui, and Cristopher Moore, \emph{The {K}ikuchi
  hierarchy and tensor {PCA}}, 60th {IEEE} Annual Symposium on Foundations of
  Computer Science, {FOCS} 2019, Baltimore, Maryland, USA, November 9-12, 2019,
  2019, pp.~1446--1468.

\bibitem[WF09]{Watanabe2009graph}
Yusuke Watanabe and Kenji Fukumizu, \emph{Graph zeta function in the bethe free
  energy and loopy belief propagation}, Advances in Neural Information
  Processing Systems \textbf{22} (2009).

\end{thebibliography}

%%%% APPENDIX
\clearpage
\appendix
%Include appendixes here
\section{Missing proofs}\label{section:deferred-proofs}

This section contains proofs deferred troughout the paper.

\subsection{Deferred proofs of \cref{section:warm-up}}
We upper bound the number of $100(\log\log  n)$ tangle free canonical paths for sparse graphs.

\begin{lemma}[Enumeration of canonical paths, restatement of \cref{lemma:enumeration-binary-canonical-paths}]
	Let $\cW^{2q,z}(v,e)$ be the set of canonical paths with $v$ vertices and $e$ distinct edges. We have
	\begin{align*}
		\Card{\cW^{2q,z}(v,e)}\leq \Paren{2^{2t}z}^{2qt}\cdot (2zq)^{6tq\cdot (e-v+1)}\,.
	\end{align*}
	\begin{proof}	%\Tnote{Expand if we have time}
		Our proof is closely related to  Lemma $17$ in \cite{BordenaveLM15}, thus we only specify where it differs.
		Using similar notation, we may represent each walk $W\in \cW^{2q,z}(v,e)$ as a sequence $A_{1,1},\ldots, A_{1,z},\ldots,A_{2q,z}$. We explore the sequence in lexicographic order and think of index $(i,\ell)$ as a time. We say $A_{i,\ell}$ is a first time if the target endpoint of the edge did not appear in the sequence before. The set of first time edges form a tree with vertex set $\Set{1,\ldots, v}$. The distinct edges in $W$ not in the tree are thus $\epsilon=e-v+1$. We use the same encoding of \cite{BordenaveLM15}.
		In particular we encode  long cycling times in the same way. For short cycling times we use them same encoding, however as we may have up to $2^t$ cycles in each subsequence $A_{i,1},\ldots, A_{i,z}$, we further need to specify which among the possible paths we are going to take for each short cycling time, there are $2^{2t}$ possible such paths. Finally, we may have up to $t$ short cycling times in each walk and up to $\epsilon \cdot t$ long cycling times. The result follows.
	\end{proof}
\end{lemma}

The next lemma shows that graphs sampled from a distribution in $\cD_{d, \gamma}$ for $\gamma\geq O(\log^2n)$ have small average degree.%\Tnote{Is it better than simply markov with expectation?}

\begin{lemma}\label{lemma:avg-degree-sparse-dependent-graph}
	Let $n$ be an integer, $d>0$, $\gamma \geq O(d+\log n)$ and consider a distribution $P_{d,\gamma}\in \cD_{d,\gamma}$. Then for $\mathbf{A}\sim \mathcal{P}_{d,\gamma}$, with probability $0.999$
	\begin{align*}
		\Tr D(\mathbf{A})\leq O(nd)\,.
	\end{align*}
	\begin{proof}
		The expected average degree of the graph $\mathbf{G}$ associated with $\mathbf{A}$ is $d$. By linearity of expectation we thus have $\E 	\Tr D(\mathbf{A})= nd$. By Markov's inequality the result follows.
		% Previous proof
		% Was it unnecessariyl complicated?
		%It suffices to compute the average degree of the graph $\mathbf{G}$ associated with $\mathbf{A}$.
		%Now for any $t>0$ and $v\in [n]$,
		%\begin{align}\label{eq:appendix-concentration-degree}
		%	\bbP\Brac{\deg_{\mathbf{G}}(v)\geq t}\leq \bbP\Brac{\exists v_i,\ldots, v_t\,:\, \prod_{s\in [t]}\mathbf{A}_{v v_s}\neq 0}\leq \Paren{\frac{d}{n}}^t\cdot \binom{n}{t}\leq \Paren{\frac{ed}{t}}^t\,.
		%\end{align}
		%Now fix $t=10000$. For any $i=1,2,\ldots, \gamma/t$ denote by $\cE_i$ the event that 
		%$$\Card{\Set{v\in[n]\suchthat t\cdot (i-1)\leq\deg_{\mathbf{G}}(v)\leq t\cdot i }} \geq n\Paren{\frac{ed}{\sqrt{t\cdot i}}}^{t\cdot i}\,.$$ By Markov's inequality for any $i\leq i\leq \gamma/t$ we have $\bbP \Brac{\cE_i}\leq \Paren{t\cdot i}^{\frac{t\cdot i}{2}}\,.$
		%Thus by union bound the event $\cE:=\bigcup_{i\leq \gamma/t}\cE_i$ is verified with probability at most $0.0001$. As by \cref{eq:appendix-concentration-degree} no vertex has degree larger than $O(\log n)$, it follows that with probability $0.999$ the average degree of $\mathbf{G}$ is at most $O(d)$.
	\end{proof}
\end{lemma}

\subsection{Deferred proofs of \cref{section:random-3xor}}
We start by proving the rough bound on $\mathbf{A}''$ of \cref{lemma:cut-norm-residual-A}.

\begin{lemma}[Restatement of \cref{lemma:cut-norm-residual-A}]
	Consider the settings of \cref{lemma:cut-norm-A}. 
	Let $\mathbf{A}'$ as defined in \cref{eq:preprocessed-A} and let $\mathbf{A}''= \mathbf{A}-\mathbf{A}'$.
	Then with probability $1-o(1)$ 
	\begin{align*}
		\Normio{\mathbf{A}''}\leq n^{k-1-\Omega(1)}\cdot O(p\cdot n^{k/2})\,.
	\end{align*}
	\begin{proof}
		Since each entry in $\mathbf{A}''$ is either zero or at least $\Omega(1)$ in absolute value, we have $\Normio{\mathbf{A}''}\leq O\Paren{\Snorm{\mathbf{A}''}_F}$.
		We may decompose $\Normf{\mathbf{A}''}^2$ as
		\begin{align}\label{eq:decompose-A2}
			\Normf{\mathbf{A}''}^2 =& \sum_{\substack{\alpha_1,\alpha_2,\beta_1,\beta_2\in[n]^{(k-1)/2}\\\text{s. t. }S(\alpha_1,\alpha_2)=S(\beta_1,\beta_2)}} \mathbf{A}''_{(\alpha_1,\beta_1),(\alpha_2,\beta_2)} + \sum_{\substack{\alpha_1,\alpha_2,\beta_1,\beta_2\in[n]^{(k-1)/2}\\\text{s. t. }k-1>\Card{S(\alpha_1,\alpha_2)\cap S(\beta_1,\beta_2)}>1\\
   }} \mathbf{A}''_{(\alpha_1,\beta_1),(\alpha_2,\beta_2)}
		\end{align}
		We have less than $n^{k-1}(k-1)!$ elements in the first sum and less than $n^{2k-3}(k-1)!$ elements in the second one.
		To bound the first sum, notice that for any $t\geq 1$ and 
		and $\alpha,\beta \in [n]^{(k-1)/2}$
		\begin{align*}
			\bbP \Paren{\Paren{\mathbf{A}''_{(\alpha,\alpha),(\beta,\beta)}}^2=t^2}&= \bbP \Paren{\Paren{\sum_{\ell\in[n]}T_{(\alpha,\beta,\ell)}^2}^2=t^2} = \\
			&= \bbP \Paren{\Abs{\sum_{\ell\in[n]}T_{(\alpha,\beta,\ell)}^2}=t} \\
			&\leq \binom{n}{t}(1-p)^{n-t}p^t\\
			&\leq \binom{n}{t}p^t\\
			&\leq \Paren{\frac{e\cdot n\cdot p}{t}}^t\,.
		\end{align*}
		We split the contribution of entries based on their magnitude and bound their number using Markov's inequality. For any $t\geq 1$, denote by $\cE_t$ the event that 
		\[
		\sum_{\substack{\alpha_1,\alpha_2,\beta_1,\beta_2\in[n]^{(k-1)/2}\\\text{s. t. }S(\alpha_1,\alpha_2)=S(\beta_1,\beta_2)}} \Paren{\mathbf{A}''_{(\alpha_1,\alpha_2),(\beta_1,\beta_2)}}^2\iverson{\mathbf{A}''_{(\alpha_1,\alpha_2),(\beta_1,\beta_2)}=t}\leq 100\cdot n^{k-1}\cdot(k-1)! \cdot p^{1/10}\cdot 2^{-t}\,,
		\]
		where the Iverson brackets denote the indicator function.
		Let $\bar{\cE}_t$ be its complement event. 
		Notice that if $\cE_t$ is verified for all $t\geq 1$, then
		\begin{align*}
			\sum_{\substack{\alpha_1,\alpha_2,\beta_1,\beta_2\in[n]^{(k-1)/2}\\\text{s. t. }S(\alpha_1,\alpha_2)=S(\beta_1,\beta_2)}} \Paren{\mathbf{A}''_{(\alpha_1,\beta_1),(\alpha_2,\beta_2)}}^2
			&\leq 10^3\cdot  n^{k-1}\cdot(k-1)! \cdot p^{1/10}\\
			&\leq n^{(k-1)/2-\Omega(1)}(k-1)!\cdot O(p\cdot n^{k/2})\,,
		\end{align*}
		where in the last step we used the assumption that $p\geq n^{-k/2}$. We  verify that with high probability this intersection event happens.	By Markov's inequality
		\begin{align*}
			\bbP \Paren{\bar{\cE}_t}\leq \Paren{2e\cdot n \cdot p}^t\cdot \frac{100}{p^{1/10}}\,.
		\end{align*}
		Thus 
		\begin{align*}
			\bbP \Paren{\underset{t\geq 1}{\bigcup}\bar{\cE}_t}\leq O(n\cdot p^{9/10})\leq n^{-\Omega(1)}\,.
		\end{align*}
		We focus next onto the second sum in \cref{eq:decompose-A2}. Let $\alpha_1,\alpha_2,\beta_1,\beta_2\in[n]^{(k-1)/2}$ chosen accordingly, then
		\begin{align*}
			\E \Brac{\Paren{\mathbf{A}''_{(\alpha_1,\beta_1),(,\alpha_2,\beta_2)}}^2} &= \sum_{\ell, \ell'\in [n]}\E \mathbf{T}_{\alpha_1\alpha_2\ell}\mathbf{T}_{\beta_1\beta_2\ell}\mathbf{T}_{\alpha_1\alpha_2\ell'}\mathbf{T}_{\beta_1\beta_2\ell'}\\
			&= \sum_{\ell\in [n]}\E \mathbf{T}_{\alpha_1\alpha_2\ell}^2\mathbf{T}_{\beta_1\beta_2\ell}^2\\
			&\leq np^2\,.
		\end{align*}
		where in the second step we used independence of the entries of $\mathbf{T}$.
		As before, by Markov's inequality the result follows.
		%\begin{align*}
		%	\bbP \Paren{\sum_{\substack{\alpha_1,\alpha_2,\beta_1,\beta_2\in[n]^{(k-1)/2}\\\text{s. t. }\Card{S(\alpha_1,\alpha_2)\cap S(\beta_1, \beta_2)}=\frac{k-3}{2}}} \Paren{\mathbf{A}''_{(\alpha_1,\beta_1),(\alpha_2,\beta_2)}}^2\geq t\cdot n^k(k-1)!}\leq \frac{np^2}{t}\,.
		%\end{align*}
		%Choosing $t= n^{1.1}p^2$ the result follows.
	\end{proof}
\end{lemma}

Next we show that the trace of $D(\mathbf{A})$, for $\mathbf{A}'$ as in defined in \cref{eq:preprocessed-A} and the associated $D$ as defined in \cref{section:general-ihara-bass}, concentrates around its expectation.

\begin{lemma}\label{lemma:avg-degree-sparse-hyper-graph}
	Consider the settings of \cref{lemma:cut-norm-preprocessed-A}. Let $\mathbf{A}'$ as defined in \cref{eq:preprocessed-A}  and let $D$ be the associated matrix as defined in \cref{section:general-ihara-bass}. Then with probability at least $1-10^4$
	\begin{align*}
		\Tr D(\mathbf{A}')\leq O\Paren{p^2\cdot n^{2k-1}}\,.
	\end{align*}
	\begin{proof}
		By linearity
		\begin{align*}
			\E \Tr D(\mathbf{A}') &= \Tr \E D(\mathbf{A}')= \sum_{(\alpha_1,\beta_1)\in [n]^{k-1}}\sum_{(\alpha_2,\beta_2)\in [n]^{k-1}}\E\Abs{\mathbf{A}'_{(\alpha_1,\beta_1)(\alpha_2,\beta_2)}}\,.
		\end{align*}
		As $\E \Abs{\mathbf{A}'_{(\alpha_1,\beta_1)(\alpha_2,\beta_2)}}=O(p^2n)$, by Markov's inequality the result folllows.
	\end{proof}
\end{lemma}

We restate and prove \cref{lemma:rewriting-backtracking-xor}.

\begin{lemma}[Restatement of \cref{lemma:rewriting-backtracking-xor}]
	Consider the settings of \cref{theorem:bound-preporcessed-A-via-Ihara-Bass}. Let $W\in \bnbw{}{2q,z}$. Then for any term in $\overline{\annoying}(W)$
	\begin{align}
		\E &\prod_{i=1}^{2q} \Brac{\Abs{\sum_{\ell_1^i}\mathbf{T}_{(\alpha_1^i\alpha_2^i\ell_1^i)}\mathbf{T}_{(\beta_1^i\beta^i_2\ell^i_1)}}\Paren{\prod_{\substack{s=2}}^{z-1}\mathbf{T}_{(\alpha^i_s \alpha_{s+1}^i\ell_{s}^i)} \mathbf{T}_{(\beta_{s}^i \beta_{s+1}^i\ell_{s}^i)}}}\nonumber\\
		&\leq C^{2q} \cdot \E \prod_{i=1}^{2q} \Brac{\Paren{\sum_{\ell_1^i}\mathbf{T}_{(\alpha_1^i\alpha_2^i\ell_1^i)}\mathbf{T}_{(\beta_1^i\beta^i_2\ell^i_1)}}^2\Paren{\prod_{\substack{s=2}}^{z-1}\mathbf{T}_{(\alpha^i_s\alpha_{s+1}^i\ell_{s}^i)} \mathbf{T}_{(\beta_{s}^i \beta_{s+1}^i\ell_{s}^i)}}}\,,\nonumber
	\end{align}
	for some constant $C>0$.
	\begin{proof}
		We may rewrite the left hand side
		\begin{align*}
			\E &\prod_{i=1}^{2q}\underbrace{\Abs{\sum_{\ell_1^i\in[n]}\mathbf{T}_{(\alpha_1^i\alpha_2^i\ell_1^i)}\mathbf{T}_{(\beta_1^i\beta^i_2\ell^i_1)}}}_{=:\mathbf{L}_1^i}
			\cdot\underbrace{\Paren{\prod_{\substack{s=2}}^{z-1}\mathbf{T}_{(\alpha^i_s\alpha_{s+1}^i \ell_{s}^i)} \mathbf{T}_{(\beta_{s}^i \beta_{s+1}^i\ell_{s}^i)}}}_{=:\mathbf{L}^i_s}
		\end{align*}
		and the right hand side
		\begin{align*}
			\E &\prod_{i=1}^{2q}\underbrace{ \Paren{\sum_{\ell_1^i\in [n]}\mathbf{T}_{(\alpha_1^i\alpha_2^i\ell_1^i)}\mathbf{T}_{(\beta_1^i\beta^i_2\ell^i_1)}}^2}_{=:\mathbf{R}_1^i}
			\cdot\underbrace{\Paren{\prod_{\substack{s=2}}^{z-1}\mathbf{T}_{(\alpha^i_s \alpha_{s+1}^i\ell_{s}^i)} \mathbf{T}_{(\beta_{s}^i \beta_{s+1}^i\ell_{s}^i)}}}_{=:\mathbf{L}^i_s}\,.
		\end{align*}
		Opening up the sums in $\mathbf{L}_2^1,\ldots,\mathbf{L}_{z-1}^{2q}$ we get sums of terms of the form
		\begin{align*}
			\text{on the LHS:}&\quad \mathbf{L}_1^1\cdots \mathbf{L}_1^{2q}\cdot \underbrace{\mathbf{T}_{(\alpha\alpha'\ell)}\cdots}_{=:\mathbf{S}}\\
			\text{on the RHS:}&\quad  \mathbf{R}_1^1\cdots \mathbf{R}_1^{2q}\cdot \underbrace{\mathbf{T}_{(\alpha\alpha'\ell)}\cdots}_{=\mathbf{S}}\,.
		\end{align*}
		By linearity of expectation we may consider each such element independently. 
		By independence of the entries of $\mathbf{T}$ if there is a term in $\mathbf{S}$ of odd degree that does not appear in $\mathbf{L}^1_1\cdots\mathbf{L}_1^{2q}$ we have 
		\begin{align*}
			\E \mathbf{L}^1_1\cdots\mathbf{L}_1^{2q}\cdot \mathbf{S}\leq 0\leq \E \mathbf{R}^1_1\cdots\mathbf{R}_1^{2q}\cdot \mathbf{S}\,.
		\end{align*}
		It remains to consider the case in which each term in $\mathbf{S}$ has even degree.
		But then for any possible realization $T$ of $\mathbf{T}$, as all entries are non-zero entries are bounded away from $0$, we have
		\begin{align*}
			\E \Brac{\mathbf{L}_1^1\cdots \mathbf{L}_1^{2q}\cdot\mathbf{S}\suchthat\mathbf{T}=T}\leq C^{2q}\E \Brac{\mathbf{R}_1^1\cdots \mathbf{R}_1^{2q}\cdot\mathbf{S}\suchthat\mathbf{T}=T}\,,
		\end{align*}
		for some constant $C>0$.
		This concludes the proof.
	\end{proof}
\end{lemma}

\section{Additional tools}\label{section:additional-tools}

\begin{fact}\label{fact:symmetric-tensor-equivalent}
	Let $T\in \Paren{\R^n}^{\otimes k}$ be a tensor. Let $\tilde{T}\in \Paren{\R^n}^{\otimes k}$ be its symmetrization, that is for any multi-index $\alpha \in [n]^k$
	\begin{align*}
		\tilde{T}_{\alpha} = \frac{1}{k!}\sum_{\alpha'\in \Pi(\alpha)} T_{\alpha'}\,,
	\end{align*} 
	where $\Pi(\alpha)$ is the set of permutations of $\alpha$.
	Then for any $x\in\R^n$
	\begin{align*}
		\iprod{T, \tensorpower{x}{k}} = \iprod{\tilde{T}, \tensorpower{x}{k}}\,.
	\end{align*}
	\begin{proof}
		Fix a mapping $\pi:[k]\rightarrow [k]$, then we have
		\begin{align*}
				\iprod{T, \tensorpower{x}{k}} = \sum_{\alpha\in [n]^k}T_\alpha x^{\alpha} = \sum_{\alpha\in [n]^k}T_{\pi(\alpha)} x^{\alpha}\,.
		\end{align*}
		Repeating the reasoning for all $k!$ possible permutations the result follows.
	\end{proof}
\end{fact}

Next we provide some matrix concentration inequalities.

\begin{theorem}[Matrix Bernstein, \cite{Tropp12}]\label{theorem:matrix-bernstein}
	Consider a finite sequence $\Set{\mathbf{M}_\ell}$ of independent, random, matrices with dimensions $n_1\times n_2$. Assume that each random matrix satisfies
	\begin{align*}
		\E \mathbf{M}_\ell = \mathbf{0}\quad \text{ and }\quad \Norm{\mathbf{M}_\ell}\leq R \text{ almost surely.}
	\end{align*}
	Define 
	\begin{align*}
		\sigma^2 :=\max \Set{\Norm{\sum_\ell\E \mathbf{M}_\ell\transpose{\mathbf{M}}_\ell}\,, \Norm{\sum_\ell\E \transpose{\mathbf{M}}_\ell\mathbf{M}_\ell}}\,.
	\end{align*}
	Then, for all $t\geq 0$,
	\begin{align*}
		\bbP \Brac{\Norm{\sum_\ell \mathbf{M}_\ell}\geq t}\leq (n_1+n_2)\cdot \exp\Paren{\frac{-t^2/2}{\sigma^2+Rt/3}}\,.
	\end{align*}
\end{theorem}

%\begin{lemma}\label{lemma:matrix-concentration}
%	Let $\mathbf{M}$ be a sum of $m$ independent matrices $\mathbf{M}_1,\ldots, \mathbf{M}_m$ with dimension $n_1\times n_2$ each satisfying
%	\begin{align*}
%		\E \Brac{\Paren{\mathbf{M}_\ell}_\ij} &= 0\\
%		\bbP \Brac{\Paren{\mathbf{M}_\ell}_\ij\neq 0}&\leq q\\
%		\Abs{\Paren{\mathbf{M}_\ell}_\ij}&\leq 1\\
%		\text{for }(i,j)\neq ('i,j')\quad \E \brac{\Paren{\mathbf{M}_\ell}_\ij\Paren{\mathbf{M}_\ell}_{i'j'}}&=0
%	\end{align*} 
%	Then with probability $1-n^{-\Omega(1)}$
%	\begin{align*}
%		\Norm{\mathbf{M}}\leq O\Paren{1+\sqrt{m\cdot q\cdot \max\Set{n_1\,,n_2}}}\sqrt{\log(n_2)}\,.
%	\end{align*}
%	\begin{proof}
%		Without loss of generality let's assume $n_1\leq n_2$.
%		For any $\ell\in [m]$ and distinct $i, j\in [n_2]$ we have $\E\iprod{\Paren{\mathbf{M}_\ell}_{i,-}, \Paren{\mathbf{M}_\ell}_{j,-}}$. On the other hand $\E\iprod{\Paren{\mathbf{M}_\ell}_{i,-}, \Paren{\mathbf{M}_\ell}_{i,-}}=q\cdot n_2$. Similarly for $i\in [n_1]$ we have $\E\iprod{\Paren{\mathbf{M}_\ell}_{-,i}, \Paren{\mathbf{M}_\ell}_{-,i}}=q\cdot n_1$. We thus have
%		\begin{align*}
%			\sigma^2 :=\max \Set{\Norm{\sum_\ell\E \mathbf{M}_\ell\transpose{\mathbf{M}}_\ell}\,, \Norm{\sum_\ell\E \transpose{\mathbf{M}}_\ell\mathbf{M}_\ell}}\leq m\cdot q\cdot n_2\,.
%		\end{align*}
%		Applying  \cref{theorem:matrix-bernstein} with $t=10\sqrt{\log(n_2)}\Paren{1+\sqrt{m\cdot q\cdot n_2}}$ the result folllows.
%	\end{proof}
%\end{lemma}

\end{document}